\documentclass[sn-vancouver,default,iicol]{sn-jnl}

\usepackage{natbib}

\jyear{2022}%

\theoremstyle{thmstyleone}%
\usepackage{mathtools}
\usepackage{amssymb}
\usepackage{multirow}
\usepackage{longtable}
\usepackage{arydshln}
\usepackage{amsmath}
\usepackage{braket}
\usepackage{mathrsfs}
\usepackage{graphicx}
\usepackage{cuted}
\theoremstyle{thmstyletwo}%

\theoremstyle{thmstylethree}%

\renewcommand{\vec}[1]{\mathbf{#1}}
\newcommand{\beq}{\begin{equation}}
\newcommand{\eeq}{\end{equation}}
\newcommand{\qv}{{\bf q}}

\begin{document}



\title{Model-free interpretation of X-ray Thomson scattering measurements}


\author*[1]{\fnm{Thomas} \sur{Gawne}}
\email{t.gawne@hzdr.de}


\author[2]{\fnm{Jan} \sur{Vorberger}}

\author[2]{\fnm{Zhandos}~\sur{Moldabekov}}

\author[1,3]{\fnm{Hannah~M.}~\sur{Bellenbaum}}

\author*[2,1]{\fnm{Tobias} \sur{Dornheim}}\email{t.dornheim@hzdr.de}

\affil[1]{\orgname{Center for Advanced Systems Understanding (CASUS) at Helmholtz-Zentrum Dresden-Rossendorf (HZDR)}, \orgaddress{\city{G\"orlitz}, \postcode{D-02826}, \country{Germany}}}

\affil[2]{\orgname{Institute of Radiation Physics at Helmholtz-Zentrum Dresden-Rossendorf (HZDR)}, \orgaddress{\city{Dresden}, \postcode{D-01328},  \country{Germany}}}

\affil[3]{\orgname{Institut f\"ur Physik, Universit\"at Rostock}, \orgaddress{\city{Rostock}, \postcode{D-18057},  \country{Germany}}}



\abstract{ 
X-ray Thomson scattering (XRTS) has emerged as a widely used diagnostics for extreme states of matter in a great variety of situations, and over a broad range of parameters. The standard approach for the interpretation of XRTS measurements is given by the forward modeling approach, where the electronic dynamic structure factor $S_{ee}(\mathbf{q},\omega)$ is computed from a suitable theoretical model, convolved with the combined source-and-instrument function, and then matched with the experimental observation, treating a-priori unknown parameters such as the mass density, temperature and ionization state as free fit parameters. Very recently, it has been suggested that this inherent model dependence can be avoided by analyzing XRTS spectra in terms of the imaginary-time correlation function (ITCF) $F_{ee}(\mathbf{q},\tau)$ [Dornheim \textit{et al.}, \textit{Nature Commun.}~\textbf{13}, 7911 (2022)], giving one model-free access to the temperature, normalization, Rayleigh weight, as well as a number of other properties.
Here, we present a comprehensive review article on these developments, including accessible discussions of the method's theoretical background in terms of Feynman's imaginary-time path integral picture of statistical mechanics as well as its remaining limitations, in particular with respect to the source-and-instrument function of the experimental set-up. In addition, we discuss new chances for the further development of this framework by utilizing emerging capabilities for high-repetition XRTS experiments with meV resolution over spectral ranges of tens of eV at state-of-the-art x-ray free electron laser (XFEL) facilities such as the European XFEL in Germany.
}




\maketitle

\section{Introduction}

Understanding matter at extreme densities, temperatures, and pressures has emerged as a highly active frontier at the interface of plasma, solid state and atomic physics, quantum chemistry, and material science~\cite{vorberger2025roadmapwarmdensematter,fortov_review,drake2018high}.
Such \emph{warm dense matter} (WDM)~\cite{wdm_book} and related extreme states are ubiquitous throughout our universe and naturally occur in a great variety of astrophysical objects, including giant planet interiors~\cite{wdm_book,Benuzzi_Mounaix_2014,Brygoo2021}, brown dwarfs~\cite{becker,saumon1} and white dwarf atmospheres~\cite{Kritcher_Nature_2020,SAUMON20221}. Indeed, material properties such as the equation of state (EOS) of various materials~\cite{Militzer_PRE_2021} play an important role for the interpretation of various observations, e.g., the Juno mission measuring various observables of Jupiter such as its gravitational moments~\cite{Bolton2017}.
On Earth, extreme states of matter are becoming increasingly important for cutting-edge technological applications. One branch of examples comes from material science, e.g., the man-made creation of nanodiamonds from ambient plastic at high pressures and moderate temperatures~\cite{Kraus2016,Kraus2017,Frost2024}.
Moreover, theoretical simulations have predicted the formation of novel materials with exotic properties for hitherto unattained parameters~\cite{vorberger2025roadmapwarmdensematter}, such as the meta-stable BC8 phase of carbon that is expected to be harder than ordinary diamond~\cite{Ramakrishna_2020,Nguyen-Cong2024}.
The holy grail of high-pressure physics is arguably given by the liquid--liquid insulator to metal phase transition in hydrogen, which remains an active subject of investigation~\cite{Pierleoni_PNAS_2016,Dias_Silvera_Science_2017,Karasiev2021,vorberger2026vanderwaalsexchangecorrelationfunctionalshigh}, see also the recent review by Bonitz \textit{et al.}~\cite{Bonitz_POP_2024} and references therein.
Finally, we mention the recent breathtaking progress in the burgeoning field of inertial fusion energy~\cite{AbuShawareb_PRL_2022,AbuShawareb_PRL_2024,Gopalaswamy2024,Hurricane_RevModPhys_2023,Zylstra2022}, where in traditional set-ups both the fusion fuel and the surrounding ablator material have to traverse the WDM regime in a controlled way to reach ignition~\cite{hu_ICF}.

Due to all of these applications, WDM is nowadays very actively realized in large research facilities using different methods~\cite{falk_wdm}, including isochoric x-ray heating~\cite{Pascarelli_NatureReviews_2023,Sperling_PRL_2015,kraus_xrts,Glenzer_PRL_2007,DOPPNER2009182}, optically driven shocks~\cite{bespalov2026experimentalevidencebreakdownuniformelectrongas,Preston_APL_2019,Fletcher2015}, spherical implosions~\cite{Tilo_Nature_2023,Kraus_PRE_2016,Fletcher_PRL_2014,Kritcher_PRL_2011}, particle heating~\cite{Fortov_FAIR_2012,Luetgert_MRE_2024}, and Z-pinches~\cite{Valenzuela_SciRep_2018,Yang_MRE_2025}.
In this regard, a great challenge is given by the accurate diagnostics. In particular, a suitable diagnostics must be capable of simultaneously dealing with extreme densities and temperatures as well as the ultrafast time scales, which are typically on the order of $t\sim10^{-15}-10^{-9}\,$s, depending on whether one is interested in the electronic time scales (femtoseconds), ionic timescales (picoseconds) or in the collective behavior evolving on hydrodynamic scales (nanoseconds).
Over a quarter of a century ago, Landen \textit{et al.}~\cite{LANDEN2001465} have suggested to use x-ray Thomson scattering (XRTS)~\cite{siegfried_review}, which, in principle, meets all of these demands.
By probing the dynamic structure factor $S_{ee}(\mathbf{q},\omega)$ of the electrons (see Eq.~(\ref{eq:DSF}) below), where $\hbar\omega$ and $\hbar q$ are the energy loss and photon transfer of the scattered photon, respectively, XRTS gives one detailed insights into the nanophysics of the probed sample, which is useful to obtain various components of the EOS such as mass density, temperature and ionization~\cite{Falk_PRL_2014,Tilo_Nature_2023,Gregori_PRE_2003,bohme2023evidencefreeboundtransitionswarm,Kasim_POP_2019} as well as dedicated physics effects such as miscibility~\cite{Frydrych_NatComm_2020} and collisional damping~\cite{Neumayer_PRL_2010,Sperling_2017,bespalov2026experimentalevidencebreakdownuniformelectrongas}.
On the one hand, XRTS has emerged as a de-facto standard diagnostics for WDM and is actively being used at a very diverse collection of facilities, including the European XFEL~\cite{bespalov2026experimentalevidencebreakdownuniformelectrongas,Smid_SciRep_2026,Gawne_PRB_2024}, FLASH~\cite{Zastrau,Toleikis_2010,Faustlin_PRL_2010} and GSI~\cite{Luetgert_MRE_2024,Hesselbach_MRE_2024,Kraus_POP_2015} in Germany, LCLS~\cite{Fletcher2015,Fletcher_Frontiers_2022,Martin_POP_2025}, the National Ignition Facility (NIF)~\cite{Tilo_Nature_2023,Kraus_PRE_2016,MacDonald_POP_2023}, the OMEGA laser facility~\cite{Glenzer_POP_2003,Glenzer_PRL_2007,Poole_PRR_2024}, and the Sandia Z-machine~\cite{AO201626} in the US, the VULCAN laser~\cite{Kugland_PRE_2009,GarciaSaiz2008,Gregori_AIP_2002} in the UK, and the Shenguang-II laser facility in China~\cite{shi2025firstprinciplesanalysiswarmdense,15-20220361,Lv_POP_2019}; see also the recent comprehensive overview by Dornheim \textit{et al.}~\cite{dornheim2026overviewxraythomsonscattering}.
In addition to being interesting in its own right, XRTS is nowadays also often used as a secondary diagnostics for the measurement of other properties, e.g., stopping power~\cite{Lahmann_PPCF_2023}, x-ray transmission~\cite{Tilo_Nature_2023} and novel x-ray photon correlation spectroscopy techniques~\cite{Heaton_PRR_2025}.
On the other hand, the interpretation of a given XRTS measurement was usually based on a particular theoretical model for $S_{ee}(\mathbf{q},\omega)$. Unfortunately, the complexity of describing WDM~\cite{wdm_book,new_POP,review,vorberger2025roadmapwarmdensematter} made it necessary to employ de-facto uncontrolled approximations on different levels of sophistication, starting with semi-empirical chemical \emph{Chihara} models~\cite{Chihara_1987,Gregori_PRE_2003,bohme2023evidencefreeboundtransitionswarm} that rely on an effective decomposition into separate populations of \emph{bound} and \emph{free} electrons and going up to computationally substantially more involved time-dependent density functional theory (TD-DFT) simulations~\cite{Baczewski_PRL_2016,baczewski2021predictionsboundboundtransitionsignatures,Schoerner_PRE_2023,Moldabekov_MRE_2025,Mo_PRL_2018,Moldabekov_filter}, which are not based on the chemical decomposition but have to approximate the dynamic exchange--correlation (XC) potential (real-time TD-DFT) or the static XC-functional and dynamic XC-kernel (linear-response TD-DFT).
The extracted system parameters of the probed sample can thus strongly depend on the employed model, see, e.g., Refs.~\cite{Sperling_PRL_2015,Mo_PRL_2018,Dornheim_PRL_2020_ESA,Dornheim_T_2022} and Refs.~\cite{Tilo_Nature_2023,bohme2023evidencefreeboundtransitionswarm,Dornheim_NatComm_2025,Dornheim_POP_2025,schwalbe2025staticlineardensityresponse,Dharma_wardana_PRE_2025} for different interpretations of LCLS and NIF data sets.

This rather unsatisfactory situation has sparked important new developments in various aspects of XRTS, most notably the development of the model-free imaginary-time correlation function (ITCF) framework presented in Refs.~\cite{Dornheim_T_2022,Dornheim_T2_2022,Dornheim_MRE_2023,Dornheim_PTRS_2023,Dornheim_review}.
The key idea is to switch from the usual frequency representation in terms of the dynamic structure factor $S_{ee}(\mathbf{q},\omega)$ to the ITCF $F_{ee}(\mathbf{q},\tau)$ [given by the usual intermediate scattering function~\cite{sheffield2010plasma} $F(\mathbf{q},t)$ evaluated at an imaginary time $t=-i\hbar\tau$ with $\tau=\in[0,\beta]$ and $\beta=1/k_\textnormal{B}T$ the inverse temperature], which are related by a two-sided Laplace transform, see Eq.~(\ref{eq:ITCF}) below.
This combines a number of key advantages: (i) the Laplace transform allows for a very stable deconvolution [Eq.~(\ref{eq:deconvolution})] with respect to the source-and-instrument function (SIF). This is crucial for model-free access to physical properties of the probed system without the need for any forward modeling; (ii) the Laplace transform acts as an effective noise filter. The ITCF method is, thus, available even for comparably noisy data, see, e.g., the ITCF analysis of the pioneering plasmon measurement in isochorically heated beryllium by Glenzer \textit{et al.}~\cite{Glenzer_PRL_2007} in Refs.~\cite{Dornheim_T_2022,Dornheim_T2_2022} and the ITCF analysis of the XRTS data set for CH foams by Martin \textit{et al.}~\cite{Martin_POP_2025} in Ref.~\cite{bohme2026correlationfunctionmetrologywarm}; (iii) the ITCF $F_{ee}(\mathbf{q},\tau)$ contains, by definition, exactly the same information as $S_{ee}(\mathbf{q},\omega)$, only in a different representation. In fact, some aspects such as the temperature [Eq.~(\ref{eq:ITCF_symmetry})] and static linear density response [Eq.~(\ref{eq:static_chi})] can be extracted more directly in the $\tau$-domain, whereas other aspects such as a plasmon width are substantially more obvious in the $\omega$-domain~\cite{Dornheim_MRE_2023}.

Having been introduced as a model-free thermometry approach in 2022~\cite{Dornheim_T_2022}, the ITCF method has been further developed to also give one model-free access to the normalization and electronic static factor $S_{ee}(\mathbf{q})$~\cite{Dornheim_SciRep_2024}, to the degree of non-equilibrium of the system~\cite{Vorberger_PLA_2024}, to the Rayleigh weight $W_\textnormal{R}(\mathbf{q})$ [Eq.~(\ref{eq:Wr}) below] that describes the degree of electronic localization around the ions~\cite{Dornheim_POP_2025}, and to the static linear density response [Eq.~(\ref{eq:static_chi})]~\cite{schwalbe2025staticlineardensityresponse}.
The ITCF method is being actively used by groups around the world, and it has been applied to data sets measured at the NIF~\cite{Dornheim_NatComm_2025,Dornheim_SciRep_2024,schwalbe2025staticlineardensityresponse}, at OMEGA~\cite{Dornheim_T_2022,Dornheim_T2_2022,Schoerner_PRE_2023}, at the European XFEL~\cite{Dornheim_SciRep_2024,Smid_SciRep_2026}, at LCLS~\cite{Dornheim_T_2022,Dornheim_T2_2022,Bellenbaum_APL_2025} and at Shenguang-II~\cite{shi2025firstprinciplesanalysiswarmdense}.
Recent work by Gawne~\textit{et al.}~\cite{Gawne_ITCF_Ratio} has outlined the intriguing possibility for a model-free ITCF analysis \emph{without the need for explicit knowledge even of the SIF} based on simultaneous XRTS measurements at two (or more) separate scattering angles.
Finally, further extensions of the ITCF method, e.g., to the estimation of the frequency moments of $S_{ee}(\mathbf{q},\omega)$~\cite{Dornheim_moments_2023,Dornheim_MRE_2023} and for the study of other physics effects such as the \emph{roton type} feature~\cite{Chuna_JCP_2025,Dornheim_MRE_2023} that has been predicted to be observable, e.g., in XRTS measurements on heated hydrogen jet targets~\cite{Hamann_PRR_2023}, have already been suggested.

In the present work, we present a comprehensive review of all these aspects of the ITCF method, with an additional emphasis on its theoretical background in terms of Feynman's imaginary-time path integral picture of statistical mechanics~\cite{kleinert2009path} and on its remaining limitations.
We hope that this will help experienced practitioners to place the method into the proper context, and to decide if its application makes sense for a particular experimental scenario.
In addition, we hope that this overview will provide a valuable reference for newcomers to the field, and that it might help to give a fresh perspective onto a well established method.

The paper is organized as follows: In Sec.~\ref{sec:relevant}, we introduce relevant parameters such as the reduced temperature, density parameter, Fermi wavenumber, etc.
Sec.~\ref{sec:process} is devoted to the discussion of the XRTS measurement process, including details on the scattering cross section (\ref{sec:crosssection}), different mechanisms for the experimental broadening (\ref{sec:SIF}), as well as the idea behind realistic raytracing simulations (\ref{sec:RayTracing}).
In Sec.~\ref{sec:models}, we give a brief overview of the most important models for the dynamic structure factor $S_{ee}(\mathbf{q},\omega)$ on various levels of complexity, touching upon linear response theory (\ref{sec:LRT}), the Chihara decomposition (\ref{sec:Chihara}), time-dependent density functional theory (\ref{sec:TDDFT}) and the analytic continuation of path integral Monte Carlo simulation results (\ref{sec:AC}).
Sec.~\ref{sec:model_free} then discusses the gamut of aspects of the titular topic of this review article: the model-free XRTS diagnostics, including the deconvolution in the Laplace domain (\ref{sec:deconvolution}), the physics directly encoded into the ITCF $F_{ee}(\mathbf{q},\tau)$ itself (\ref{sec:ITCF}),
thermometry (\ref{sec:thermometry}), detection and quantification of non-equilibrium (\ref{sec:non_eq}), the determination of the a-priori unknown absolute intensity and normalization (\ref{sec:moments}), the Rayleigh weight (\ref{sec:Rayleigh_weight}) and finally the static linear density response (\ref{sec:density_response}).
To further test the underlying assumptions, limits and capabilities of our approach, we present extensive raytracing results in Sec.~\ref{sec:RT_Examples}, focusing on the model-free thermometry (\ref{sec:RT_thermometry}), the extraction of the full ITCF (\ref{sec:RT_ITCF}), the normalization (\ref{sec:RT_SSF}), the detection of non-equilibrium (\ref{sec:RT_neq}), and on a novel idea for model-free thermometry that automatically eliminates source-and-instrument function effects (\ref{sec:noSIF}).
The paper is concluded by a summary and outlook in Sec.~\ref{sec:summary}.

\subsection{Relevant parameters\label{sec:relevant}}

For convenience and an approximate overview of the physical state of the system that is getting probed, a number of simple parameter exist that allow one to classify the different effects and parameter regimes~\cite{Ott2018}.

The Brueckner parameter $r_s=d/a_B$ compares the mean particle distance $d=(3/4\pi n)^{1/3}$ ($n$ being the particle density) to the Bohr radius $a_B$. For large densities, for which $r_s \sim 1$, the Brueckner parameter serves as a coupling parameter, weakly coupled states are accessed when $r_s\ll 1$. 

In general, and for WDM in particular, temperature influences the coupling strength between particles as well, and one introduces the coupling parameter $\Gamma=\langle V\rangle/\langle K\rangle$ as the ratio between mean potential and mean kinetic energy. For high densities and low temperatures, the parameter $\Gamma$ gives similar boundaries for weak and strong coupling, resp., as the Brueckner parameter. For low densities or high temperatures, it simplifies to the well-known classical expression $\Gamma=e^2/k_BTd$ with elementary charge $e$ and Boltzmann constant $k_B$.

Systems feature either Fermi-Dirac- or Bose-Einstein-statistics (also fictitious fractional statistics have enjoyed very recent popularity among theorists~\cite{Xiong_JCP_2022,Dornheim_JCP_2023,m1,dornheim2025taylorseriesperspectiveab}) but these can simplify to Boltzmann statistics for high enough temperatures or low enough densities. A useful parameter is therefore the ratio of temperature to Fermi-temperature $\Theta=T/T_F$. The system can be considered as non-degenerate for $\Theta\gg 1$. The Fermi temperature is derived from the Fermi energy $E_F=k_BT_F$, and the Fermi energy has the usual established connection to Fermi wavenumber and density $E_F=\hbar^2q_F^2/2m$, $q_F=(3\pi^2n)^{1/3}$. An alternative measure for the degeneracy of the system is given by a comparison of the mean particle distance and the particle's deBroglie wavelength $n\lambda_\beta^3=n(2\pi\hbar/k_BTm)^{3/2}$. $n\Lambda^3\ll 1$ signifies non-degenerate conditions, fo $n\Lambda^3\gg 1$ states are fully degenerate.

Even for non-degenerate states, the (binary) collisions between particles in the system might need a quantum description. The relevant quantum scattering parameter $\xi=e^2\sqrt{m/\hbar^2k_BT}$ becomes greater than unity if this is the case.

A typical energy scale in connection with the penetration and scattering of radiation in matter is given by the plasma frequency of the electrons $\omega_{pl}^2=4\pi e^2 n/ m_e$. Electromagnetic fields that shall penetrate the sample need a frequency higher that the plasma frequency. Also, the plasma frequency serves as a rough estimator of the location of the plasmon peaks in the scattering spectrum.

Whether a scattering experiment mainly probes single particle behavior in the target or is more susceptible to collective effects can be estimated by the parameter $\alpha=1/q\lambda_s$ that compares the wavelength of the probing radiation to a typical length scale of correlations. $\lambda_s$ is the screening length, which in a non-degenerate case is equal to the Debye length $\lambda_s^2=k_BT/4\pi e^2 n$. The wavenumber $q$ is related to the scattering angle $\theta$ probed in the experiment using radiation of wavelength $\lambda_i$: $q=4\pi/\lambda_i\sin(\theta/2)$. For large $q$, the collectivity parameter becomes small $\alpha\ll 1$, thus single particle properties are probed. For small $q$, i.e., small angles $\theta$, $\alpha$ becomes large ($\alpha>1$), meaning the particles in the system scatter the radiation collectively.


\section{XRTS measurement process\label{sec:process}}

\begin{figure}
    \centering
    \includegraphics[width=0.85\linewidth]{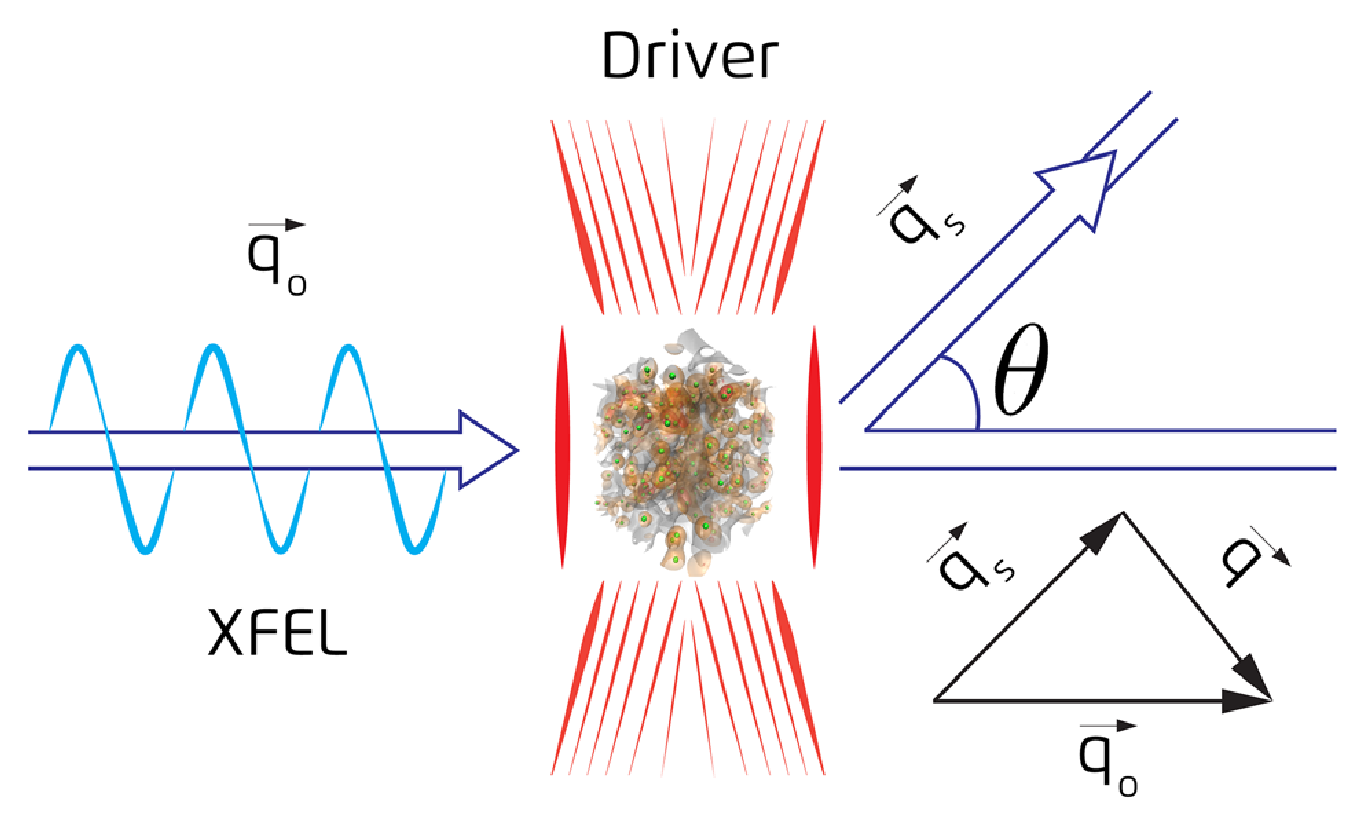}
    \caption{Schematic illustration of XRTS experiments. An x-ray source (here an XFEL, but in principle also a backlighter source, synchrotron, etc.) provides photons of incoming momentum $\mathbf{q}_0$. They get scattered in the sample (here a solid irradiated with an optical drive laser), and the scattering angle $\theta$ determines the momentum transfer $\mathbf{q}$, Eq.~(\ref{eq:approx_q}).
    Taken from Ref.~\cite{Dornheim_T2_2022} with the permission of the authors.}
    \label{fig:xrts_scheme}
\end{figure}

\subsection{Scattering cross section\label{sec:crosssection}}

The basic idea behind XRTS is sketched in Fig.~\ref{fig:xrts_scheme}: x-ray photons approach from the left side, penetrate the (potentially warm and dense) sample, and get scattered under a scattering angle $\theta$.
Throughout this work, we assume the scattering process to be non-relativistic. For a fully relativistic treatment and a derivation of the non-relativistic case from it see Ref.~\cite{Uwe_Events_2026}.
The energy arriving at a pixel of the detector after the incoming monochromatic x-rays have been scattered is given by~\cite{Dornheim_T2_2022}
\begin{align}
    E_{pixel}&=\int\limits_{0}^{t_{probe}}dt
    \int\limits_{E_s/\hbar}^{(E_s+\Delta E)/\hbar}
    d\omega_s\; \frac{\partial P_s}{\partial \omega_s}\nonumber\\
    &\approx \frac{\Delta E}{\hbar}\int\limits_{0}^{t_{probe}}dt\;
    \frac{\partial P_s}{\partial \omega_s}
    \Bigg\vert_{\hbar\omega_s=E_s+\Delta E/2}\,.
\end{align}
Here, we integrate over the time the detector is responsive (open) or over which the probe beam is switched on and account for the fact that each pixel is receptive to a small but finite energy range $\Delta E$. If the scattered power per unit frequency $\partial P_s/\partial \omega_s$ is constant over each pixel, the simplification of the 2nd line is possible.

The differential power spectrum might be expressed by the more general generalized scattering cross section
\begin{align}
    \frac{\partial P_s}{\partial \omega_s}=\int d{\cal V} d\Omega
    \left( I_0n_e\frac{\partial^2\tilde{\sigma}}{\partial\Omega\partial \omega_s}\right)\,.
\end{align}
Here, $I_0$ is the intensity of the incoming x-rays, $n_e$ is the mean electron number density in the volume element $\cal V$, and $\partial^2\tilde{\sigma}/\partial\Omega\partial \omega_s$ is the double-differential generalized scattering cross section per unit solid angle and per unit frequency. The latter might be expressed in the non-relativistic case as
\begin{align}
\frac{\partial^2\tilde{\sigma}}{\partial\Omega\partial \omega_s}=
\frac{\omega_s}{\omega_0}\frac{\partial\tilde{\sigma}}{\partial\Omega}
\Big\vert_T S_{ee}(\vec{q},\omega)\;,
\end{align}
in terms of the differential Thomson cross section $\partial\tilde{\sigma}/\partial\Omega\Big\vert_T$ and the total dynamic electron structure factor $S_{ee}(\vec{q},\omega)$.

The differential Thomson cross section features the classical electron radius $r_e=e^2/m_ec^2$, the geometric factor $G$ and the ratio of the frequencies of scattered and incoming x-ray beam
\begin{align}\label{eq:xrts_dxsec}
    \frac{\partial\tilde{\sigma}}{\partial\Omega}\Big\vert_T=
    \left(\frac{\omega_s}{\omega_0}\right)^nr_e^2G(\theta,\phi)\,.
\end{align}
The power $n$ of the ratio $(\omega_s/\omega_0)^n$ depends on the characterization of the detector as quantum or classical~\cite{Crowley_2013}. The nature of the employed detectors usually necessitates $n=2$. Thus, the entire relation is
\begin{align}\label{eq:xrts_power}
    \frac{\partial P_s}{\partial \omega_s}\approx I_0r_e^2G(\theta,\phi)d\Omega n_e{\cal V}\left(\frac{\omega_s}{\omega_0}\right)^2S_{ee}(\vec{q},\omega)\,.
\end{align}
It was used here that the scattering volume $\cal V$ is small in comparison to the distances to detector and probe source as well as that the state of matter in the volume is homogeneous. Thus the integrations reduce to simple multiplications with volume and solid angle $d\Omega$. The geometric factor $G$ accounts for the projection of the incident to the scattered x-ray polarizations
\begin{align}
    G(\theta,\phi)&=(\hat{\bf e_0}\cdot \hat{\bf e_s})^2\,,\\
    G(\theta,\phi)&=
    \begin{cases}
        1-\sin^2\theta\cos^2\phi\;: \mbox{(linearly polarized)}\,,\\
        \frac{1}{2}(1+\cos^2\theta)\;: \mbox{(unpolarized)}\,.
    \end{cases}
\end{align}

The general relation between the wavenumbers of the incoming and scattered radiation is
\begin{align}
    q=q_{full}=\sqrt{q_0^2+q_s^2-2q_0q_s\cos(\theta)}\,.
\end{align}
In the case of x-ray scattering with incoming photon energies in the range of several keV, the x-ray energy being also very much larger than the plasma frequency, and the energy of the scattered photons deviating from the incoming keV photons by merely several $10$ or $100$~eV, this can be simplified to
\begin{align}\label{eq:approx_q}
    q\approx q_{approx}=2q_0\sin(\theta/2)\,.
\end{align}
In this case, the scattering angle and spectral range are decoupled as they need to be for the imaginary time thermometry to work.


\subsection{Experimental Broadening of XRTS\label{sec:SIF}}

A measured scattering spectrum does not just reproduce the scattering power defined in Eq.~(\ref{eq:xrts_power}). Rather, the spectrum is broadened by the energy spectrum of the probing beam (the source function), and the instrument function of the spectrometer.

\subsubsection{The Source Function}

The source broadening is due to the probing beam having a finite spectral width. From Eq.~(\ref{eq:xrts_power}), it is trivial to see that if the probing beam has a spectrum $B(E_0)$, then the scattering spectrum $I_s(E_s)$ emerging from the target is
\begin{equation}
    \begin{aligned}
    I_s(E_s, \theta, \phi) &= I_0 r_e^2 G(\theta,\phi)d\Omega n_e {\cal V} E_s^2 \\
    &\times \int_0^\infty dE_0 \, B(E_0) \frac{S_{ee}({\bf q}, (E_0 - E_s)/\hbar)}{E_0^2}  \, ,
    \end{aligned}
\end{equation}
remembering that ${\bf q} = {\bf q}(\omega_0, \omega_s, \theta)$. Under the approximation~(\ref{eq:approx_q}), the scattering spectrum is just the convolution of $B(\hbar\omega_0)$ with $\omega_0^{-2} S_{ee}({\bf q}, \omega)$.

It is worth considering briefly the different types of probes used in WDM XRTS experiments, since they cover a range of resolutions, and reflect the practicalities of the facilities they are used at.

Backlighters have long been used as a source of hard x-rays, and today are most commonly seen at laser facilities~\cite{Kritcher_PRL_2011, Doeppner_RSI_2014, Fletcher_PRL_2014, MacDonald_RSI_2018, MacDonald_POP_2022, Tilo_Nature_2023, Luetgert_MRE_2024, Poole_PRR_2024, Dornheim_NatComm_2025}. They consist of a high-Z foil which is heated with a laser so that it emits specific x-ray emission lines -- for example, Zn He$\alpha$ backlighters are a popular choice at the NIF~\cite{Doeppner_RSI_2014,Tilo_Nature_2023,Dornheim_NatComm_2025} and OMEGA~\cite{Kritcher_PRL_2011,Fletcher_PRL_2014,MacDonald_RSI_2018}. Since the foil emits photons in all directions, to improve the resolution and reduce $q$-blurring (discussed in more detail below), a slit or pinhole is often use to quasi-collimate the divergent source before it is incident on the target (see e.g. Refs.~\cite{MacDonald_POP_2022,Tilo_Nature_2023}). The spectral width of the source depends on the particular lines being used, but they are generally several 10s~eV wide.
To minimise the spectral bandwidth, a filter with an appropriate binding energy can be used to absorb the photons above the edge; e.g. using a Cu filter (K-edge at 8979 eV), MacDonald~\textit{et al.}~\cite{MacDonald_POP_2022} developed a Zn He$\alpha$ backlighter with a spectral resolution of 48~eV, versus 87~eV without the filter.

At XFEL facilities, a number of different beam modes are available, but all generally provide higher resolution than a backlighter. The standard beam mode of an XFEL is self-amplified spontaneous emission (SASE), which is stochastic shot-to-shot with an approximately Gaussian envelope. SASE bandwidths for hard x-ray facilities are usually $\Delta E/E \sim$0.2--0.3\% (i.e. an 8.5~keV beam has a FWHM $\sim$17--25~eV)~\cite{Nagler_JSI_2015,Glenzer_JPB_2016,Zastrau_JSI_2021}, and up to $\sim0.5$\%~\cite{Yabashi_JSI_2015,Yabashi_AS_2017}. Beam energies in SASE mode vary from around 100s of~$\mu$J up to $\sim2$~mJ~\cite{Nagler_JSI_2015,Glenzer_JPB_2016,Zastrau_JSI_2021,Yabashi_JSI_2015,Yabashi_AS_2017}.
Using the self-seeding technique~\cite{Saldin_NIMPRSA_2001,Geloni_JMO_2011,Amann_NP_2012,Geloni_HXRSS_2019,Liu_PRAB_2019}, a narrow energy band of the SASE pulse can be selected for amplification. This produces a $\sim$1--3~eV FHWM spike that sits on top of a much weaker SASE pedestal. This greatly improves the resolution of the spectrum, at the cost of beam energy (generally $< 1$~mJ). The SASE pedestal also still contributes to the broadening, and can hide low intensity inelastic scattering features within the SASE width. To improve the resolution further, a monochromator can be used to greatly limit the range of photons that can reach the target. Depending on the monochromator setup used, beam widths ranging from $\sim 500$~meV~\cite{Gawne_PRB_2024} down to $\sim 30$~meV~\cite{McBride_RSI_2018,Wollenweber_RSI_2021}, and even $\sim8$~meV~\cite{Descamps_JSR_2022}, have been reported. Of course, the narrower the acceptance range of the monochromator, the fewer photons that will be incident on target, so there is a balance to be struck between the resolution and the signal-to-noise ratio of the spectrum within a given number of shots. This can be offset by first self-seeding the beam, and aligning in energy the monochromator and seeded spike.

XFELs provided a very flexible platform to produce beams covering a wide range of resolutions. However, they can require a large number of shots to produce spectra with good photon statistics; but their high repetition rates (1--100~Hz~\cite{Zastrau_JSI_2021,Nagler_JSI_2015,Glenzer_JPB_2016,Yabashi_JSI_2015,Yabashi_AS_2017}) can overcome these difficulties, provided the rest of the setup (e.g. laser drivers) can match it. Backlighters, on the other hand, lean into the efficiency of x-ray emission, and produce enough photons to generate XRTS spectra in single shots. For high energy laser facilities, this is a necessity given the time between shots (up to several hours). Backlighters are also more flexible: since they only require a laser powerful enough to induce x-ray emission from a foil, they can be fielded almost anywhere (see e.g. Ref.~\cite{Luetgert_MRE_2024}), whereas current hard x-ray FELs require kilometer-long linear accelerators, followed by several meters of undulators and focusing optics.

\subsubsection{Dispersive Crystal Spectrometer Instrument Functions}

The instrument function is much more complicated. It represents the physical processes in a spectrometer/experiment that give rise to a finite resolution versus an idealized, perfect setup. The vast majority of XRTS experiments use dispersive crystal spectrometers to measure spectra -- since they are so common, the instrument function will be considered for these types of spectrometers. Much of the discussion here is informed from \textit{Theory of X Ray Diffraction in Crystals} by W. H. Zachariasen~\cite{Zachariasen_XrayDiffraction}, which provides an excellent and detailed description of its titular topic.

Dispersive crystal spectrometers use diffraction by a crystal to scatter photons of different energies to different angles.
Specifically, the diffraction pattern for a photon of energy $E$ has a maximum point when the Laue condition is satisfied~\cite{Laue_Condition_1,Laue_Condition_2}
\begin{equation}\label{eq:LaueEq}
    {\bf k}_H - {\bf k}_0 = {\bf B}_H = h {\bf b_1} + k {\bf b_2} + l {\bf b_3} \, ,
\end{equation}
where ${\bf k}_0$ and ${\bf k}_H$ are the incident and diffracted wave vectors, $H=(hkl)$ are the Miller indices of the lattice plane, ${\bf b_i}$ are the reciprocal lattice vectors, and ${\bf B}_H$ is normal to the lattice plane.
The geometry of the reciprocal lattice will generally restrict this to a single point~\cite{Zachariasen_XrayDiffraction}.
The magnitude of ${\bf B}_H$ is $1/d_H$, i.e. the inverse of the spacing between two lattice planes. Typically, but not always, spectra are measured in Bragg geometry, i.e. photons are reflected out of the crystal (in contrast to Laue geometry, where photons are scattered through the crystal).
Bragg's law simplifies this idea, and says that in order for a photon of energy $E \ge hc/2d_H$ to be reflected, it needs to be incident on the crystal at an angle $\alpha_B$~\cite{Braggs_Law_1912,Braggs_Law_1913}:
\begin{equation}\label{eq:BraggLaw}
    \sin(\alpha_B) = \frac{hc}{2d_H} \frac{1}{E} \,.
\end{equation}
Since Eq.~(\ref{eq:LaueEq}) is essentially a reflection equation, the photon energy on the detector can be determined by associating the Bragg angle $\alpha_B$ on the crystal with the position a photon hits the detector plane.

In reality, the scattering of x-rays through the crystal is not so simple. Inside the crystal, the incident `beam' of photons and outgoing diffracted beam interact with each other, resulting in an interference pattern called the rocking curve of the crystal, $R(\alpha-\alpha_B; {\bf \Lambda})$. More specifically, this is the power ratio of the diffracted beam to the incident beam at the relevant crystal surface (top for Bragg geometry, bottom for Laue), versus the angle of incidence $\alpha$. The parameters ${\bf \Lambda}$ describe the shape of the rocking curve, which includes the photon energy- and polarization-dependent scattering and absorbing properties of the crystal.
The key point is that a photon does not need to be incident on the crystal at exactly the Bragg angle in order to be reflected. Features will therefore appear broader on the detector since there is no longer a strict one-to-one relationship between angle of incidence and photon energy.

The rocking curve can be calculated using dynamic diffraction theory~\cite{Zachariasen_XrayDiffraction,Darwin_PMJS_1914_1,Darwin_PMJS_1914_2,Ewald_AdP_1917_1,Ewald_AdP_1917_2}. For a flat perfect crystal, where the top and bottom surfaces are parallel infinite planes, an analytical solution can be derived; see Eq.~(3.139) in Ref.~\cite{Zachariasen_XrayDiffraction}.
We will not write it here as it is ``fairly complicated'', to quote Ref.~\cite{Zachariasen_XrayDiffraction}; but it is generally asymmetric, oscillatory, and the peak reflectivity occurs away from the Bragg condition. The infinite plane approximation is good when a crystal is much larger than the x-ray wavelength~\cite{delRio_RSI_1992}.
If the crystal is instead bent, the coupled differential Takagi-Taupin equations~\cite{Takagi_1962_Dynamical, Taupin_1964_Theorie} need to be used to determine how the deformation of the planes and strain within the crystal affect the rocking curve. The bending leads to a highly asymmetric rocking curve that can be substantially broader than the flat case, even for a large ($\sim$1~m) bending radius~\cite{Uschmann_1993_Xray,Hoelzer_CRT_1998}.
Since the diffraction of a photon is still essentially a reflection for perfect crystal, the angle of incidence still determines the outgoing angle of the photon and where it will hit the detector plane. The rocking curve is then also the probability distribution function that a photon incident on the crystal at an angle $\alpha$ will be reflected.

The rocking curves of perfect crystals are typically very narrow, so they only reflect photons near to their peak. Perfect crystals have therefore been used in ultrahigh resolution setups to achieve $\sim 10$~meV instrument function broadening~\cite{McBride_RSI_2018,Wollenweber_RSI_2021,Gawne_PRB_2024,Descamps_JSR_2022}. But, this also means most of the photons at a given energy are not scattered to the detector, so their reflectivity is relatively low. Combined with the small Thomson scattering cross-section, this can lead to very few photons detected in a single shot (e.g. Ref.~\cite{Gawne_PRB_2024} measured a peak of $\lt 0.1$ inelastic photons per shot per energy bin).

To increase the number of photons measured in each shot, XRTS experiments often employ mosaic crystal spectrometers~\cite{Glenzer_PRL_2003,Glenzer_PRL_2007,Doeppner_RSI_2014,Kraus_HEDP_2012,Kraus_PRE_2016,kraus_xrts,Preston_JoI_2020,Tilo_Nature_2023,bespalov2026experimentalevidencebreakdownuniformelectrongas}. A mosaic crystal consists of small, perfect crystallites whose surface normals ${\hat n}$ are randomly oriented to the surface normal of the bulk crystal ${\hat N}$~\cite{Zachariasen_XrayDiffraction,Darwin_PMJS_1922}.
The angular distribution of the crystallites is described by the mosaic distribution function, $W(\Theta, \Gamma)$, where $\Theta = \arccos({\hat n} \cdot {\hat N})$ is the crystallite angle and $\Gamma$ is the width of the distribution. Because there is a finite probability of finding a crystallite oriented such that its rocking curve is maximised, a photon can be reflected anywhere in a mosaic crystal, hence the entire crystal becomes reflective to all photon energies $E \ge hc/2d$.
However, this increase in reflectivity comes at the cost of resolution. First, the mosaic distribution function broadens the spectrum asymmetrically towards higher photon energies~\cite{Gawne_JAP_2024,Gawne_CPC_2026}. Second, the mosaicity strongly enhances depth broadening~\cite{Zastrau_JOI_2012,Schlesiger_JAC_2017,Gawne_JAP_2024}: whereas perfect crystals tend to be millimeters thick to improve their resolution, mosaic crystals are often only 10s--100s of microns thick. 

As with flat crystals, the rocking curve mosaic crystals, ${\cal R}(\alpha-\alpha_B, \Gamma, {\bf \Lambda})$, is also the power ratio of the diffracted to incident beams at the crystal surface. It depends on both the mosaic distribution function, and the rocking curves of the crystallites.
An analytic expression for parallel plane flat mosaic crystals does exist; see Eq.~(4.52) in Ref.~\cite{Zachariasen_XrayDiffraction}.
However, the random orientation of the crystallites means there is no longer a direct relationship between the angle of incidence and the angle a diffracted photon emerges from the crystal. This means the mosaic crystal rocking curve is just the total probability that a photon incident at angle $\alpha$ will reflect, but not where it will go.
For calculating the crystallites' rocking curves, the infinite plane approximation can still usually be used for flat crystallites~\cite{delRio_RSI_1992,Freund_SPIE_1996,Ohler_JAC_2000,Zastrau_JOI_2012}.
It is also worth noting that Ref.~\cite{Zachariasen_XrayDiffraction} and many other publications only consider the case that the crystallites restricted within the plane containing ${\bf k}_0$ and ${\bf k}_H$. In reality, ${\hat n}$ is oriented on a unit sphere, which changes the form of the mosaic distribution function and the scattering cross-section~\cite{Wuttke_Acta_2014,Bornemann_Acta_2020,Gawne_CPC_2026}.

The x-ray detector after the crystal also has an instrument function, $D(E; {\bf \Upsilon})$.
There are a wide class of detectors in operation today, such as hybrid-pixel~\cite{Blaj_SPIE_2016,Mozzanica_JOI_2016} and microchannel plate (MCP)~\cite{Oertel_RSI_2006,Rochau_RSI_2006, Farley_MIMPRSA_2013} detectors. The basic operating principle is the same: a photon hits a layer of the detector to excite electrons, and the generated charge is measured. Unsurprisingly then, the detector response parameters ${\bf \Upsilon}$ also depend on photon energy.
The detector also bins the photons into discrete pixels, and defines the observable spectral range.
The detector response is largely neglected, except to account for its quantum efficiency (QE). Since it is possible to reproduce spectra without needing a detailed detector model (see e.g. Ref.~\cite{Gawne_CPC_2026}), this appears to be reasonable. However, the thermal noise and charge spreading between pixels play a role in the measured spectrum by changing the spectral uncertainty in each bin from Poissonian or Gaussian, which may impact Markov chain Monte Carlo (MCMC)-based analyses, which does warrant further investigation.

Finally, the instrument function has a strong dependency on the geometry of the setup. This is especially for mosaic crystals due to the entire crystal being reflective~\cite{Gawne_JAP_2024,Gawne_CPC_2026}.
First, the accessible angles of incidence depends upon the position of the crystal in space relative to the source. For mosaic crystals, reaching the physical edges of the crystal leads to pronounced drops in the instrument function.
The shape of the crystal (i.e. flat, cylindrical, conical, spherical, toroidal) determines image measured on the detector. For example, the focusing von H\'amos (cylindrical)~\cite{vonHamos_Geometry} and Hall (conical)~\cite{Hall_Geometry} geometries produce a line on the detector, while flat crystals spread photons over circular arcs.
The position of the detector relative to the crystal also affects how broadening is observed, particularly in focusing geometries. This is especially noticeable on the MACS spectrometer of the NIF~\cite{Doeppner_RSI_2014}, which uses a cylindrical crystal in Hall geometry, leading to `butterfly' defocusing across the spectrometer. The instrument function therefore varies substantially in shape and peak intensity across the spectral range.
Finally, additional geometric broadening comes from the physical size and shape of the photon source, and the position of the source relative to the crystal.

\subsubsection{$q$-vector blurring}

Since the crystal optic is finite in size, it covers a range of scattering vectors. The detected spectrum is therefore integrated over a range of vectors, resulting in an effect known as $q$-blurring (or $k$-blurring). This can have a substantial effect on the interpretation of a spectrum by deforming or broadening spectral features compared to those expected from the nominal scattering vector.

As an additional complication, since the angle of incidence is used to disperse photons, the mean scattering vector covered by a given energy bin varies across the spectrometer. This means $q$-blurring is also tied to the instrument function since this determines the signal a given scattering vector can contributes to each energy bin. The dispersive $q$-blurring from the spectrometer comes on top of the scattering vector itself being dispersive; i.e. ${\bf q} = {\bf q}(\omega_i,\omega_s, \theta)$.
Large physical sources (e.g. capsule implosions), divergent probe sources (e.g. backlighters), and broad beam profiles also contribute to the $q$-blurring effect.

A few publications have examined the theoretical impact of $q$-blurring from backlighters probing physically large targets~\cite{Golovkin_HEDP_2013,Chapman_POP_2014,Poole_POP_2022,Poole_PPCF_2025}, though to our knowledge these have neglected spectrometer effects.
As the most recent example, Poole~\textit{et al.}~\cite{Poole_PPCF_2025} performed a MCMC analysis of theoretical ray traced capsule spectra using a single scattering angle and set of conditions, and found the density-averaged capsule conditions could be modelled to within the 2$\sigma$ uncertainty.

Kraus~\textit{et al.}~\cite{Kraus_PRE_2016} used ray tracing simulations -- both for the photon transport through a 3D capsule and the spectrometer -- to compare the predictions of a radiation-hydrodynamic simulation to an experimental XRTS spectrum of an imploding CH capsule. These were compared against fits using a typical XRTS approach at a single scattering angle. Nominally, agreement is seen between both approaches, however the very large uncertainties due to the background subtraction make it difficult to ascertain whether the $q$-blurring actually mattered or not.

Additionally, a few recent publications of ultrahigh resolution measurements of plasmons in Al and Si have found $q$-blurring does affect the interpretation of theoretical predictions versus experiment~\cite{Gawne_PRB_2024,Gawne_ElectronicStructure_2025,gawne2025orientationaleffectslowpair}. However, in these works, the spectrometer geometry -- spherical diced crystal analysers with a narrow spectral range -- means the spectrometer dispersion does not affect the $q$-blurring. Instead, the $q$-burring is determined by the number of dice covering a given scattering vector.
A number of papers have also highlighted the role of $q$-blurring in the orientation of the scattering vector through perfect crystals~\cite{Gawne_ElectronicStructure_2025}; as well as polycrystalline samples~\cite{Sternemann_PRB_1998,Larson_JPCS_2000,gawne2025orientationaleffectslowpair}, where the random orientation of the perfect crystalline domains alone can cause a $q$-blurring effect by randomly sampling the orientation of the scattering vector through a domain.

\subsubsection{The Source-and-Instrument Function}

The instrument function of dispersive crystal spectrometers is evidently complicated: it depends on the crystal and detector properties, the spectrometer's geometric layout, and the photon energy and polarization.
Provided the angles of incidence $\alpha$ can be associated to a particular pixel $p$, the measured spectrum will be of the form
\begin{equation}\label{eq:sif_broad}
    \begin{split}
        {\cal I}_p = \int_{E_{l,p}}^{E_{u,p}} dE &\int_0^\infty dE_d \int_0^\infty dE_s D(E-E_d; {\bf \Upsilon}) \\
        &\times {\cal R}(E_d-E_s; {\bf \Lambda}, \Gamma) I_s(E_s)
    \end{split}
\end{equation}
where $E_s$ are the actual photon energies, and $E_i$ is the central energy of the pixel based on the detector calibration. The width parameters ${\bf \Lambda}$ and ${\bf \Upsilon}$ generally depend on $E_s$. The crystal and detector response functions are also assumed to account for any filters applied added ahead of each of the components. The outermost integral bins the photons into the appropriate pixel on the detector. This function is somewhat simplified in that the spectrometer is actually three dimensional, so one would really need to work with the ingoing and outgoing wavevectors.
Regardless, this is sufficiently complicated to provide ample difficulty in the experimental broadening. While some works have attempted to consider the instrument function along these lines (e.g. Refs.~\cite{Schlesiger_JAC_2017,Gawne_JAP_2024}), in practice a number of simplifying approximations still need to make the problem more manageable.

In any case, apart from a few exceptions (e.g. Refs.~\cite{Kraus_PRE_2016,Luetgert_MRE_2024}), the vast majority of XRTS analyses do not consider the instrument function in such detail. Instead, the experimental broadening is reduced to a simple convolution. To do this, a number of simplifications and approximations need to be made.

An easy first step is to ignore the pixel binning, and assume that the measured spectrum is defined by the central energy of the pixel $E_p$. This approximation is generally applicable since spectral features tend to be much broader than the pixel width. However, some caution is needed in high resolution setups where this may not be true, and features may appear distorted by the binning process.

Second, detector effects are generally neglected entirely, except for the energy-dependent QE, which is done by dividing the experimental spectrum by the QE. Similarly, the experimental spectrum is divided by the total transmission of any filter stacks. The solid angle of the crystal is sometimes corrected for in a similar way. Note that all of these energy-dependent corrections are applied assuming the ideal setup; i.e. using the central energy of each bin in the spectrum.

This leaves only the crystal's rocking curve which, at this point, can be considered `the instrument function'. It is often modelled using simple functions, such as Gaussians or Voigt profiles~\cite{Tilo_Nature_2023,bespalov2026experimentalevidencebreakdownuniformelectrongas}; or, it is measured in experiment using a known line profile~\cite{Voigt_POP_2021}. Either way, it is assumed that this instrument function is the same across the entire spectrometer, which removes any photon parameterization. This enables the final measured spectrum to be approximated as two convolutions:
\begin{equation}
    \begin{split}
    {\cal I}(E_p) &\propto \int dE_s {\cal R}(E_p-E_s) E_s^2 \\
    &\times \int_0^\infty dE_0 \, B(E_0) \frac{S_{ee}({\bf q}, (E_0 - E_s)/\hbar)}{E_0^2}  \, .
    \end{split}
\end{equation}
A final simplification can be made using Eq.~(\ref{eq:approx_q}) and assuming $(E_s/E_0)^2 \approx 1$ over the spectral range. In this case, the source and instrument functions can be combined into a single source-and-instrument function (SIF) ${\cal B}(E)$ that is convolved with the scattering power to give the measured XRTS spectrum,
\begin{equation}\label{eq:xrts_as_conv}
    {\cal I}(E) \propto \int dE_0 {\cal B}(E_0) S_{ee}\left({\bf q}, \frac{E_0 - E}{\hbar} \right) \, .
\end{equation}
This approximation of the XRTS spectrum, where a SIF is assumed to convolve with the DSF, is widely used in XRTS analysis. It also forms the basis of the model-free ITCF analysis, since it implies SIF can be removed via deconvolution.

Of course, after all the emphasis about the photon-energy-dependence of the instrument function, it is natural to wonder whether Eq.~(\ref{eq:xrts_as_conv}) is even a reasonable way to treat the actual experimental broadening. The answer transpires to be yes, at least within some spectral range that depends on what is being measured~\cite{Gawne_JAP_2024}.
For the purposes of the model-free thermometry, the approximation does appear to hold sufficiently well, with recent ray tracing simulations suggesting it is applicable to at least 200~eV away from the elastic feature~\cite{Gawne_ITCF_Ratio}. We will also demonstrate this point later in Section~\ref{sec:RT_Examples} with some examples of the model-free methods applied to ray traced spectra.

\subsection{Raytracing}\label{sec:RayTracing}

An alternative approach to treating the instrument function is to simulate the transport of photons through the spectrometer. This is done in ray tracing simulations, for which there are a number of public codes available~\cite{Gawne_CPC_2026,SHADOW,SHADOW3,SHADOW4,XRT_code,McXtrace,XICSRT,Smid_CPC_2021}.

The basic idea of ray tracing is to represent the photons or incident beam as rays -- i.e. straight lines representing the path of the photons -- and to trace them from their origins until their end point (i.e. detection, absorption in the crystal, or lost to free space).
The response functions of the crystal and detector then act as probability distribution functions for the reflection and detection of the x-rays (e.g. in Ref.~\cite{Gawne_CPC_2026}); or they can be used to weight the ray hits on the detector to represent bunches of photons (e.g. in Ref.~\cite{McXtrace}).
Aside from the response functions, all the geometric effects of the spectrometer are inherently accounted for in three dimensional ray tracers.
Additional effects such as filter stacks and their orientation can also be included in more detail -- in particular, the actual path lengths of the rays through filters can be determined, rather than relying on the single path length used in spectral post-processing (e.g. in Ref.~\cite{Doeppner_RSI_2014}).
The detector image is then developed by randomly sampling rays incident on the crystal.

For perfect crystals, the rocking curve for each of the incident rays can be evaluated to accurately calculate the instrument function; e.g. in the \textsc{SHADOW} ray tracers~\cite{SHADOW,SHADOW3}. Absorption in and transmission through the crystal are automatically accounted for within the rocking curve~\cite{Zachariasen_XrayDiffraction}. The tracing of the ray vectors is then quick to evaluate since the rays just reflected against the reciprocal lattice vector; see Eq.~(\ref{eq:LaueEq}).

Mosaic crystals need to be handled differently. While one can calculate their rocking curve versus angle of incidence, this is only useful for determining their total reflectivity at a given angle. Since the crystallites are oriented randomly, the direction a ray travels after reflection is also random. Moreover, the rays can undergo multiple reflections within the crystal~\cite{Zachariasen_XrayDiffraction,Schlesiger_JAC_2017,Gawne_JAP_2024}. Multiple reflections also occurs in perfect crystals, however the fixed reflection normal means this is automatically accounted for in the rocking curve (although modelling layers can be useful for bent crystals~\cite{Chantler_JAC_1992_1,Chantler_JAC_1992_2}). A proper treatment of mosaic crystals requires tracing a ray's full path through the crystal until termination~\cite{Alianelli_SPIE_2004,Zastrau_JOI_2012,Gawne_CPC_2026}. Fortunately, the random orientation of the crystallites means that the photons in a mosaic crystal lose coherence with one another, so the individual photons can be traced independently of each other. This makes mosaic crystal ray tracing very well-suited to Monte Carlo methods. A mosaic crystal ray tracer then involves determining the distance a photon travels before interacting (absorption or scattering) and, if it is still in the crystal and scatters, what is the orientation of the crystallite; see Refs.~\cite{Gawne_CPC_2026} and~\cite{Wuttke_Acta_2014,Bornemann_Acta_2020} for further details.
The crystallite rocking curves can be treated to different levels of approximation, depending on the balance between performance and accuracy; e.g. Ref.~\cite{Alianelli_SPIE_2004} uses dynamic diffraction theory, whereas Ref.~\cite{Gawne_CPC_2026} approximates them as Voigt profiles.

A key benefit of ray tracing is that the calculation is performed on a ray-by-ray basis, meaning it is an embarrassingly parallelizable problem. Parallelization over threads, MPI and on GPUs is therefore relatively trivial, so ray tracers are computationally efficient.
However, compared to doing a convolution with a single SIF, ray tracers are much slower. It is not surprising then that, aside from Ref.~\cite{Luetgert_MRE_2024}, ray tracing is not used in MCMC analysis frameworks for XRTS given the thousands of simulations required, nor for parameter optimization via least squares fitting. Ray tracing has therefore largely been limited to theoretical studies of XRTS~\cite{Golovkin_HEDP_2013,Chapman_POP_2014,Poole_POP_2022,Poole_PPCF_2025,Gawne_ITCF_Ratio,Uwe_Events_2026,Bellenbaum_xdave}.

Ray tracing provides a powerful method to model the instrument function broadening. The modelling can be made as arbitrarily detailed as required to compare against experiment, but the problem of calculating the instrument function overall remains manageable. Furthermore, by treating photons on an individual level, the photon counting noise inherent to ray tracing simulations becomes representative of the experimental counting noise at similar signal levels. The impact of photon statistics can therefore be investigated on diagnostic methods using ray tracing. The complexity contained in ray traced instrument functions means they serve a useful testbed for analysis methods, including the model-free methods. While not currently widely-used for MCMC analysis, they would certainly benefit the method by reducing biases in inferred conditions from using simple SIFs, and by allowing e.g. filter and solid angle corrections to be properly handled as well.
Finally, Monte Carlo ray tracing is well-suited for integration into a recently-proposed event generation scheme for XRTS analysis~\cite{Uwe_Events_2026}.

\section{Dynamic structure factor models\label{sec:models}}

The electronic dynamic structure factor $S_{ee}(\mathbf{q},\omega)$ is usually defined as the Fourier transform of the intermediate scattering function $F_{ee}(\mathbf{q},t)=\braket{\hat{n}_e(\mathbf{q},t)\hat{n}_e(-\mathbf{q},0)}$,
\begin{eqnarray}\label{eq:DSF}
    S_{ee}(\mathbf{q},\omega) = \frac{1}{2\pi}\int_{-\infty}^\infty\textnormal{d}t\ F_{ee}(\mathbf{q},t)\ e^{i\omega t} ,
\end{eqnarray}
and correlates electronic density fluctuations at different wavelengths $\lambda=2\pi/\vert \mathbf{q}\vert$~\cite{sheffield2010plasma}.
An alternative and often useful representation of the dynamic structure factor is given by the well-known spectral representation~\cite{quantum_theory}
\begin{eqnarray}\label{eq:spectral_representation}
    & &S_{ee}(\mathbf{q},\omega) =\\ & &  \sum_{l,m} P_m \vert\langle l\vert \hat{n}(\mathbf{q}) \vert m \rangle\vert^2 \delta\left(\omega - \frac{E_l-E_m}{\hbar} \right)\ ,\ \nonumber
\end{eqnarray}
with $l,m$ being the eigenstates of the full many-body Hamiltonian $\hat{H}$, $E_{l}$ and $E_m$ being the corresponding eigenenergies, and $P_m$ being the probability to occupy an initial eigenstate $m$.
In other words, the full DSF is given by the sum over all possible transitions between electronic eigenstates induced by the single-electron density operator $\hat{n}(\mathbf{q})$.
From Eq.~(\ref{eq:spectral_representation}), it also immediately follows that $\omega>0$ ($\omega<0$) corresponds to transitions in which the system gains (looses) energy due to the transition.
In an XRTS measurement, on the other hand, we focus on the energy balance of the scattered photon, which has the oppose sign; the photon looses (gains) energy when the system is excited (de-excited) into an energetically higher (lower) state.
In addition, Eq.~(\ref{eq:spectral_representation}) also directly leads to the important detailed balance relation~\cite{quantum_theory}
\begin{eqnarray}\label{eq:detailed_balance}
   S_{ee}(\mathbf{q}, -\omega) = S_{ee}(\mathbf{q},\omega)\ e^{-\beta \hbar\omega}\ ,
\end{eqnarray}
which holds universally in thermal equilibrium.

\subsection{Linear response theory\label{sec:LRT}}

Linear response theory~\cite{nolting,quantum_theory,Dornheim_review} is one of the most successful and fundamental methodologies over many domains of physics and related fields of research.
For the present work, we consider the electronic density response to an external harmonic perturbation of wavevector $\mathbf{q}$ and frequency $\omega$, which, in the limit of an infinitesimal perturbation amplitude, is fully characterized by the dynamic density response function $\chi_{ee}(\mathbf{q},\omega)$. Importantly, the dynamic density response is directly related to the dynamic structure factor $S_{ee}(\mathbf{q},\omega)$ via the celebrated fluctuation--dissipation theorem~\cite{quantum_theory}
\begin{eqnarray}\label{eq:FDT}
S_{ee}(\mathbf{q},\omega) = - \frac{\textnormal{Im}\chi_{ee}(\mathbf{q},\omega)}{\pi n (1-e^{-\beta\hbar\omega})}\ .
\end{eqnarray}
Eq.~(\ref{eq:FDT}) directly implies that $\chi_{ee}(\mathbf{q},\omega)$ and $S_{ee}(\mathbf{q},\omega)$ contain precisely the same information as $\textnormal{Im}\chi(\mathbf{q},\omega)$ and $\textnormal{Re}\chi(\mathbf{q},\omega)$ can be computed from each other via Kramers-Kronig relations.
In other words, one can potentially utilize insights from density response theory to explain physical effects observed in $S_{ee}(\mathbf{q},\omega)$, e.g., Refs.~\cite{Dornheim_CommPhys_2022,Chuna_JCP_2025}.

For a uniform electron gas, the dynamic density response is often expressed as~\cite{kugler1}
\begin{eqnarray}\label{eq:define_G}
    \chi_{ee}(\mathbf{q},\omega) = \frac{\chi_0(\mathbf{q},\omega)}{1-\frac{4\pi}{q^2}\left(1-G_{ee}(\mathbf{q},\omega)\right)\chi_0(\mathbf{q},\omega)}\ ,
\end{eqnarray}
where $\chi_0(\mathbf{q},\omega)$ denotes the Lindhard function that describes the density response of an ideal (i.e., non-interacting) Fermi gas at the same conditions.
The complete frequency- and wavenumber-resolved information about electronic XC-effects is encoded into the dynamic local field correction $G_{ee}(\mathbf{q},\omega)$, which is subject to continued active investigation~\cite{Dabrowski_PRB_1986,moroni2,cdop,Panholzer_PRL_2018,Adrienn_PRB_2020,Kaplan_PRB_2022,dornheim_ML,Hou_PRB_2022,Dornheim_PRL_2020_ESA,Dornheim_PRB_2021,Tolias_JCP_2023,dornheim_dynamic,dynamic_folgepaper,Hamann_PRB_2020}.
Setting $G_{ee}(\mathbf{q},\omega)\equiv0$ in Eq.~(\ref{eq:define_G}) thus corresponds to the widely used \emph{random phase approximation} (RPA), which describes the electronic density response on a mean-field level.
We note that $G_{ee}(\mathbf{q},\omega)$ is also formally equivalent to the dynamic XC-kernel $K_\textnormal{xc}(\mathbf{q},\omega)$ that plays a key role in linear-response TDDFT, see Eq.~(\ref{eq:define_Kxc}) in Sec.~\ref{sec:TDDFT} below.

Describing the dynamic density response of multi-component systems is substantially more involved, but conceptually straightforward~\cite{ichimaru_bookII,Dornheim_PRR_2022,Dornheim_MRE_2024}. A pragmatic ansatz is given by the Born-Mermin dielectric function~\cite{mermin_prb_1970,chuna2026merminsdielectricfunctionfsum}, which can also be combined with electronic local field corrections~\cite{Fortmann_PRE_2010} and with dynamic collision frequencies obtained from DFT calculations~\cite{hentschel_PoP_2023,hentschel_PoP_2025,Witte_PRL_2017}.

\subsection{Chihara decomposition\label{sec:Chihara}}

A common approach to analyse XRTS spectra makes use of the Chihara decomposition~\cite{siegfried_review}, where scattering contributions are split into free and bound electrons~\cite{Chihara_1987}.
The total dynamic structure factor is initially split into quasi-elastic and inelastic contributions as
\begin{equation}\label{eq:elastic}
    S_{ee}(\mathbf{q},\omega) = W_\textnormal{R}(\mathbf{q}) \delta(\omega) + S_\textnormal{inel}(\mathbf{q},\omega) \,.
\end{equation}
The quasi-elastic feature, also known as the Rayleigh weight, is defined as
\begin{equation}\label{eq:Wr}
    W_R(\mathbf{q}) = \frac{S_{en}^2(\mathbf{q})}{S_{nn}(\mathbf{q})}\,,
\end{equation}
with $S_{en}$ and $S_{nn}$ measuring static correlations between electrons and nuclei, as well as between the nuclei themselves.
In the chemical picture for a multi-component system, $W_R(\mathbf{q})$ can be expressed in terms of the ionic form factors $f(\mathbf{q})$, the screening cloud $q(9)\mathbf{q})$ and the partial structure factors $S_{ab}(\mathbf{q})$ as~\cite{Wuensch_PRE_2008}
\begin{align}\label{eq:Rayleigh}
    W_R(\mathbf{q}) = \sum_{a,b} \sqrt{x_a x_b} & [f_a(\mathbf{q}) + q_a(\mathbf{q})] \\ \nonumber
    & [f_b(\mathbf{q}) + q_b(\mathbf{q})] S_{ab}(\mathbf{q}) \,.
\end{align}
The inelastic scattering is then further split into free-free and bound-free contributions as~\cite{Chihara_1987}
\begin{equation}
    S_{\text{inel}}(\mathbf{q},\omega) = S^{\text{ff}}(\mathbf{q},\omega) + S^{\text{bf}}(\mathbf{q},\omega) \,.
\end{equation}
The reverse process of the bound-free, the free-bound, is estimated using the detailed balance relation~\cite{bohme2023evidencefreeboundtransitionswarm}
\begin{equation}
    S_{ee}^\text{fb}(\mathbf{q},\omega) = S_{ee}^\text{bf}(\mathbf{q},\omega) \exp{\left[-\frac{\hbar \omega}{k_B T_e} \right]} \,.
\end{equation}

The different components in the Chihara decomposition are often calculated using analytic models that are cheap and accessible.
The bound-free contribution can be analytically estimated using the impulse approximation~\cite{Schumacher_JPB_1975}.
Continuum lowering is accounted for using ionization potential depression models like Stewart-Pyatt~\cite{Stewart_AJ_1966} or Crowley~\cite{Crowley_2014_ipd}.
The free-free contribution can be estimated using the Linhard dielectric function in the RPA, including local field corrections to account for electronic correlations~\cite{Fortmann_PRE_2010}.
Electron-ion collisional effects can also be included by using the Mermin dielectric function~\cite{mermin_prb_1970}.
Static structure factors for a multi-component system can be modeled using the hypernetted-chain (HNC) approximation, which has been shown to match DFT in the WDM regime~\cite{Wuensch_PRE_2008}.
Screening cloud caluclations are commonly done using linear response theory, for example in the finite wavelength screening model~\cite{Gericke_PRE_2010}.
Simplified form factor calculations are taken from the Pauling-Sherman model~\cite{Pauling_1932}.






\subsection{Time dependent density functional theory\label{sec:TDDFT}}

TDDFT is one of the most accurate ab initio methods for modeling XRTS spectra. Two main approaches are commonly used to compute the dynamic structure factor (DSF) within TDDFT: (1) real-time propagation of Kohn–Sham orbitals under a time-dependent external perturbation (RT-TDDFT) \cite{Sakko_jpc_2010, Baczewski_PRL_2016, White_2025}, and (2) a linear-response formulation based on Kohn–Sham orbitals (LR-TDDFT) \cite{Mo_PRL_2018, Mo_prb_2020, Moldabekov_PRR_2023, Schoerner_PRE_2023}. The LR-TDDFT approach requires significant computational resources due to memory-intensive matrix inversions and multiplications. Moreover, accurate calculation of the DSF at high frequencies and wavenumbers—relevant for typical detector geometries and parameters in XRTS experiments—requires a large number of empty bands. As a result, conventional LR-TDDFT becomes impractical for simulations of matter under extreme conditions.
In contrast, RT-TDDFT avoids these numerical bottlenecks but lacks the explicit separation of the noninteracting response function, local-field effects (arising from density inhomogeneities), and the exchange–correlation (XC) kernel. Such decomposition, naturally available in linear-response theory, is valuable for interpreting simulation results, improving accuracy, and developing approaches that incorporate dynamical effects in the XC kernel \cite{Byun_2020}. An alternative approach that avoids the memory limitations associated with matrix inversions and eliminates the need of empty bands, while retaining the linear-response framework, is the Liouville–Lanczos approach to TDDFT, also known as time-dependent density-functional perturbation theory \cite{Walker_PRL_2006, Motornyi_prb_2020, Timrov_prb_2017, TIMROV2015460}. Recently,  the computational efficiency and suitability of the the Liouville–Lanczos method for XRTS modeling has demonstrated in Refs~\cite{Moldabekov_MRE_2025, Moldabekov_filter, bespalov2026experimentalevidencebreakdownuniformelectrongas, Gawne_ElectronicStructure_2025, gawne2025orientationaleffectslowpair}. 

In practice, the TDDFT calculations are performed with fixed ions, where ionic configurations are obtained from separate DFT-MD simulations. This approach allows TDDFT to capture electron-density inhomogeneities arising from electron–ion interactions. For disordered systems—typical of isothermal warm dense matter—averaging over multiple ionic snapshots results in an effectively homogeneous electron density \cite{Moldabekov_jpc_2023}.
Within LR-TDDFT, the noninteracting density response function is given by the Kohn–Sham (KS) response function, $\chi_{\rm KS}(\vec q,\omega)$. Analogous to Eq.~\eqref{eq:define_G}, the TDDFT dynamic response function for disordered systems can be written as \cite{Moldabekov_PRR_2023}
\begin{eqnarray}\label{eq:define_Kxc}
    \chi_{ee}(\mathbf{q},\omega) = \frac{\chi_\textnormal{KS}(\mathbf{q},\omega)}{1-\left(\frac{4\pi}{q^2}+K_\textnormal{xc}(\mathbf{q},\omega)\right)\chi_\textnormal{KS}(\mathbf{q},\omega)}\,
\end{eqnarray}
where $K_\textnormal{xc}(\mathbf{q},\omega)$ is the XC kernel.
In general, for any material, the XC kernel is defined as the second-order functional derivative of the XC functional with respect to density. Eq.~(\ref{eq:define_Kxc}) is derived from a more general formulation of the LR-TDDFT for inhomogeneous systems~\cite{ullrich2012time}.  

The XC kernel, together with the underlying XC functional, represents a key approximation in TDDFT. Benchmarks against experimental XRTS data for shock-compressed and heated aluminum \cite{bespalov2026experimentalevidencebreakdownuniformelectrongas}, various metals and semiconductors at ambient conditions \cite{Gawne_ElectronicStructure_2025, Gawne_PRB_2024, gawne2025orientationaleffectslowpair, Motornyi_prb_2020, Timrov_prb_2017}, as well as comparisons with exact PIMC results for warm dense hydrogen \cite{Moldabekov_MRE_2025, Moldabekov_filter, Moldabekov_MRE_2026, Moldabekov_PPNP_2025, Moldabekov_jctc_2024}, demonstrate that XC kernels at the level of the local density approximation (LDA) and the generalized gradient approximation (GGA), such as using the Perdew-Burke-Ernzerhof (PBE) functional \cite{PBE}, often offer highly satisfactory accuracy. That being said, the XC kernel for XRTS applications is not limited to LDA and GGA level functionals and can be computed using any available XC functional on any level of complexity \cite{Moldabekov_jctc_2023, Moldabekov_PRR_2023, Moldabekov_JPCL_2023}.

The analysis of XRTS measurements using TDDFT, in combination with other model-free diagnostic tools, enhances the diagnostics of thermodynamic parameters and allows a better understanding of the observed spectral features. In addition, high-resolution XRTS spectra with sufficiently well-constrained density and temperature serve as valuable benchmarks for XC kernel approximations in TDDFT. This, in turn, supports further methodological developments for calculating other dynamical response properties, such as absorption.

\subsection{Analytic continuation\label{sec:AC}}

The gold standard of WDM theory is currently given by \emph{ab initio} path integral Monte Carlo (PIMC) simulations~\cite{cep}, which are, in principle, capable of providing exact results for the full quantum many-body problem of interest.
In practice, a severe limitation is given by the notorious fermion sign problem~\cite{troyer,dornheim_sign_problem}, which limits the range of application of PIMC to relatively light elements (quasi-exact simulations up to beryllium have been demonstrated so far~\cite{Dornheim_JCP_2024,Dornheim_NatComm_2025}) at moderate to high temperatures; see also the discussion of recent methodological advances in Refs.~\cite{Bonitz_POP_2024,Xiong_JCP_2022,Dornheim_JCP_2024,dornheim2025taylorseriesperspectiveab} and references therein.
A second limitation of PIMC is given by its formulation in the imaginary-time domain, which precludes the straightforward computation of dynamic properties.
Instead, PIMC simulations give one access to a variety of imaginary-time correlation functions~\cite{Dornheim_JCP_ITCF_2021,Rabani_PNAS_2002,Filinov_PRA_2012,efremkin2026computationthermalconductivitybased,hamann2026abinitiopathintegralmonte}, which are connected to various dynamical spectral properties via integral transforms.
For the present work, a key relation is given by the connection between the imaginary-time density--density corelation function $F_{ee}(\mathbf{q},\tau)$ and the dynamic structure factor $S_{ee}(\mathbf{q},\omega)$,
\begin{eqnarray}\label{eq:ITCF}
F_{ee}(\mathbf{q},\tau) &=& \mathcal{L}\left[S_{ee}(\mathbf{q},\omega)\right] \\ &=& \int_{-\infty}^\infty\textnormal{d}\omega\ S_{ee}(\mathbf{q},\omega)\ e^{-\tau \hbar\omega}\ ,
\end{eqnarray}
which is given by a two-sided Laplace transform.
The task at hand is thus the numerical inversion of Eq.~(\ref{eq:ITCF}) to reconstruct $S_{ee}(\mathbf{q},\omega)$ from a given PIMC dataset for $F_{ee}(\mathbf{q},\tau)$.
This so-called \emph{analytic continuation} is a notoriously difficult, exponentially ill-posed problem~\cite{Jarrell_PhysRep_1996,Chuna_JPA_2025,chuna2025noiselesslimitimprovedpriorlimit}; indeed, small statistical errors in $F_{ee}(\mathbf{q},\tau)$ can lead to vast deviations between different trial solutions for $S_{ee}(\mathbf{q},\omega)$.

In 2018, Dornheim \textit{et al.}~\cite{dornheim_dynamic,dynamic_folgepaper} have presented the first reliable PIMC results for $S_{ee}(\mathbf{q},\omega)$ by further constraining the inversion of Eq.~(\ref{eq:ITCF}) based on the stochastic sampling of the dynamic local field correction, cf.~Eq.~(\ref{eq:define_G}) above.
These efforts have initiated a surge of analytical continuation based activity in the study of the dynamic structure factor (and related properties) UEG~\cite{Dornheim_Vorberger_finite_size_2020,Hamann_PRB_2020,Hamann_CPP_2020,Filinov_PRB_2023,Filinov_CPP_2026,Chuna_JCP_2025,Chuna_PRB_2025}.
Moreover, recent novel, more general methods and implementations of the AC~\cite{Chuna_JPA_2025,BENEDIXROBLES2026109904,SmoQyDEAC,Nichols2022,Otsuki_JPSJ_2020,ShaoSandvik_PhysRep_2023}
open up promising avenues to reconstruct the full $S_{ee}(\mathbf{q},\omega)$ from existing ITCF results for warm dense hydrogen~\cite{Dornheim_MRE_2024,Dornheim_JCP_2024} and beryllium~\cite{Dornheim_JCP_2024,Dornheim_NatComm_2025,schwalbe2025staticlineardensityresponse}.
In the mean time, raw PIMC results for $F_{ee}(\mathbf{q},\tau)$ are already highly valuable as a benchmark for existing approximate models of $S_{ee}(\mathbf{q},\omega)$, see the recent comparisons to Chihara models~\cite{Bellenbaum_PRR_2025} and TD-DFT~\cite{Moldabekov_MRE_2025}.

\begin{figure}
    \centering
    \includegraphics[width=0.999\linewidth]{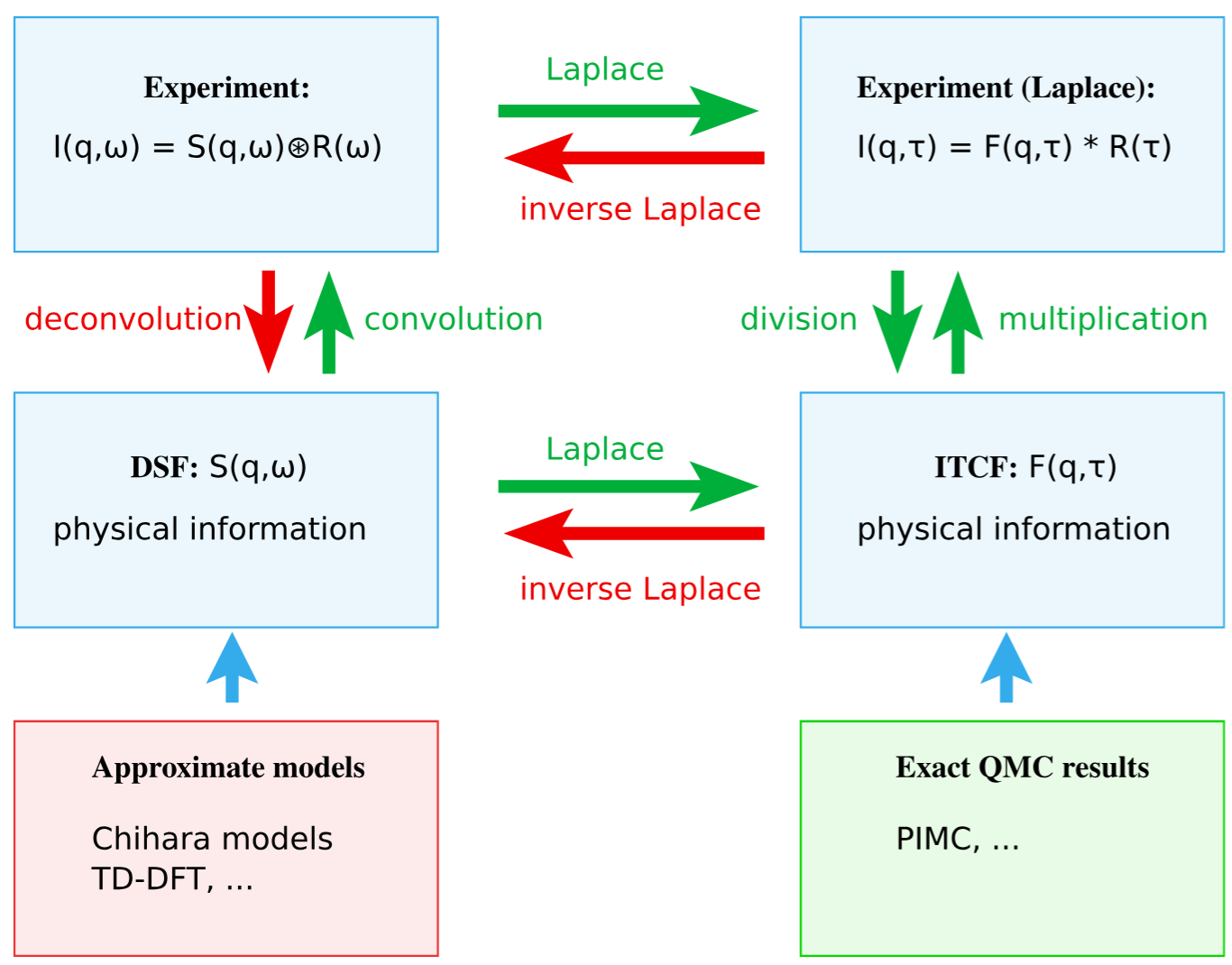}
    \caption{Analyzing XRTS measurements in the frequency- and imaginary-time domains. Deconvolution in the frequency-domain (left side) is highly unstable, which usually precludes direct access to the physics information of interest; instead, approximate models can be convolved with the SIF $R(\omega)$ in forward direction, making the comparison with the measurement indirect. Switching from the frequency- to the imaginary-time domain (right) via a two-sided Laplace transform $\mathcal{L}[\dots]$ is straightforward, and allows for a stable and straightforward deconvolution. This allows for direct access to the physics information about the probed sample in the form of the ITCF $F_{ee}(\mathbf{q},\tau)$, which, in addition, can then also be compared to exact PIMC simulations (bottom right).
    Taken from Ref.~\cite{Dornheim_MRE_2023} with the permission of the authors.
    \label{fig:problem}}
\end{figure}

\section{Model-free XRTS diagnostics\label{sec:model_free}}

\subsection{Deconvolution\label{sec:deconvolution}}

A key problem in the analysis of XRTS experiments is the fact that we do not directly measure the electronic dynamic structure factor, but its convolution with the combined source-and-instrument function $R(\omega)$, see Sec.~\ref{sec:SIF}.
This is illustrated by the top left corner in Fig.~\ref{fig:problem}.
The numerical deconvolution is very unstable and, thus, generally not feasible or reliable.
Consequently, XRTS does give one direct access to the physics information of interest in the frequency-domain, i.e., the left
side of Fig.~\ref{fig:problem};
direct thermometry using the detailed balance relation [Eq.~(\ref{eq:detailed_balance})] is thus precluded.
Instead, the traditional way of interpreting was to construct a theoretical model for $S_{ee}(\mathbf{q},\omega)$, which could then be convolved in forward direction with $R(\omega)$ and subsequently be compared with the experimental observation, $I(\mathbf{q},\omega)$.
A priori unknown parameters such as the temperature or mass density can be inferred in this way either from a best fit between model and experiment, or by more sophisticated Bayesian inference via Markov chain Monte Carlo methods~\cite{Kasim_POP_2019}.
As mentioned above, the quality of the extracted parameters will, by necessity, depend on the employed level of approximation, with more sophisticated methods such as TDDFT requiring a substantial amount of computational expense as well as dedicated convergence studies of various parameters.

In 2022, Dornheim \textit{et al.}~\cite{Dornheim_T_2022} have proposed to switch from the usual frequency domain to the imaginary-time domain.
In particular, the direct connection between $S_{ee}(\mathbf{q},\omega)$ and the ITCF $F_{ee}(\mathbf{q},\tau)$ had routinely been utilized in the quantum Monte Carlo community across diverse fields of research~\cite{Boninsegni1996,Jarrell_PhysRep_1996,Filinov_PRA_2012,Vitali_PRB_2010,Chuna_PRB_2025,dornheim_dynamic,Saccani_Supersolid_PRL_2012}, but in the opposite direction: to carry out an analytic continuation from the ITCF to the dynamic structure factor, see Sec.~\ref{sec:AC} above.
While the inverse two-sided Laplace transform $\mathcal{L}^{-1}[\dots]$ is highly unstable (red leftwards arrows in Fig.~\ref{fig:problem}), the two-sided Laplace transform $\mathcal{L}[\dots]$ itself acts as a noise filter and works well in most situations.
We can then utilize the well-known deconvolution theorem
\begin{eqnarray}\label{eq:deconvolution}
    \mathcal{L}\left[S_{ee}(\mathbf{q},\omega)\right] = \frac{\mathcal{L}\left[S_{ee}(\mathbf{q},\omega)\circledast R(\omega)\right]}{\mathcal{L}\left[R(\omega)\right]}\ ,
\end{eqnarray}
giving us direct access to the physical information as it is encoded into the ITCF $F_{ee}(\mathbf{q},\tau)$, see Eq.~(\ref{eq:ITCF}).
While the corresponding analytic continuation to reconstruct the deconvolved $S_{ee}(\mathbf{q},\omega)$
is precluded by the aforementioned ill-conditioned nature of $\mathcal{L}^{-1}[\dots]$, we argue that this step is not necessary; the two-sided Laplace transform is a unique transformation, which means that $F_{ee}(\mathbf{q},\tau)$ contains exactly the same information as $S_{ee}(\mathbf{q},\omega)$, only in an a-priori unfamiliar representation.
It is now our task to understand how this information is encoded into and can be extracted from $F_{ee}(\mathbf{q},\tau)$, which is the subject for the remainder of this section.

\begin{figure}
    \centering
    \includegraphics[width=0.99\linewidth]{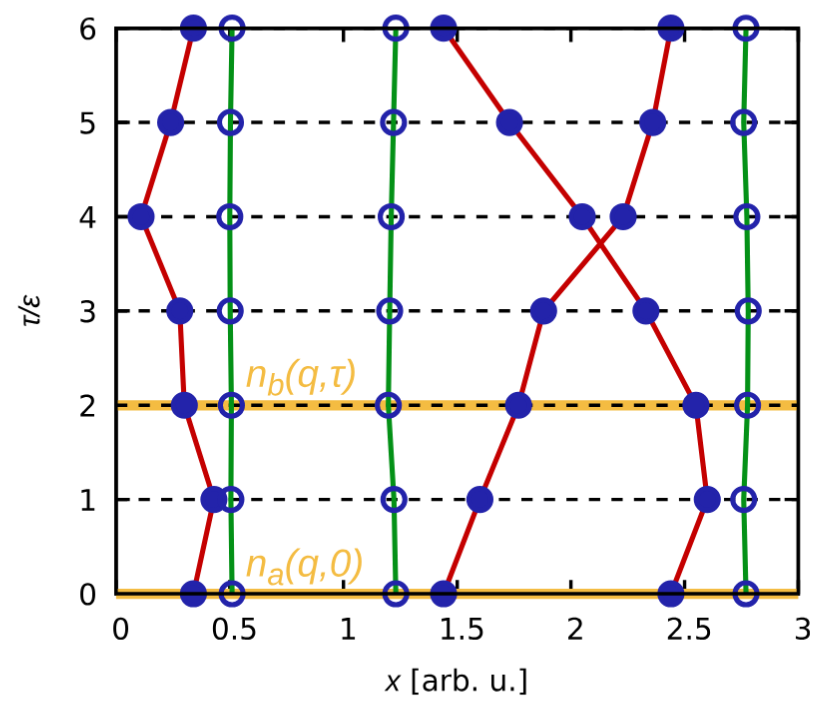}
    \caption{Schematic illustration of the imaginary-time path integral representation of statistical mechanics that gives rise to the ITCF. Shown is a configuration of $N=3$ hydrogen atoms, with the green and red paths representing the protons and electrons respectively. Note the reduced imaginary-time diffusion of the protons, which resemble classical point particles (straight lines in the $\tau$-$x$-plane) due to their heavier mass. The horizontal yellow lines indicate the estimation of the ITCF $F_{ab}(\mathbf{q},\tau)$ by correlating two single-particle density operators at an imaginary-time difference $\tau$. Taken from Ref.~\cite{Dornheim_MRE_2024} with the permission of the authors.
    \label{fig:paths}}
\end{figure}

\begin{figure}
    \centering
    \includegraphics[width=0.99\linewidth]{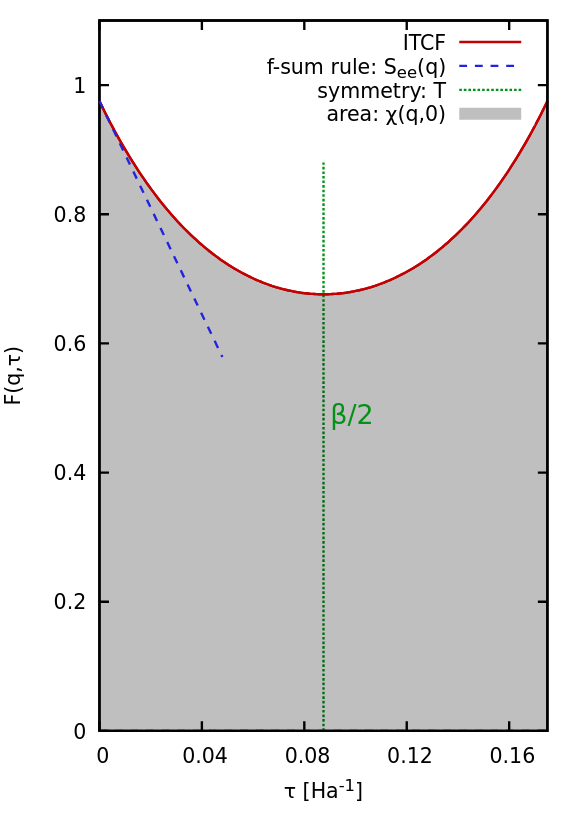}
    \caption{\emph{Ab initio} PIMC results for the electronic ITCF $F_{ee}(\mathbf{q},\tau)$ of warm dense beryllium at $T=155.5\,$eV and $\rho=7.5\,$g/cc at wavenumber $q=7.68\,$\AA$^{-1}$ (solid red curve). In thermal equilibrium, the ITCF is symmetric around $\tau=\beta/2$ (vertical dotted green line), see the detailed balance Eq.~(\ref{eq:ITCF_symmetry}). The first derivative of $F_{ee}(\mathbf{q},\tau)$ around $\tau=0$ is governed by the f-sum rule, Eq.~(\ref{eq:fsum}) (dashed blue line). The area under the ITCF is directly related to the static linear electronic density response function $\chi_{ee}(\mathbf{q},0)$ via the imaginary-time version of the fluctuation--dissipation theorem, Eq.~(\ref{eq:static_chi}) (shaded gray).  
    Taken from Ref.~\cite{schwalbe2025staticlineardensityresponse} with the permission of the authors.
    \label{fig:ITCF_PIMC}}
\end{figure}

\subsection{Imaginary-time correlation function\label{sec:ITCF}}

The imaginary-time density--density correlation function
\begin{eqnarray}\label{eq:define_ITCF}
    F_{ee}(\mathbf{q},\tau) = \braket{\hat{n}_e(\mathbf{q},0)\hat{n}_e(-\mathbf{q},\tau)}\ ,
\end{eqnarray}
has traditionally been considered in the context of \emph{ab initio} PIMC simulations, the basic idea of which is illustrated in Fig.~\ref{fig:paths}.
Specifically, PIMC is based on the celebrated \emph{classical isomorphism}~\cite{Chandler_JCP_1981}, where the interacting quantum many-body system of interest is mapped onto an effectively classical system of interacting ring polymers: the eponymous paths depicted in Fig.~\ref{fig:paths}, where we show a configuration of $N=3$ hydrogen atoms in the $x$-$\tau$-plane, which $x$ being an arbitrary spatial coordinate and $\tau\in[0,\beta]$ being partitioned into $P=6$ discrete time slices of length $\epsilon=\beta/P$.
We note that this $\tau$-discretization is a typical feature of PIMC, but it does not affect the model-free analysis of XRTS measurements for which we can compute the two-sided Laplace transform on any arbitrary $\tau$-grid. 
The extension of the paths along the diffusion throughout the imaginary time is proportional to the thermal deBroglie wavelength
\begin{eqnarray}\label{eq:deBroglie}
    \lambda_\beta = \left( \frac{2\pi\beta\hbar^2}{m} \right)^2\ .
\end{eqnarray}
Consequently, the heavier protons exhibit much less quantum delocalization than the electrons.
The green paths in Fig.~\ref{fig:paths} corresponding to the former resemble straight lines, which are the path integral representation of classical point particles; the red paths corresponding to the electrons noticeably diffuse throughout the imaginary time, with $\lambda_\beta$ being comparable to the average interparticle distance, leading to the formation of fermionic exchange, i.e., permutation cycles~\cite{Dornheim_permutation_cycles} with more than a single particle in them, see the two electronic paths on the right side of Fig.~\ref{fig:paths}.
The basic idea of the PIMC method is to randomly sample all possible paths using modern implementations of the Metropolis algorithm~\cite{metropolis}.

The PIMC computation of an ITCF $F_{ab}(\mathbf{q},\tau)$ [with $a$ and $b$ corresponding to either electrons or nuclei~\cite{Dornheim_MRE_2024}] is illustrated by the two horizontal yellow lines, indicating the 
evaluation of single-particle densities $n_{a}(\mathbf{q})$ at two separate imaginary times with a difference $\tau$.
In Fig.~\ref{fig:ITCF_PIMC}, we show PIMC results for the electronic ITCF $F_{ee}(\mathbf{q},\tau)$ for warm dense beryllium at $T=155.5\,$eV and $\rho=7.5\,$g/cc and at a wavenumber of $7.68\,$\AA$^{-1}$.
These conditions are relevant for the interpretation of an XRTS experiment carried out at the NIF~\cite{Tilo_Nature_2023,Dornheim_NatComm_2025,Dornheim_POP_2025,schwalbe2025staticlineardensityresponse,Dharma_wardana_PRE_2025,bohme2023evidencefreeboundtransitionswarm}, with the wavenumber corresponding to a backscattering geometry ($\theta\approx120^\circ$ at $E_0=9\,$keV).
First, we note that the ITCF is symmetric around $\tau=\beta/2$ (vertical dotted green line),
\begin{eqnarray}\label{eq:ITCF_symmetry}
    F_{ee}(\mathbf{q},\tau) = F_{ee}(\mathbf{q},\beta-\tau)\ ,
\end{eqnarray}
which directly follows from the translation of the detailed balance relation Eq.~(\ref{eq:detailed_balance}) into the Laplace domain~\cite{Dornheim_T_2022}.
Equivalently, Eq.~(\ref{eq:ITCF_symmetry}) also naturally follows from the imaginary-time translation invariance in thermal equilibrium.
In any case, the symmetry of the unconvolved ITCF is the key ingredient to both the model-free thermometry approach (Sec.~\ref{sec:thermometry}) as well as the model-free non-equilibrium detection approach (Sec.~\ref{sec:non_eq}).
A second class of physics information that is encoded into the ITCF are the frequency moments of the dynamic structure factor $S_{ee}(\mathbf{q},\omega)$~\cite{Dornheim_moments_2023}
\begin{eqnarray}\label{eq:moment_DSF}
   M^S_l(q)  = \int_{-\infty}^\infty \textnormal{d}\omega\ S_{ee}(\mathbf{q},\omega)\ (\hbar\omega)^l\ .
\end{eqnarray}
It is easy to see that these moments can also be directly obtained as the derivatives of the ITCF via~\cite{Dornheim_MRE_2023,Dornheim_moments_2023}
\begin{eqnarray}\label{eq:moment_ITCF}
    M^S_l(q) = \left. (-1)^l \frac{\partial^l}{\partial \tau^l} F_{ee}(\mathbf{q},\tau)\right\rvert_{\tau=0}\ ,
\end{eqnarray}
with the first moment being governed by the universal f-sum rule~\cite{chuna2026merminsdielectricfunctionfsum,quantum_theory}
\begin{eqnarray}\label{eq:fsum}
    M^S_1(\mathbf{q}) = \frac{\hbar^2\mathbf{q}^2}{2m_e}\ ;
\end{eqnarray}
see the dashed blue line in Fig.~\ref{fig:ITCF_PIMC}.
Absolute knowledge of $M^S_1(q)$ then allows one to infer the usually a-priori unknown normalization of the XRTS signal, which is discussed in detail in Sec.~\ref{sec:moments} below.
Thirdly, the area under the ITCF $L(\mathbf{q})$ (shaded gray) determines the static limit of the linear density response function $\chi(\mathbf{q},\omega)$~\cite{Dornheim_MRE_2023},
\begin{eqnarray}\label{eq:static_chi}
    \chi_{ee}(\mathbf{q}) &=& - n_e \underbrace{\int_0^\beta \textnormal{d}\tau\ F_{ee}(\mathbf{q},\tau)}_{L(\mathbf{q})}\\ &=& -2 n_e\ M_{-1}^S(\mathbf{q})\ ,
\end{eqnarray}
which is often denoted as the imaginary-time version of the fluctuation--dissipation theorem in the literature~\cite{Dornheim_MRE_2023,bowen2}; it is also directly related to the inverse frequency moment $M_{-1}^S(\mathbf{q})$, which is the only convergent inverse moment of the dynamic structure factor.
The utility of Eq.~(\ref{eq:static_chi}) is discussed in more detail in Sec.~\ref{sec:density_response}.
For completeness, we note that Eq.~(\ref{eq:static_chi}) is of great importance for PIMC based investigations of the static density response~\cite{dynamic_folgepaper,dornheim_ML,Dornheim_PRR_2022,Dornheim_review,Dornheim_MRE_2024,dornheim_HEDP,svensson2026reweightingestimatorsdensityresponse}, as it allows for the evaluation of the full range of wavenumbers from a single simulation of the unperturbed system; non-linear extensions and a generalization to the Matsubara frequency domain have been discussed in Refs.~\cite{Dornheim_JCP_ITCF_2021,Dornheim_CPP_2022,Dornheim_review,svensson2026reweightingestimatorsdensityresponse} and Refs.~\cite{Tolias_JCP_2024,Dornheim_PRB_2024,Dornheim_EPL_2024,Dornheim_CPP_2025}, respectively.

\begin{figure}
    \centering
    \includegraphics[width=0.99\linewidth]{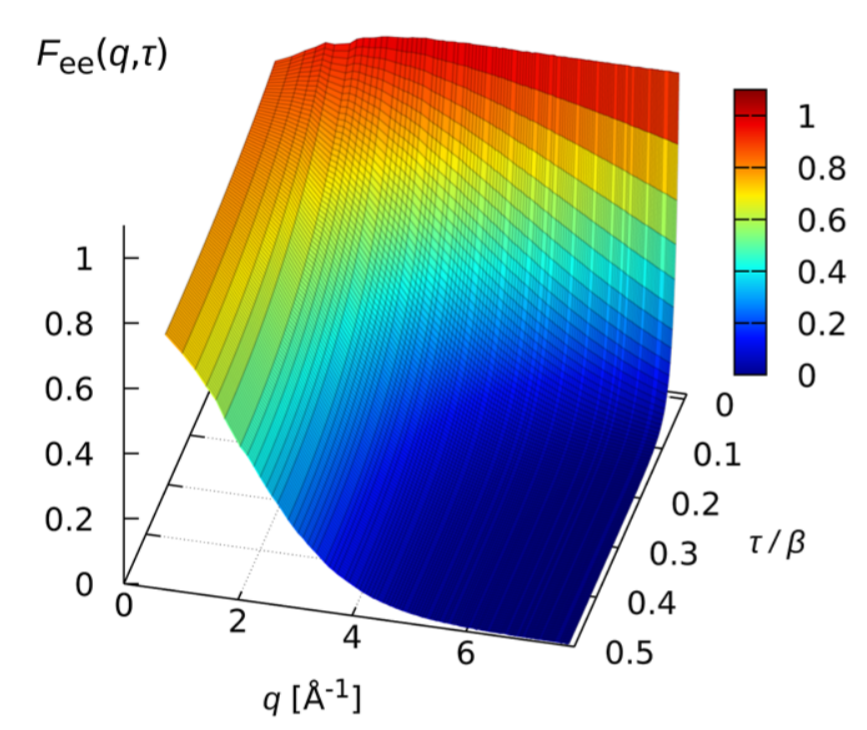}
    \caption{\emph{Ab initio} PIMC results for the electronic ITCF $F_{ee}(\mathbf{q},\tau)$ of warm dense hydrogen at $\rho=0.8\,$g/cc and $T=4.8\,$eV in the $q$-$\tau$-plane. Note that it is sufficient to show the interval $\tau\in[0,\beta/2]$ because of the symmetry of the ITCF, Eq.~(\ref{eq:ITCF_symmetry}).
    Taken from Ref.~\cite{Dornheim_MRE_2024} with the permission of the authors.
    \label{fig:ITCF_Hjet}}
\end{figure}

Let us next consider the dependence of the ITCF on the wavenumber $q$, which is shown for warm dense hydrogen at $T=4.8\,$eV and $\rho=0.08\,$g/cc in Fig.~\ref{fig:ITCF_Hjet}~\cite{Dornheim_MRE_2024}. These conditions can be realized in experiments with heated cryogenic hydrogen jets~\cite{Zastrau,Toleikis_2010,Faustlin_PRL_2010,Fletcher_Frontiers_2022} and are located in a particularly interested regime that is characterized by approximately half ionization~\cite{Bellenbaum_PRR_2025,Bonitz_POP_2024} and that might give rise to a roton type feature in the dynamic structure factor at intermediate wavenumbers.
To understand the main trends of the full $F_{ee}(\mathbf{q},\tau)$, we first recall that $F_{ee}(\mathbf{q},0)=S_{ee}(\mathbf{q})$,
with the static structure factor 
\begin{eqnarray}\label{eq:SSF}
    S_{ee}(\mathbf{q}) = \int_{-\infty}^\infty\textnormal{d}\omega\ S_{ee}(\mathbf{q},\omega)\ ,
\end{eqnarray}
being defined as the normalization of the dynamic structure factor.
It always holds $\lim_{q\to\infty}S_{ee}(\mathbf{q})=1$ (i.e., perfect correlation of a particle exclusively with itself in the limit of small length scales $\lambda_q=2\pi/q$).
For the UEG (and, hence, in the limit of fully ionized systems at very high temperatures), it holds $\lim_{q\to0}S_{ee}(q)\sim q^2$, which is sometimes known as the \emph{perfect screening sum rule}~\cite{kugler_bounds};
however, $S_{ee}(\mathbf{q})$ generally tends towards a finite value in the limit of $q\to0$ for multi-component systems, which is governed by the compressibility sum rule~\cite{CHATURVEDI198111}.
The dependence of $F_{ee}(\mathbf{q},\tau)$ on $\tau$ strongly depends on the wavenumber $q$, which can be understood in multiple ways.
As mentioned above, the slope of $F_{ee}(\mathbf{q},\tau)$ with respect to $\tau$ around $\tau=0$ scales as $\sim q^2$ [Eq.~(\ref{eq:fsum})], leading to the increasingly steep decay for large wave numbers observed in Fig.~\ref{fig:ITCF_Hjet}.
This trend also becomes intuitively clear by re-calling the calculation of the ITCF within PIMC simulations sketched in Fig.~\ref{fig:paths}.
As mentioned before $S_{ee}(\mathbf{q})=F_{ee}(\mathbf{q},0)$ is equal to unity for large $q$ because of self correlations; these, however, slowly decay along the diffusion throughout the imaginary time when the paths are not just classical straight lines. In this sense, the decay of $F_{ee}(\mathbf{q},\tau)$ is a direct measure for the degree of quantum delocalization of the electrons, measured on the length scale $\lambda_q=2\pi/q$. For large $q$, it is $\lambda_q \ll \lambda_\beta$ and hardly any self correlations remain for finite $\tau$, thus explaining the observed decay of the ITCF in the bottom right corner of Fig.~\ref{fig:ITCF_Hjet}.
An in-depth discussion between the imaginary-time diffusion process and the $\tau$-decay of $F_{ee}(\mathbf{q},\tau)$ has been presented in Ref.~\cite{Dornheim_PTRS_2023}.

\begin{figure}
    \centering
    \includegraphics[width=0.99\linewidth]{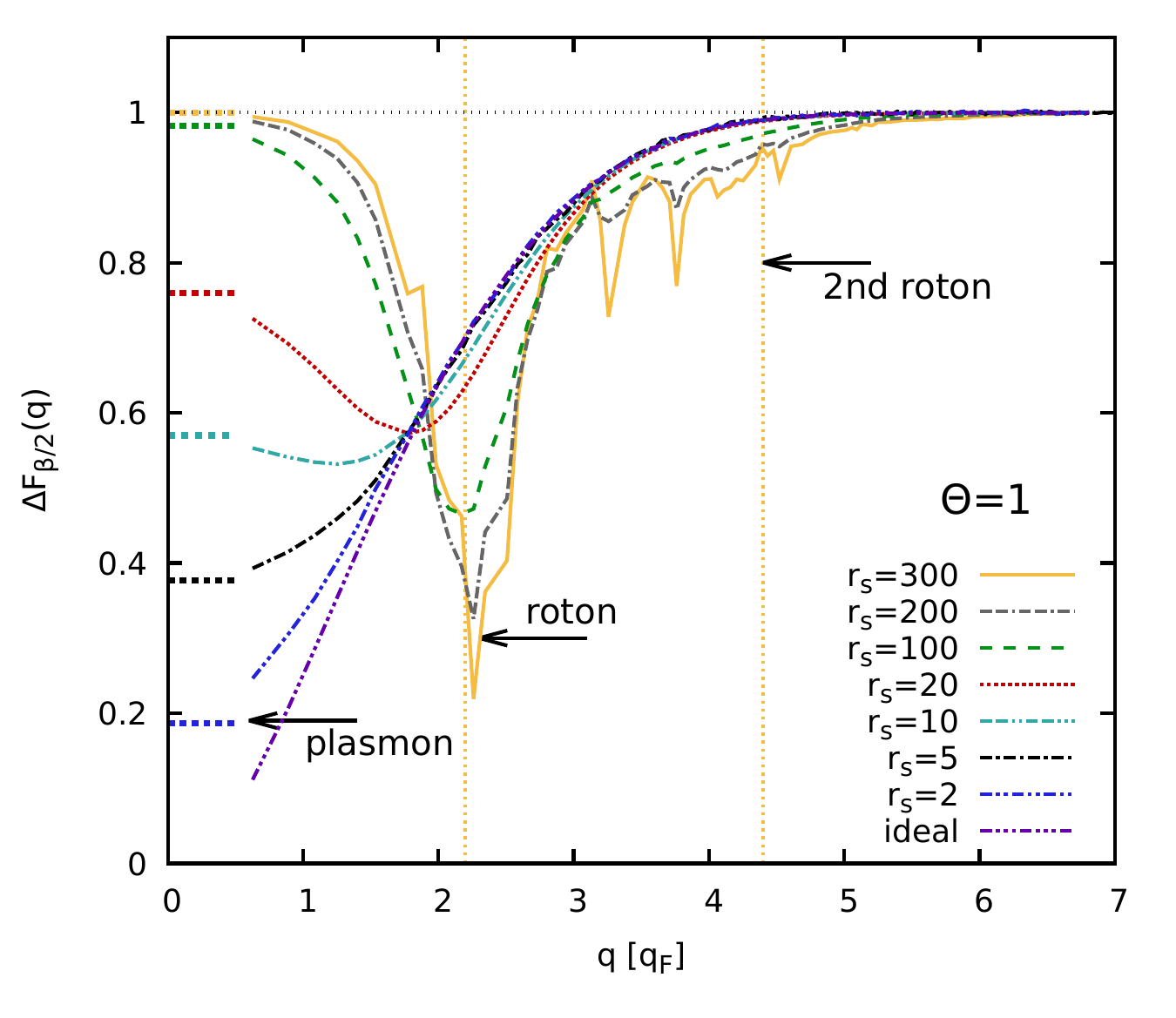}
    \caption{\emph{Ab initio} PIMC results for the relative $\tau$-decay of the ITCF [Eq.~(\ref{eq:decay_measure})] of the uniform electron gas at the electronic Fermi temperature for different density parameters $r_s$.
    Taken from Ref.~\cite{Chuna_JCP_2025} with the permission of the authors.
    \label{fig:roton}}
\end{figure}

Let us conclude our discussion of the physical behavior of the ITCF with an explicit consideration of its relation to the dynamic structure factor $S_{ee}(\mathbf{q},\omega)$, i.e., the two-sided Laplace transform defined in Eq.~(\ref{eq:ITCF}) above. 
In practice, Eq.~(\ref{eq:ITCF}) governs how the distribution of spectral weight shapes the $\tau$-dependence of $F_{ee}(\mathbf{q},\tau)$ for any given $q$,
with spectral weight at low frequencies leading to a small decay with $\tau$ (and vice versa).
This straightforward mechanics has inspired Dornheim and co-workers~\cite{Dornheim_MRE_2023,Chuna_JCP_2025} to define a measure for the relative decay of the ITCF with $\tau$,
\begin{eqnarray}\label{eq:decay_measure}
    \Delta F_{ee}^\tau(\mathbf{q}) = \frac{F_{ee}(\mathbf{q},0) - F_{ee}(\mathbf{q},\tau)}{F_{ee}(\mathbf{q},0)}\ ,
\end{eqnarray}
which is shown in Fig.~\ref{fig:roton} based on extensive \emph{ab initio} PIMC simulations of the UEG at the electronic Fermi temperature over a broad range of density parameters $r_s$.
This depiction phenomenologically resembles a dispersion relation of $S_{ee}(\mathbf{q},\omega)$ and gives one access to similar physical information.
In the limit of long wavelengths (i.e., $q\to0$), the dynamic structure factor of the UEG is given by a single plasmon excitation at the plasma frequency, in which case the ITCF is simply given by a a combination of two exponential functions, $F_{ee}^\textnormal{pl}(q\to0,\tau)\sim e^{-\tau\hbar\omega_\textnormal{pl}} + e^{-(\beta-\tau)\hbar\omega_\textnormal{pl}}$; this limit is indicated by the dotted bars on the left side of Fig.~\ref{fig:roton}.
At strong coupling, the dynamic structure factor of the UEG exhibits a non-monotonous dispersion relation~\cite{dornheim_dynamic,Takada_PRB_2016,Koskelo_PRL_2025,Dornheim_CommPhys_2022,Dornheim_JCP_2022}, with a minimum at intermediate wavenumbers that phenomenologically resembles the roton feature that is well known from ultracold quantum liquids~\cite{Godfrin2012,Kalman_2010,Dornheim_SciRep_2022,Ferre_PRB_2016} when $\lambda_q\sim d$.
Such a redshift of spectral weight in $S_{ee}(\mathbf{q},\omega)$ inevitably manifests as a reduced $\tau$-decay in the ITCF, which is the origin of the first roton feature at $q\approx2q_\textnormal{F}$ in Fig.~\ref{fig:roton}.
Interestingly, Chuna \textit{et al.}~\cite{Chuna_JCP_2025} have reported a second roton feature at around the second harmonic of the first roton at very strong coupling strengths, see the second dip in Fig.~\ref{fig:roton}.
This prediction in terms of $\Delta F^\tau_{ee}(\mathbf{q})$ has been fully confirmed by analytic continuation results for $S_{ee}(\mathbf{q},\omega)$,
thus further highlighting the potential of the ITCF for direct physical analyses.

\begin{figure*}
    \centering
    \includegraphics[width=0.79\linewidth]{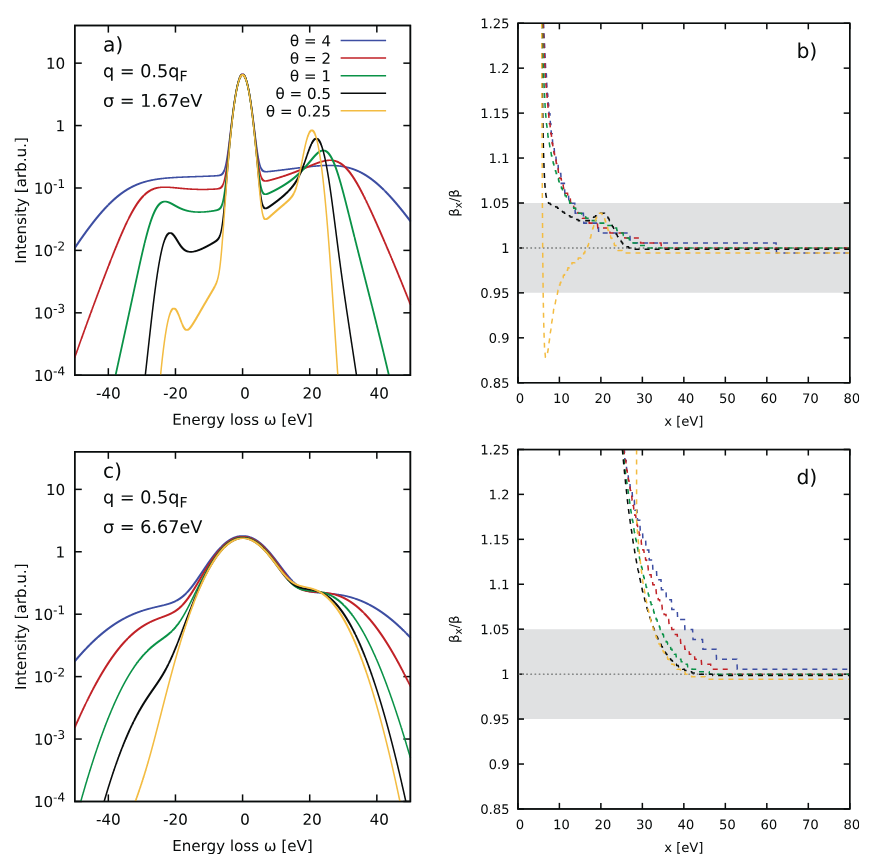}
    \caption{a) and c): synthetic XRTS spectra computed for different temperatures (given in units of the electronic Fermi temperature of $T=12.53\,$eV at a density parameter $r_s=2$) for two widths of a Gaussian model SIF, $\sigma=1.67\,$eV and $\sigma=6.67\,$eV in a forward scattering geometry of $q=0.5q_\textnormal{F}$  ($q\approx0.91$\AA$^{-1}$); b) and d): convergence of the extracted inverse temperature from the position of the minimum of the truncated deconvolved ITCF $F^x_{ee}(\mathbf{q},\tau)$ [Eq.~(\ref{eq:ITCF_x})] as a function of the spectral integration range $x$.
    Taken from Ref.~\cite{Dornheim_T2_2022} with the permission of the authors.
    \label{fig:artificial_4}}
\end{figure*}

\subsection{Thermometry\label{sec:thermometry}}

\textbf{Idea:} The basic idea behind the model-free thermometry approach that has been introduced in Refs.~\cite{Dornheim_T_2022,Dornheim_T2_2022} and can be decomposed into the following steps:
(1) compute the Laplace transforms of the measured XRTS intensity $I(\mathbf{q},\omega)$ and of the SIF $R(\omega)$. Strictly speaking, the two-sided Laplace transform [Eq.~(\ref{eq:ITCF})] and its deconvolution theorem [Eq.~(\ref{eq:deconvolution})] would require one to integrate from $\omega\to-\infty$ to $\omega\to\infty$, whereas the spectral range on any given detector is always finite. 
Since the SIF $R(\omega)$ is often characterized using analytical models~\cite{Gawne_JAP_2024}, we will assume in the following that we can evaluate the full Laplace transform $\mathcal{L}\left[R(\omega)\right]$ and instead focus on the Laplace transform of the XRTS intensity $I(\mathbf{q},\omega)$.
For this purpose, we define the symmetrically truncated Laplace transform
\begin{eqnarray}\label{eq:Laplace_x}
    \mathcal{L}^x\left[ f(\omega) \right] = \int_{-x}^x\textnormal{d}\omega\ f(\omega)\ e^{-\tau\hbar\omega}\ ,
\end{eqnarray}
and the corresponding symmetrically truncated deconvolved ITCF
\begin{eqnarray}\label{eq:ITCF_x}
    F^x_{ee}(\mathbf{q},\tau) = \frac{\mathcal{L}^x\left[I(\mathbf{q},\omega)\right]}{\mathcal{L}\left[R(\omega)\right]}\ ,
\end{eqnarray}
with the trivial identity $F_{ee}(\mathbf{q},\tau) = \lim_{x\to\infty} F_{ee}^x(\mathbf{q},\tau)$.
Let us first focus on the symmetrically truncated Laplace transform of the dynamic structure factor $\mathcal{L}^x\left[S_{ee}(\mathbf{q},\omega)\right]$ and ignore the effect of the SIF.
It is easy to see that $\mathcal{L}^x\left[S_{ee}(\mathbf{q},\omega)\right]$ fulfills the detailed balance symmetry relation Eq.~(\ref{eq:ITCF_symmetry}) for any symmetric integration interval $\pm x$;
the task at hand is thus to converge Eq.~(\ref{eq:ITCF_x}) with respect to the validity of the deconvolution theorem.
A central part of the model-free ITCF thermometry approach is thus to evaluate $F^x_{ee}(\mathbf{q},\tau)$ over an entire set of integration boundaries $x$ and to check the corresponding convergence of the position of the minimum.
We further note that convergence of the thus extracted (inverse) temperature is usually achieved a lot sooner than convergence of the full ITCF $F_{ee}(\mathbf{q},\tau)$ [and derived properties such as the f-sum rule] as we do not necessarily have to take into account the full spectrum including potentially far-flung K-edge bound-free features to attain the correct symmetry.

\textbf{Results:} Let us start with a discussion of clean, synthetic XRTS spectra, which have been obtained by combining dynamic structure factors of a uniform electron gas at $r_s=2$ for different values of the reduced temperature $\Theta=k_\textnormal{B}T/E_\textnormal{F}$ (with an associated Fermi temperature of $T=12.53\,$eV) with an arbitrary elastic feature at $\omega=0$, convolved with Gaussian model SIFs of width $\sigma=1.67\,$eV and $\sigma=6.67\,$eV shown in the top and bottom row of Fig.~\ref{fig:artificial_4}~\cite{Dornheim_T2_2022}, respectively.
Specifically, we consider a wavenumber of $q=0.5q_\textnormal{F}\approx0.91\,$\AA$^{-1}$, corresponding to a forward scattering geometry with a scattering angle of $\theta=17^\circ$ for a hypothetical beam energy of $\hbar\omega_0=6\,$keV.
We note that the spectra that are shown in Fig.~\ref{fig:artificial_4}a) and c) are plotted with respect to the photon energy loss, and the up- and down-shifted sides of the spectrum are thus flipped compared to the absolute photon energy that is often shown in experimental studies.
In the idealized case of unconvolved dynamic structure factors, the temperature could be inferred directly by taking the ratio of the down-shifted (positive energy loss) to the up-shifted (negative energy loss) plasmon feature, as it has been suggested, e.g., by D\"oppner \textit{et al.}~\cite{DOPPNER2009182}.
Alas, the convolution destroys this symmetry of $S_{ee}(\mathbf{q},\omega)$.
Switching instead to the imaginary-time domain, the Laplace transform of a Gaussian model SIF $R_\sigma(\omega)$ of width $\sigma$ is given by $e^{\sigma^2\tau^2/2}$. 
In the limit of very high temperatures or very narrow SIFs, we may represent it as a Taylor expansion,
\begin{eqnarray}\label{eq:Taylor}
    \mathcal{L}\left[R_\sigma(\omega)\right] = 1 + \frac{\sigma^2\tau^2}{2} + \mathcal{O}\left(\sigma^4\tau^4\right)\ .
\end{eqnarray}
The effect of the SIF, thus, becomes negligible for $\sigma^2\tau^2\ll1$, which coincides with the naive physical intuition and which highlights the utility of state-of-the-art diced crystal analyzer set-ups with meV spectral resolution~\cite{Wollenweber_RSI_2021,McBride_RSI_2018,Descamps_SciReports_2020,Gawne_PRB_2024,Gawne_ElectronicStructure_2025,White_Nature_2025,gawne2025orientationaleffectslowpair} for model-free thermometry.

Let us next consider Figs.~\ref{fig:artificial_4}b) and d), where we show the convergence of the inverse temperature [with $\beta_x$ being defined as twice the position of the minimum of $F^x_{ee}(\mathbf{q},\tau)$] with $x$.
Interestingly, the length of the integration interval only weakly depends on the physical temperature of the probed sample, but it is mostly defined by the width of the SIF.
While it is mathematically guaranteed that one recovers the correct temperature in the limit of $x\to\infty$, which is indeed numerically verified in Fig.~\ref{fig:artificial_4}, the key question regarding the practical feasibility of the method is to what level of precision one needs to resolve the XRTS spectrum in the experiment.
Generally, the dynamic range of XRTS measurements at modern XFEL facilities such as the European XFEL in Germany~\cite{Voigt_POP_2021} or LCLS in the USA~\cite{kraus_xrts,Bellenbaum_APL_2025} is three orders of magnitude in the signal $I(\mathbf{q},\omega)$.
For the SIF with $\sigma=1.67\,$eV, we have to resolve the spectrum within an interval of at least $x\gtrsim30\,$eV, which is possible for $\Theta\gtrsim0.5$, i.e., $T\gtrsim6\,$eV.
For the broader SIF with $\sigma=6.67\,$eV, on the other hand, we have to resolve an interval of at least $x\gtrsim40\,$eV, which 
is not feasible for $T\lesssim10\,$eV.
The investigation of the synthetic data shown in Fig.~\ref{fig:artificial_4} thus directly highlights the importance of a narrow SIF to resolve lower temperatures.
This is very promising and principally opens up the possibility even for model-free XRTS thermometry in material science and discovery as well as laboratory astrophysics experiments with a direct relevance for planetary physics~\cite{wdm_book,drake2018high,vorberger2025roadmapwarmdensematter} with temperatures on the order of $T\sim1\,$eV by combining monochromated, seeded XFEL beams with the aforementioned high-resolution diced crystal analyzer set-ups~\cite{Wollenweber_RSI_2021,McBride_RSI_2018,Descamps_SciReports_2020,Gawne_PRB_2024,Gawne_ElectronicStructure_2025,White_Nature_2025,gawne2025orientationaleffectslowpair}.

While the real SIF is rarely exactly Gaussian~\cite{Gawne_JAP_2024,MacDonald_POP_2022}, the exponentially increasing nature of its two sided Laplace transform $\mathcal{L}\left[R_\sigma(\omega)\right]$ 
illuminates a second problem of broad SIFs: at low temperatures (i.e., large inverse temperatures $\beta$), we need to resolve a large $\tau$-interval to determine the position of the minimum of $F_{ee}^x(\mathbf{q},\tau)$ at $\tau=\beta/2$.
Since the physical ITCF $F_{ee}(\mathbf{q},\tau)$ is of the order of one, the two-sided Laplace transform of the measured XRTS signal $\mathcal{L}\left[I(\mathbf{q},\omega)\right]=F_{ee}(\mathbf{q},\tau) \mathcal{L}\left[R_\sigma(\omega)\right]$ is mainly determined by the SIF.
This, in turn, means that any small uncertainty in our SIF model---be it an analytical model or an experimental measurement---will be exponentially amplified with increasing $\tau$.
It is thus important to keep in mind that the application of the ITCF thermometry approach at low temperatures also poses additional requirements to our SIF modeling capabilities, in particular for broader SIFs. These stringent requirements may again be eased by high-resolution set-ups, e.g., using DCAs.

\begin{figure}
    \centering
    \includegraphics[width=0.99\linewidth]{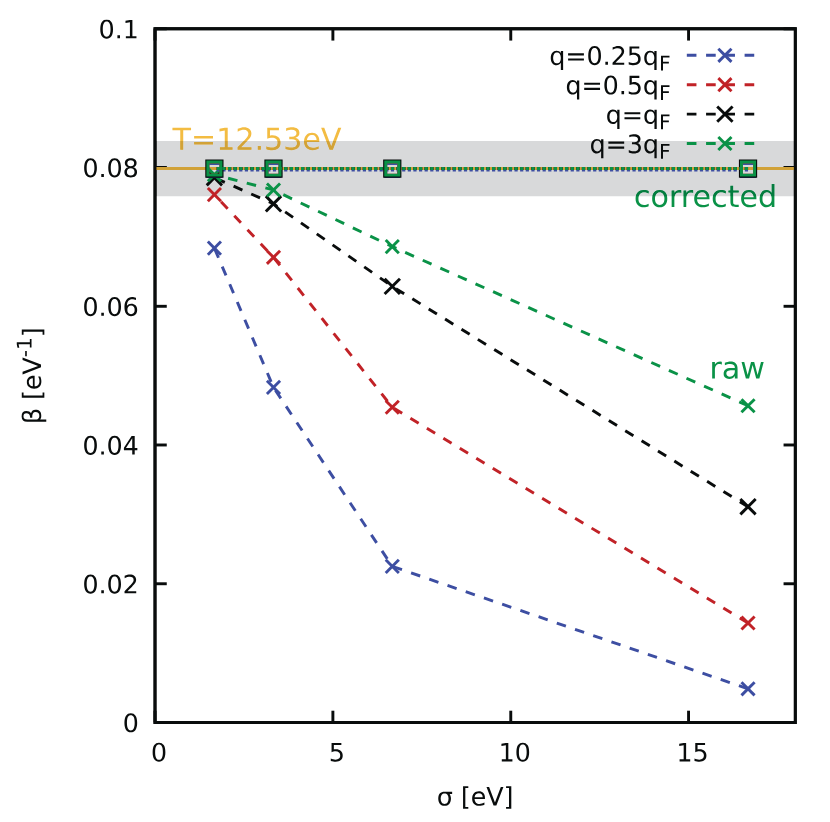}
    \caption{Impact of the width $\sigma$ of a Gaussian model SIF onto the extracted inverse temperature of synthetic spectra. The correct temperature of $T=12.53\,$eV (corresponding to the Fermi temperature at the density parameters $r_s=2$) is indicated by the vertical yellow line, and the shaded gray area indicates an uncertainty range of $\pm5\%$. The raw and properly deconvolved results are shown by the crosses and squares, respectively.
    Taken from Ref.~\cite{Dornheim_T2_2022} with the permission of the authors.
    \label{fig:artificial_sigma}}
\end{figure}

Let us conclude our investigation of synthetic data by investigating the interplay of scattering wavenumber $q$ and width $\sigma$ of the Gaussian model SIF shown in Fig.~\ref{fig:artificial_sigma} for a fixed temperature of $T=12.53\,$eV, i.e., the case of $\Theta=1$ from Fig.~\ref{fig:artificial_4}.
Specifically, the colors distinguish the different wave numbers corresponding to $\theta=7.4^\circ$ (blue), $\theta=17^\circ$ (red), $\theta=30^\circ$, and $\theta=62^\circ$ at $\hbar\omega_0=7\,$keV, and the crosses and squares show the temperatures extracted from the raw Laplace transforms of $I(\mathbf{q},\omega)$ and the properly deconvolved ITCFs $F_{ee}(\mathbf{q},\tau)$, respectively.
First, we note that the deconvolved ITCF always gives the correct temperature, as it is expected.
Second, we see that the impact of the convolution on the extracted temperature from the convolved XRTS signal is smallest for the largest wavenumber; this can be heuristically attributed to the broader dynamic structure factors at larger $q$, making the effect of the additional broadening due to the convolution with the SIF comparably less pronounced.
Third, it becomes apparent that the benefits from a narrow SIF are substantially more pronounced at larger scattering angles.
Strikingly, even the non-deconvolved Laplace transform of the full XRTS signal $I(\mathbf{q},\omega)$ gives one the correct temperature within $1\%$ for $q=3q_\textnormal{F}$ for $\sigma=1.67\,$eV.

\begin{figure*}
    \centering
    \includegraphics[width=0.99\linewidth]{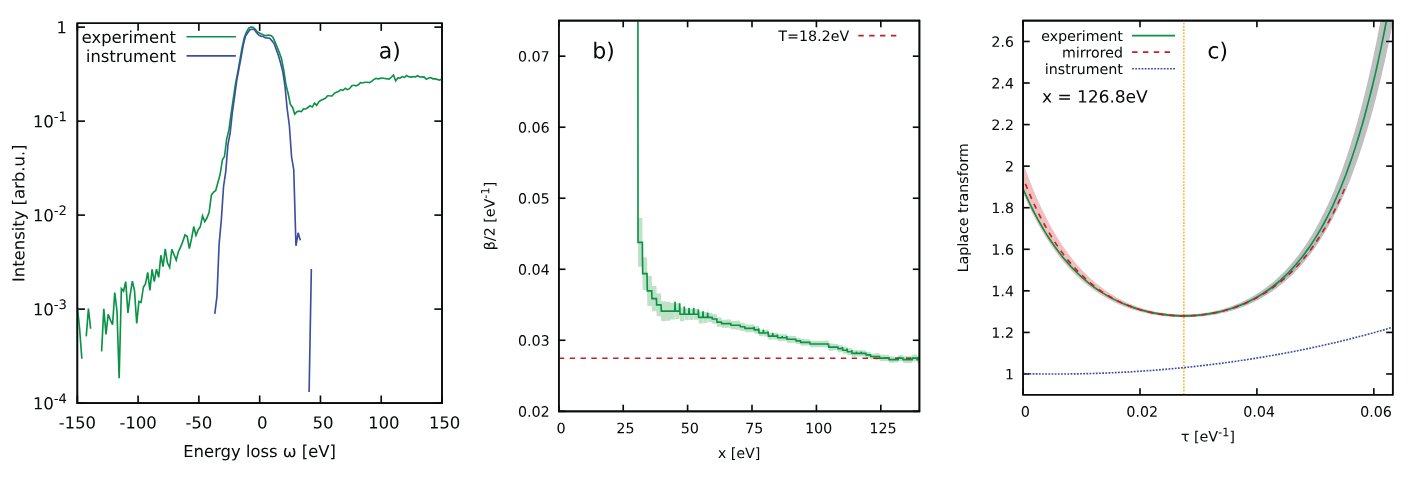}
    \caption{Model-free ITCF thermometry analysis of an XRTS measurement of isochorically heated graphite taken at LCLS by Kraus \textit{et al.}~\cite{kraus_xrts}. a) XRTS signal (green) and SIF (blue) as a function of photon energy loss; b) convergence of the extracted temperature with the symmetric integration interval $x$; c) deconvolved ITCF [arb.~units] (green) and its mirror image (red) around the position of the minimum (vertical yellow) line for $x=126.8\,$eV, as well as the Laplace transform of the SIF (blue). The gray and red shaded intervals show the associated uncertainties.
    Taken from Ref.~\cite{Dornheim_T2_2022} with the permission of the authors.
    \label{fig:Father_Son_HolyGhost}}
\end{figure*}


In Fig.~\ref{fig:Father_Son_HolyGhost}, we show results for the model-free ITCF thermometry analysis as it has been applied to an XRTS measurement of isochorically heated graphite taken at LCLC by Kraus \textit{et al.}~\cite{kraus_xrts}.
Fig.~\ref{fig:Father_Son_HolyGhost}a) shows the measured XRTS spectrum $I(\mathbf{q},\omega)$ as a function of the photon energy loss (green), as well as the SIF (blue).
Fig.~\ref{fig:Father_Son_HolyGhost}b) shows the convergence of the extracted temperature with respect to the symmetric integration boundary $x$, leading to a best estimate of $T=18\pm2\,$eV. We note that going beyond the depicted range of $x$ does not make sense as the measured XRTS signal is vanishing within the given noise level at around $\omega=-140\,$eV, on the up-shifted side.
The estimated temperature is reasonably close to the value of $T\approx21.7\,$eV estimated in the original Ref.~\cite{kraus_xrts} using a Chihara model, and within error bars to the estimate of $T=16.6\,$eV reported by B\"ohme \textit{et al.}~\cite{bohme2023evidencefreeboundtransitionswarm} using an improved Chihara model that properly takes into account free-bound transitions, see Sec.~\ref{sec:Chihara} above.
At the same, we stress again that no models were required for this result.
Fig.~\ref{fig:Father_Son_HolyGhost}c) shows the properly deconvolved ITCF (green) for an integration interval of $x=128.6\,$eV, for which the temperature extraction already has converged. 
For completeness, the dotted blue curve shows the two-sided Laplace transform of the SIF, which only mildly increases with $\tau$. The width of the SIF is thus appropriate for the application of the thermometry approach at the given conditions.
Finally, the dashed red curve corresponds to the green curve, but mirrored around the position of its minimum at $\tau=\beta/2$, i.e., the vertical yellow line.
We find that the deconvolved ITCF is indeed nicely symmetric as predicted by Eq.~(\ref{eq:ITCF_symmetry}).
This result is an important finding in itself as the role of non-equilibrium effects, which might violate the symmetry conditions, in self-heating experiments remains debated~\cite{Bellenbaum_APL_2025}, see also Sec.~\ref{sec:non_eq} below.

\begin{figure}
    \centering
    \includegraphics[width=0.99\linewidth]{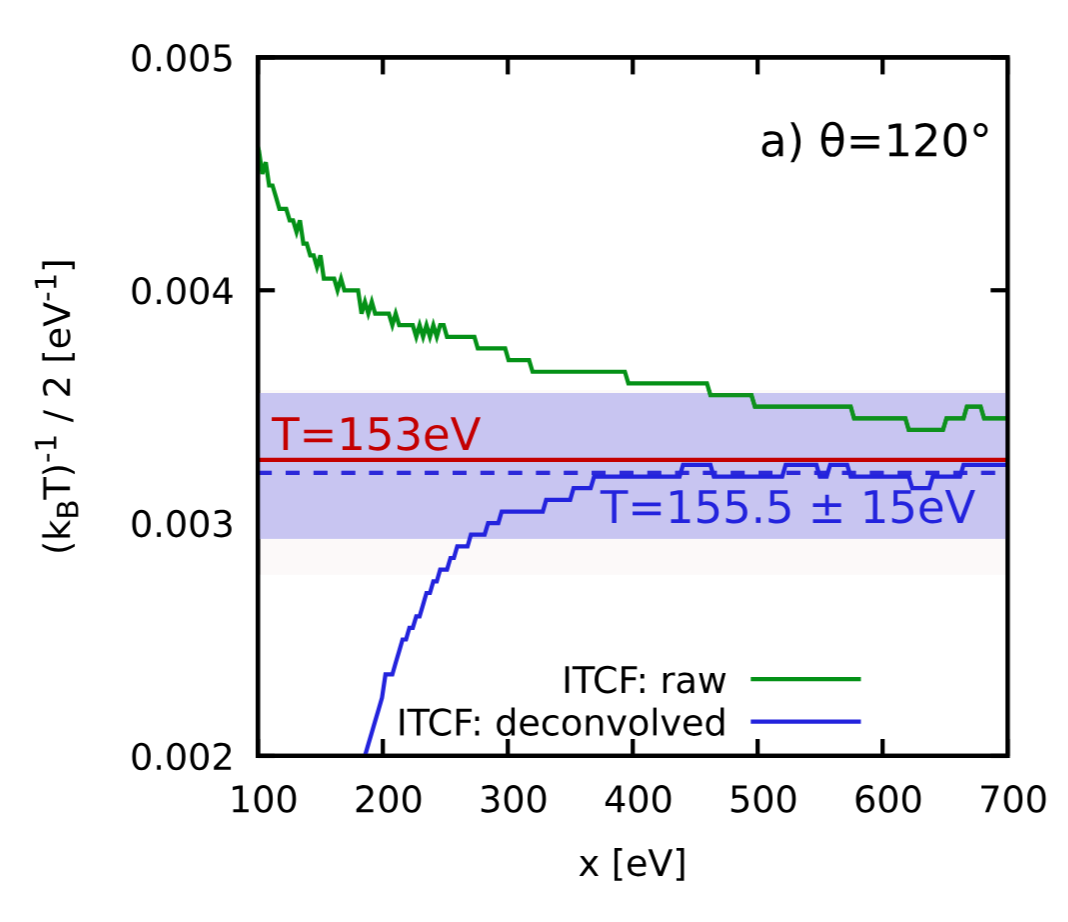}
    \caption{Convergence of the position of the minimum of $F^x_{ee}(\mathbf{q},\tau)$ [Eq.~(\ref{eq:ITCF_x})] with the symmetrically truncated integration boundary $x$ for an XRTS dataset of spherically compressed beryllium taken in a backscattering geometry with $\theta=120^\circ$ at the NIF by D\"oppner \textit{et al.}~\cite{Tilo_Nature_2023}.
    The horizontal dashed line corresponds to the model-free thermometry estimate of $T=155.5\pm15\,$eV, with the shaded blue area highlighting the associated uncertainty. The horizontal solid red line shows the best fit result of $T=153\,$eV based on the improved Chihara model by B\"ohme \textit{et al.}~\cite{bohme2023evidencefreeboundtransitionswarm}.
    Taken from Ref.~\cite{Dornheim_NatComm_2025} with the permission of the authors.
    \label{fig:tilo_converges}}
\end{figure}

In Fig.~\ref{fig:tilo_converges}, we show the convergence of the ITCF thermometry approach for a second experiment, namely an XRTS measurement of spherically compressed beryllium taken in a backscattering geometry with $\theta=120^\circ$ and $\hbar\omega_0=9\,$keV at the NIF by D\"oppner \textit{et al.}~\cite{Tilo_Nature_2023}, who have reported an estimated temperature of $T=160\pm20$\,eV
based on a Chihara model fit.
For this data set, the ITCF method converges even more smoothly than for the graphite data set shown in Fig.~\ref{fig:Father_Son_HolyGhost}, and the position of the minimum of the properly deconvolved ITCF $F^x_{ee}(\mathbf{q},\tau)$ clearly converges for $\omega\gtrsim400\,$eV; we find a best estimate of $T=155.5\pm15\,$eV, where the error bars are mostly due to some unknown parameters in the SIF, see also the discussion in Ref.~\cite{bohme2023evidencefreeboundtransitionswarm}.
For completeness, we have also included a Chihara estimate using the improved Chihara model that also takes into account the previously missing free-bound contributions (horizontal red line), with a best fit at $T=153\,$eV.

Other applications of the ITCF thermometry approach can be found in Refs.~\cite{Dornheim_T_2022,Dornheim_T2_2022,Schoerner_PRE_2023,shi2025firstprinciplesanalysiswarmdense,Smid_SciRep_2026,Dornheim_NatComm_2025,bohme2026correlationfunctionmetrologywarm}.

\textbf{Range of application and limitations:}

On the positive side, finding the correct minimum of the ITCF at $\tau=\beta/2$ does not require one to actually estimate the full, physical ITCF $F_{ee}(\mathbf{q},\tau)$ where one would need to integrate over the entire relevant spectral range.
This can be arbitrarily challenging, in particular for heavier elements and lower temperatures, where bound-free features at high excitation energies generally cannot be resolved on the up-shifted side of the XRTS spectrum due to the exponential damping mandated by the detailed balance relation Eq.~(\ref{eq:detailed_balance}) in thermal equilibrium.
On the negative side, we have to resolve the truncated ITCF $F^x_{ee}(\mathbf{q},\tau)$ over a substantial $\tau$-range, i.e., at least $\tau\gtrsim\beta/2$ to find the position of its minimum for a sufficiently large $x$.
This is problematic for low $T$ (large $\beta$), and the difficulty is further exacerbated by a broad SIF. 
Finally, we note that the ITCF thermometry method is completely agnostic with respect to the probed material, and works for arbitrary elements and even complicated mixtures.

\subsection{Detection of non-equilibrium\label{sec:non_eq}}

\textbf{Idea:} In the same way we can use the symmetry of the ITCF to determine the temperature of the system under study, we can use a lack of symmetry of the ITCF (the absence of detailed balance) as a gauge of non-equilibrium~\cite{Vorberger_PLA_2024}. A further possible check for consistency
is given by the comparison of the ITCFs obtained at different scattering angles. They might even show symmetric features as necessary for equilibrium but indicate different temperatures in which case non-equilibrium has been detected as well~\cite{Vorberger_PLA_2024}. In case the detector has a sufficiently high temporal resolution or the delay between the pump and probe beams can be reliably changed between shots, it should become possible to follow the entire temperature relaxation~\cite{Vorberger_PLA_2024}.
First re-analysis attempts by Bellenbaum \textit{et al.} of LCLS self-scattering data~\cite{kraus_xrts}, were inconclusive since the accuracy of the data and the SIF was not high enough and the experiment was not designed with this idea in mind~\cite{Bellenbaum_APL_2025}. 

In these considerations, it is useful to introduce the ratio~\cite{Vorberger_PLA_2024}
\beq
R(\qv,\tau,\beta')=F_{ab}(\qv,\tau)/F_{ab}(\qv,\beta'-\tau)
\label{deviation}
\eeq
as a measure for non-equilibrium. Here $\beta'$ is defined by $F(\tau=0)=F(\tau=\beta')$. The quantity of Eq.~(\ref{deviation}) will show an oscillating behavior for non-equilibrium situations for which the detailed balance symmetry in the ITCF $F$ is violated but approach unity the closer the system gets to equilibrium and $\beta'$ will be the inverse temperature then.


\begin{figure}
\centering
\includegraphics[width=0.48\textwidth]{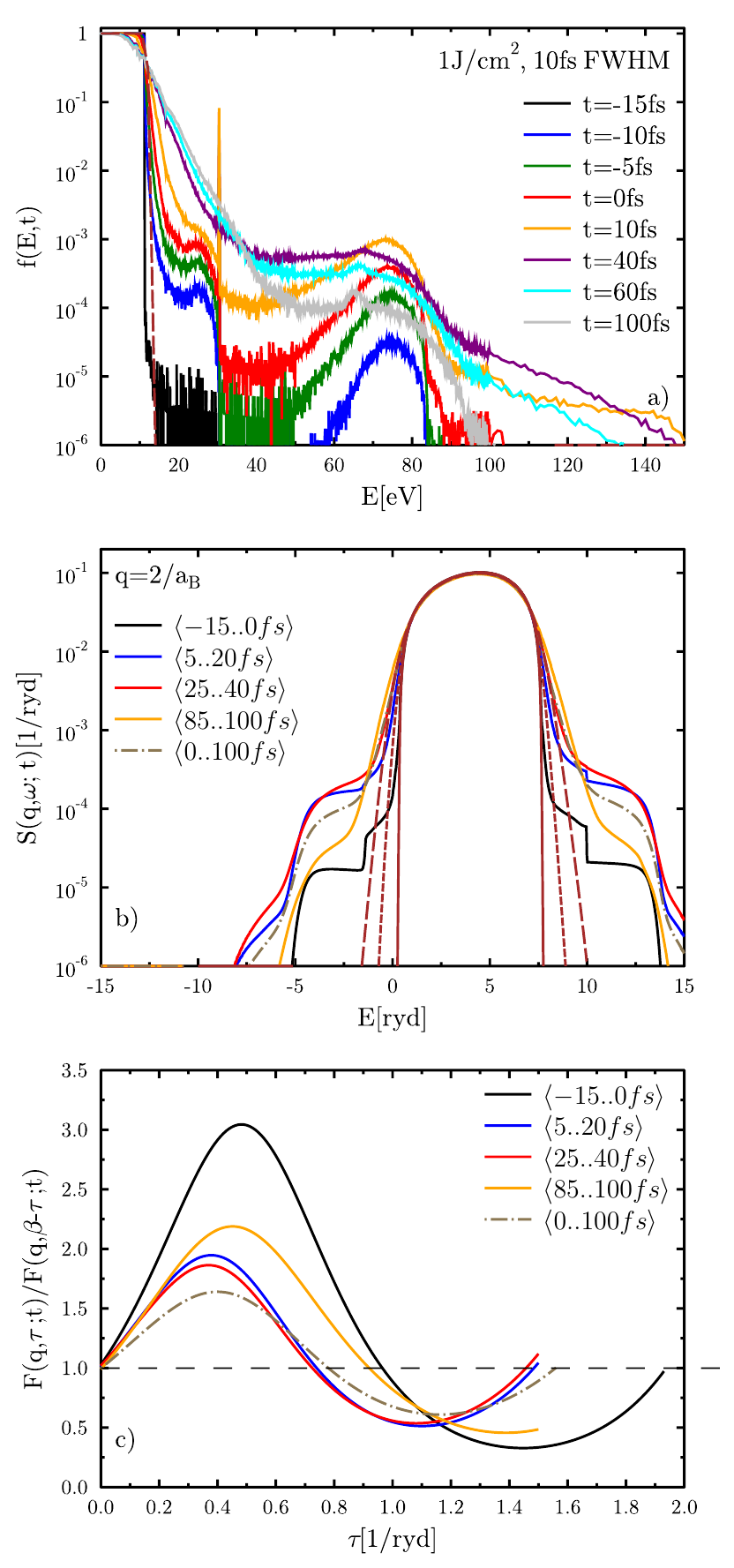}
\caption{\label{fig:nonequil_itcf} 
Time evolution of the non-equilibrium dynamic structure and the ITCF of the electrons in solid aluminium under laser irradiation. The electronic density is $n=1.7\times 10^{23}$~cm$^{-3}$. The Wigner distribution functions are shown in panel a). The dynamic structure factors averaged over different times are shown in panel b). Panel c) shows the ratio $R$ between the ITCF of two different imaginary time arguments.
}
\end{figure} 
As an example we discuss the case of a laser irradiated aluminium foil in Fig.~\ref{fig:nonequil_itcf}. The top panel shows the electron Wigner distribution functions that are to be expected when a $1$~J/cm$^2$ VUV laser ($92$~eV) with a $10$~fs FWHM strikes the foil~\cite{Medvedev2022,Vorberger_PLA_2024}. The modeled dynamic structure factor that results from these Wigner functions for a backscattering geometry is shown in panel b) of Fig.~\ref{fig:nonequil_itcf}. The main peak is the standard Compton feature and the non-equilibrium causes the wings on both sides of the Compton peak. As the relaxation of the Wigner functions proceeds non-linearly via a combination and balance of electron-electron relaxation mechanisms and the laser excitation, the wing features in the dynamic structure factor change non-monotonically as well. The final analysis is contained in panel c), where the ratio $R$ of the two ITCFs is shown. As can be easily observed, the deviation from the line of unity is substantial, indicating the expected non-equilibrium. Further, in accordance with the information gained from the evolution of the Wigner function (which in a real experiment is obviously not available), the ratio $R$ evolves in a non-trivial way towards unity.

\textbf{Range of applications and limitations:} In a realistic experimental scenario, such signal and ITCF analysis would obviously be obfuscated by the experimental noise and the SIF. Initial estimates show that a certain amount of noise does not hinder the method or drawing of conclusions~\cite{Vorberger_PLA_2024}. A high quality characterization of the SIF is however essential as many non-equilibrium features of the dynamic structure factor and thus in the ITCF are hidden in low signal wings and need a very high dynamic range of the detector to become visible.
Generally, the method faces similar (but potentially somewhat exacerbated) limitations as the thermometry approach discussed in the previous Sec.~\ref{sec:thermometry}.

\begin{figure}
\centering
\includegraphics[width=0.99\linewidth]{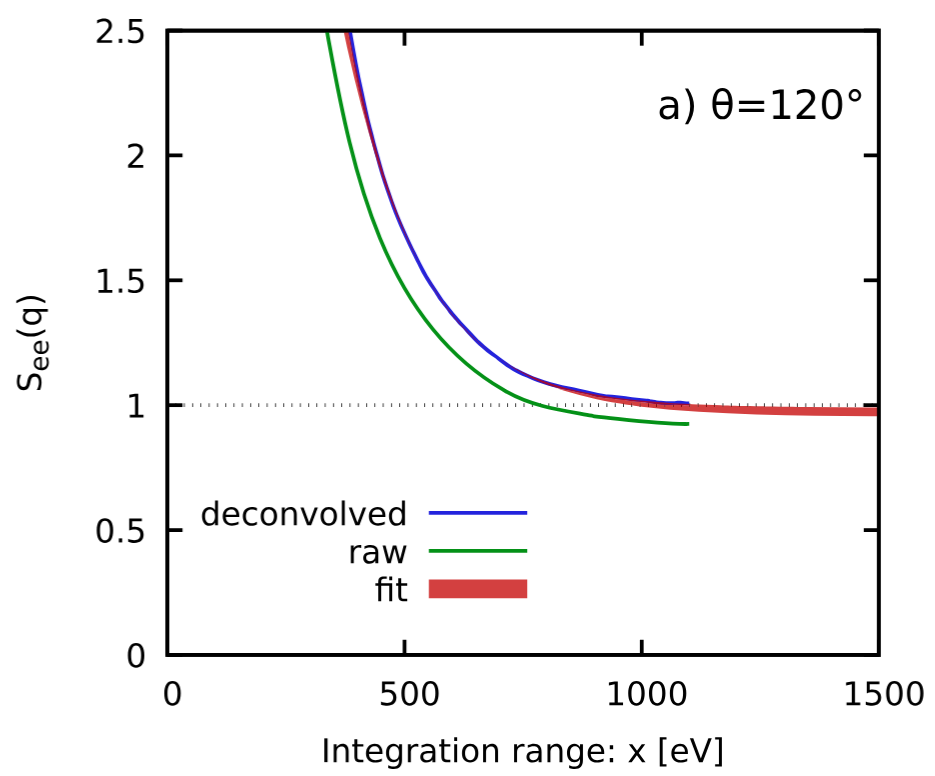}
\caption{\label{fig:SSF} Convergence of the properly normalized electronic static structure factor $S_{ee}(\mathbf{q})=F_{ee}(\mathbf{q},0)$ with the integration interval $x$. The blue and green curves correspond to the normalization inferred from the properly deconvolved, and from the convolved raw data, respectively. The shaded red area shown an exponential extrapolation according to Eq.~(\ref{eq:extrapolate_SSF}).
Taken from Ref.~\cite{Dornheim_NatComm_2025} with the permission of the authors.
}
\end{figure} 

\subsection{Normalization and absolute intensity\label{sec:moments}}

\textbf{Idea:} In practice, the measured XRTS signal is usually proportional to the photon count on a given detector plate. Therefore, the detected intensity is not normalized
\begin{eqnarray}\label{eq:unnormalized_intensity}
    I(\mathbf{q},\omega) = A\ S_{ee}(\mathbf{q},\omega)\circledast R(\omega)\ ,
\end{eqnarray}
with $A$ being the a-priori unknown normalization constant (assuming a properly normalized SIF). 
On the one hand, a constant normalization factor does not affect the location of the minimum of the ITCF or the detection of any violations of the symmetry relation Eq.~(\ref{eq:ITCF_symmetry}). 
On the other hand, other physical applications of the ITCF such as its relation to the static linear density response function [Eq.~(\ref{eq:static_chi}) in Sec.~\ref{sec:density_response} below] do require a proper normalization, which we intend to also infer directly from the measurement and without any model assumptions.
To this end, we consider the frequency moments of $S_{ee}(\mathbf{q},\omega)$, which are directly proportional to the $\tau$-derivatives of the deconvolved ITCF, see Eq.~(\ref{eq:moment_ITCF}) above.
Since the first derivative is determined by the near-universal f-sum rule [Eq.~(\ref{eq:fsum})], we can infer the normalization constant from the XRTS measurement via~\cite{Dornheim_SciRep_2024}
\begin{eqnarray}\label{eq:normalization}
    A = - \frac{2 m_e}{(\hbar q)^2} \frac{\partial}{\partial\tau} \left. \frac{\mathcal{L}\left[I(\mathbf{q},\omega)\right]}{\mathcal{L}\left[R(\omega)\right]} \right\rvert_{\tau=0}\ .
\end{eqnarray}
In contrast to the thermometry and non-equilibrium detection methods, the determination of the proper normalization requires us to integrate over the entire relevant spectral weight, including potential bound-free features at high energies.
On the other hand, since the evaluation of Eq.~(\ref{eq:normalization}) only requires us to consider the limit of $\tau\to0$, we do not have to resolve the entire range of $\tau\in[0,\beta]$, and integrating over spectral regions in which the measured intensity vanishes within the given error margins is unproblematic.
To analyze the convergence of the extracted normalization, we define the truncated normalization constant
\begin{eqnarray}\label{eq:normalization_x}
    A_x = - \frac{2 m_e}{(\hbar q)^2} \frac{\partial}{\partial\tau}  F^x_{ee}(\mathbf{q},\tau) \Big\rvert_{\tau=0}\ .
\end{eqnarray}
In Fig.~\ref{fig:SSF}, we show the convergence of the analogously defined corresponding normalized static structure factor $S_{ee}^x(\mathbf{q})$ for the same beryllium XRTS data set taken at the NIF~\cite{Tilo_Nature_2023} that has already been analyzed with the ITCF thermometry approach in Fig.~\ref{fig:tilo_converges} above; the blue and green curves show results using the properly deconvolved ITCF and the unconvolved raw data, respectively.
Evidently, the normalization converges for substantially larger integration intervals compared to the temperature extraction, as we must capture the entire down-shifted feature.
Being motivated by the expected exponential decrease of $S_{ee}(\mathbf{q},\omega)$ in the limit of large frequencies $\omega$,
Dornheim \textit{et al.}~\cite{Dornheim_NatComm_2025} have suggested the following empirical ansatz to extrapolate the calculated normalization to its converged limit:
\begin{eqnarray}\label{eq:extrapolate_SSF}
    f(x) = A_x + B_x e^{-C_x x}\ ;
\end{eqnarray}
see the shaded red area in Fig.~\ref{fig:SSF}, which fits well to the experimental data. We note that the utilization of Eq.~(\ref{eq:extrapolate_SSF}) only becomes appropriate beyond the K-edge of any given material.

\begin{figure}
\centering
\includegraphics[width=0.99\linewidth]{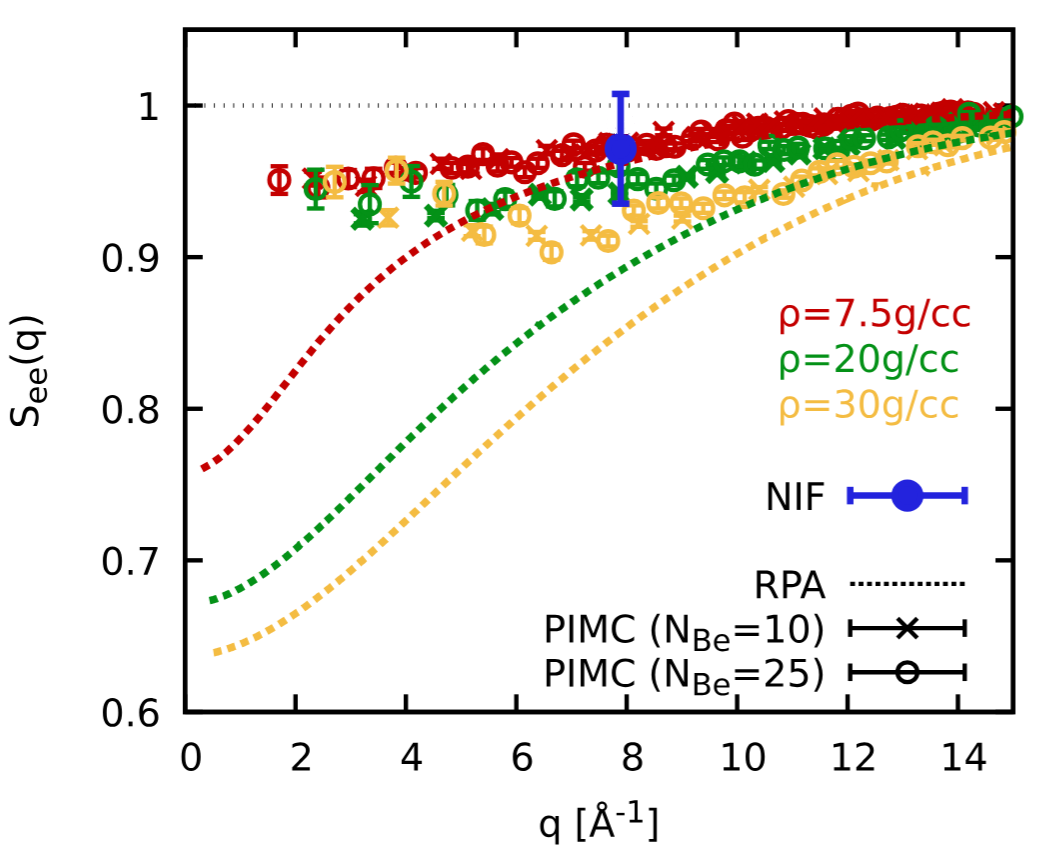}
\caption{\label{fig:SSF_PIMC} Wavenumber dependence of the electron static structure factor $S_{ee}(\mathbf{q})$ for warm dense beryllium at $T=155.5\,$eV. The crosses and empty circles show \emph{ab initio} PIMC results for $\rho=7.5\,$g/cc (red), $\rho=20\,$g/cc (green) and $\rho=30\,$g/cc (yellow), and the dotted curve approximate mean-field results (RPA) at the same conditions. The dark blue circle at $q=7.89\,$\AA${-1}$ shows the NIF data point extracted from the XRTS measurement by D\"oppner \textit{et al.}~\cite{Tilo_Nature_2023}, cf.~Fig.~\ref{fig:SSF}.
Taken from Ref.~\cite{Dornheim_NatComm_2025} with the permission of the authors.
}
\end{figure}

Putting these results into a broader context, we repeat that having the proper normalization of a measured XRTS data set is indispensable for other analyses, such as the model-free estimation of the Rayleigh weight $W_R(\mathbf{q})$ (Sec.~\ref{sec:Rayleigh_weight}) and of the electronic static linear density response $\chi(\mathbf{q},0)$ (Sec.~\ref{sec:density_response}). Moreover, we stress that the electronic static structure factor is directly related by a Fourier transform to the electron--electron pair correlation function $g_{ee}(\mathbf{r})$, which indicates the likelihood of finding two electrons in a certain distance $\mathbf{r}$ to each other.
This connection is particularly relevant in the context of upcoming XRTS experiments at modern XFEL facilities such as the European XFEL, where one can record XRTS spectra for a large number of wavenumbers $q$ within a single experiment~\cite{Gawne_PRB_2024}.
Finally, we note the possibility to compare the experimental results for $S_{ee}(\mathbf{q})$ to theoretical models and simulation data.
In Fig.~\ref{fig:SSF_PIMC}, we show the wavenumber dependence of $S_{ee}(\mathbf{q})$ for warm dense beryllium at the experimental temperature of $T=155.5\,$eV (see Fig.~\ref{fig:tilo_converges} above), with the crosses and empty circles showing quasi-exact \emph{ab initio} PIMC results for three different values of the mass density $\rho$, and with the filled blue circle at $q=7.89\,$\AA$^{-1}$ being the NIF data point extracted in Fig.~\ref{fig:SSF}.
This comparison indicates that the mass density is very likely below $\rho=30\,$g/cc (yellow symbols), even though the sensitivity of $S_{ee}(\mathbf{q})$ to the mass density is rather weak at these conditions; see also the discussion of other observables in Sec.~\ref{sec:Rayleigh_weight} and Sec.~\ref{sec:density_response}.
For completeness, we note that estimation of $S_{ee}(\mathbf{q})$ using standard DFT-MD simulations has been considered very difficult as DFT---as an effective single-electron theory---does not give one direct access to such electron--electron correlation functions.
Nevertheless, Moldabekov \textit{et al.}~\cite{Moldabekov_MRE_2026} opens up new possibilities to compute DFT based results for $S_{ee}(\mathbf{q})$, which are, in principle, available at conditions that cannot be simulation using direct PIMC methods due to the aforementioned fermion sign problem~\cite{dornheim_sign_problem}.
Finally, we mention that 
the availability of the proper normalization opens up the intriguing possibility to compare PIMC results for the full ITCF $F_{ee}(\mathbf{q},\tau)$ to the experimental result, see 
Fig.~\ref{fig:3DITCF} below.


\begin{figure*}
\centering
\includegraphics[width=0.99\linewidth]{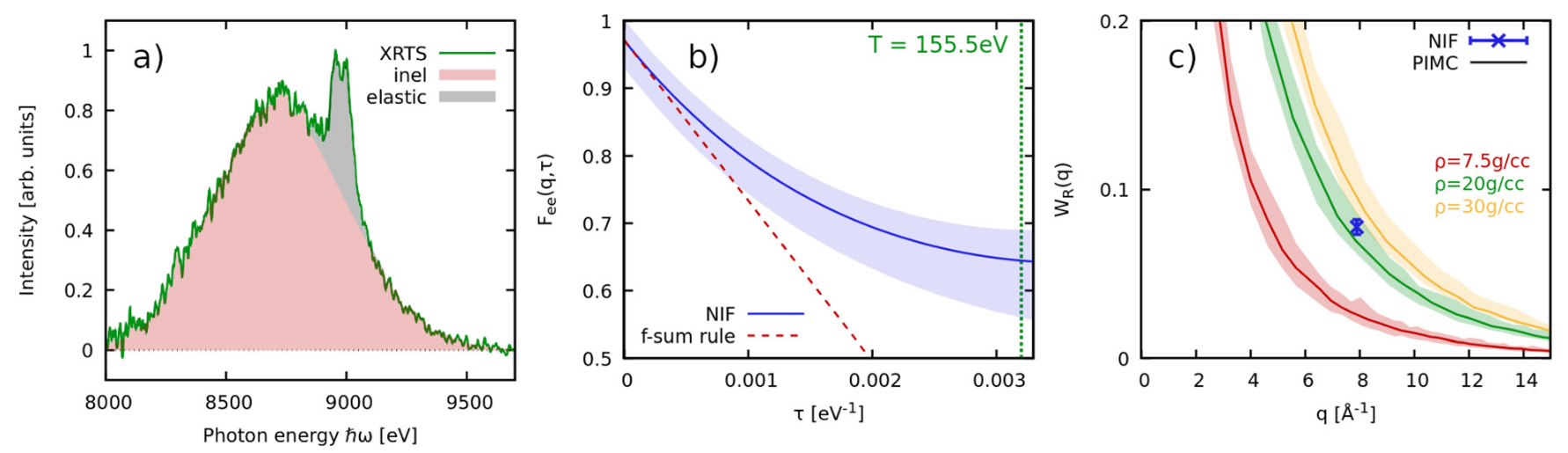}
\caption{\label{fig:Rayleigh_weight} a) XRTS spectrum measured on spherically compressed beryllium taken at the NIF in a backscattering geometry with $\theta=120^\circ$ by D\"oppner \textit{et al.}~\cite{Tilo_Nature_2023}. The red and black areas highlight the inelastic and elastic contributions to the full dynamic structure factor $S_{ee}(\mathbf{q},\omega)$, cf.~Eq.~(\ref{eq:elastic}); b) properly deconvolved and normalized ITCF $F_{ee}(\mathbf{q},\tau)$, with the dashed red line indicating the f-sum rule, Eq.~(\ref{eq:fsum}); c) wavenumber dependence of the Rayleigh weight [Eq.~(\ref{eq:Rayleigh})]. The blue cross corresponds to the experimental value extracted via Eq.~(\ref{eq:extract_Rayleigh}). The colored lines show PIMC results for the experimental temperature of $T=155.5\,$eV, cf.~Fig.~\ref{fig:tilo_converges}, for three relevant values of the mass density $\rho$, and the shaded colored areas show the propagated uncertainty in $W_\textnormal{R}(\mathbf{q})$ due to the uncertainty of $\Delta T=\pm15\,$eV in the model-free temperature estimate. 
Taken from Ref.~\cite{Dornheim_POP_2025} with the permission of the authors.
}
\end{figure*}

\textbf{Range of application and limitations:} As mentioned above, the utilization of the f-sum rule for the model-free extraction of the a-priori unknown normalization constant has, in principle, no restrictions with respect to the temperature of the probed sample and is, thus, available also for experiments in the context of laboratory astrophysics at planetary interior conditions~\cite{wdm_book,drake2018high,vorberger2025roadmapwarmdensematter} and even material science and discovery~\cite{Kraus2016,Lazicki2021} without any additional modifications.
Moreover, there are less demanding requirements with respect to the SIF, as we only have to consider the $\tau\to0$ limit and, therefore, do not have the exponential increase of the effect of the SIF for larger $\tau$.
On the negative side, the extraction of the proper normalization does, by definition, require us to integrate over the full relevant spectral range, which will become increasingly difficult for heavier elements with bound-free transition features at large energy transfers.

\subsection{Rayleigh weight\label{sec:Rayleigh_weight}}


\textbf{Idea:} The Rayleigh weight $W_\textnormal{R}(\mathbf{q})$ [Eq.~(\ref{eq:Rayleigh})] characterizes the strength of the elastic scattering, see Eq.~(\ref{eq:elastic}) above.
This concept is based on a decomposition into electronic ($\sim\,$eV) and ionic ($\sim\,$meV) energy scales, which is usually well justified.
Within the chemical picture, it combines the effects of the atomic form factor $f(\mathbf{q})$ from the bound electrons and from the screening cloud $q(\mathbf{q})$.
However, Eq.~(\ref{eq:Wr}) defines $W_\textnormal{R}(\mathbf{q})$ purely in terms of $S_{en}(\mathbf{q})$ and $S_{nn}(\mathbf{q})$, which can be computed in the physical picture without any artificial decomposition into bound and free electrons.

The first step towards the model-free extraction of the Rayleigh weight is given by the estimation of the ratio of elastic to inelastic scattering contributions,
\begin{eqnarray}\label{eq:Iratio}
    r(\mathbf{q}) &=& \frac{\int_{-\infty}^\infty\textnormal{d}\omega\ S_\textnormal{el}(\mathbf{q},\omega)}{\int_{-\infty}^\infty\textnormal{d}\omega\ S_\textnormal{inel}(\mathbf{q},\omega)}\\
    &=& \frac{W_\textnormal{R}(\mathbf{q})}{S_{ee}(\mathbf{q})-W_\textnormal{R}(\mathbf{q})} \ . \label{eq:hockenheim}
\end{eqnarray}
We note that Eq.~(\ref{eq:Iratio}) constitutes a standard observable in XRTS experiments and can often be extracted with very high accuracy~\cite{Tilo_Nature_2023}.
Solving Eq.~(\ref{eq:hockenheim}) for the Rayleigh weight then yields~\cite{Dornheim_POP_2025}
\begin{eqnarray}\label{eq:extract_Rayleigh}
    W_\textnormal{R}(\mathbf{q}) = \frac{S_{ee}(\mathbf{q})}{1+r^{-1}(\mathbf{q})}\ ,
\end{eqnarray}
with $S_{ee}(\mathbf{q})$ being estimated in a model-free way from the f-sum rule as it has been discussed in detail in Sec.~\ref{sec:moments} above.

\begin{figure}
\centering
\includegraphics[width=0.99\linewidth]{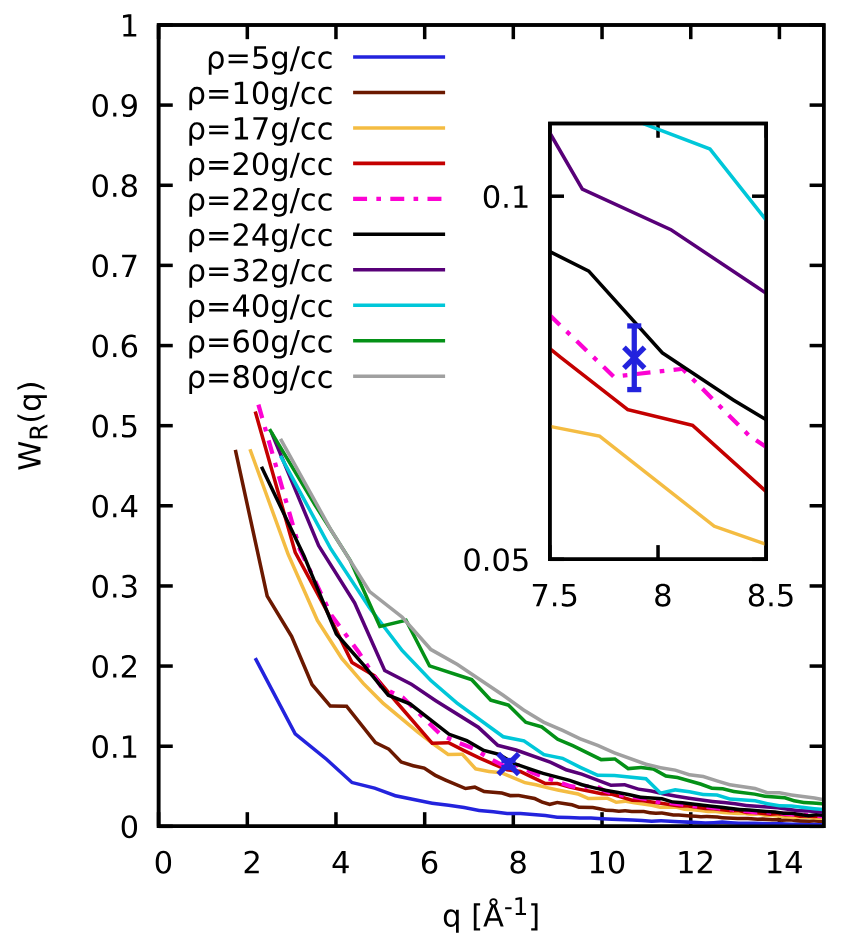}
\caption{\label{fig:Mandy} Wavenumber dependence of the Rayleigh weight $W_\textnormal{R}(\mathbf{q})$, with the curves showing DFT-MD results for different mass densities and the blue cross at $q=7.89\,$\AA$^{-1}$ being the experimental data point extracted in Fig.~\ref{fig:Rayleigh_weight}. The inset shows the best agreement between experiment and simulation for $\rho=22\pm2\,$g/cc.
Taken from Ref.~\cite{Dornheim_POP_2025} with the permission of the authors.
}
\end{figure}

\textbf{Results:} In Fig.~\ref{fig:Rayleigh_weight}, we demonstrate the work flow of the model-free extraction of $W_\textnormal{R}(\mathbf{q})$ on the example of the XRTS measurement on strongly compressed beryllium taken at the NIF by D\"oppner \textit{et al.}~\cite{Tilo_Nature_2023}, see also Fig.~\ref{fig:tilo_converges} and Fig.~\ref{fig:SSF}.
Fig.~\ref{fig:Rayleigh_weight}a) shows the spectrum itself, with the black and red areas highlighting the elastic and inelastic contributions to $S_{ee}(\mathbf{q})$, cf.~Eq.~(\ref{eq:elastic}). Fig.~\ref{fig:Rayleigh_weight}b) shows the properly deconvolved and normalized ITCF $F_{ee}(\mathbf{q},\tau)$, with the dashed red line corresponding to the first derivative at $\tau=0$ given by the f-sum rule, Eq.~(\ref{eq:fsum}).
Finally, panel c) shows the dependence of $W_\textnormal{R}(\mathbf{q})$ on the wavenumber $q$, with the blue cross at $q=7.89\,$\AA$^{-1}$ being the experimental result that has been obtained via Eq.~(\ref{eq:extract_Rayleigh}).
The colored lines show \emph{ab initio} PIMC results for the experimental temperature of $T=155.5\,$eV for three relevant values of the mass density, i.e., $\rho=7.5\,$g/cc (red), $\rho=20\,$g/cc (green), and $\rho=30\,$g/cc (yellow), with the shaded areas indicating the propagated uncertainty due to the uncertainty of $\Delta T\pm15\,$eV in the temperature.
Evidently, the experimental measurement is consistent with a mass density of $\rho\sim22\,$g/cc, although a finer grain of simulations is needed to come to a justified final verdict.

A key advantage of the Rayleigh weight is that it can be computed without any explicit knowledge of the electron--electron static structure factor $S_{ee}(\mathbf{q})$. As such, it can be straightforwardly be estimated from DFT-MD simulations~\cite{Dornheim_POP_2025,ma2026unlockingpowerorbitalfreedensity,bohme2026correlationfunctionmetrologywarm},
see Fig.~\ref{fig:Mandy} for the present NIF experiment with warm dense beryllium.
The inset shows the best agreement between \emph{ab initio} simulation and experiment for $\rho=22\pm2\,$g/cc, which is substantially lower than the nominal value of $\rho=34\pm4\,$g/cc reported in the original Ref.~\cite{Tilo_Nature_2023} on the basis of a chemical Chihara model.

\textbf{Range of application and limitations:} 
The model-free extraction of the Rayleigh weight has the same range of application and limitation like the estimation of the normalization from the f-sum rule discussed in Sec.~\ref{sec:moments} above.
The estimation of the ratio of elastic and inelastic scattering $r(\mathbf{q})$ [Eq.~(\ref{eq:ratio})] does not impose any additional constraints.

\subsection{Density response\label{sec:density_response}}



\textbf{Idea:} The basic idea behind the extraction of the electronic static linear density response function is to compute the area $L(\mathbf{q})$ under the full ITCF $F_{ee}(\mathbf{q},\tau)$ to plug it into the imaginary-time version of the fluctuation--dissipation theorem, Eq.~(\ref{eq:static_chi}).
This combines the challenges of both the model-free normalization and model-free thermometry approaches, as we (1) need to integrate $I(\mathbf{q},\omega)$ over the entire relevant spectral range and (2) we need to resolve the properly deconvolved $F_{ee}(\mathbf{q},\tau)$ at least until $\tau=\beta/2$.
In fact, one should always make use of the symmetry Eq.~(\ref{eq:ITCF_symmetry}) and only integrate over the first half-interval of the ITCF,
\begin{eqnarray}
    \chi_{ee}(\mathbf{q},0) = - n_e\ \underbrace{2 \int_0^{\beta/2} \textnormal{d}\tau\ F_{ee}(\mathbf{q},\tau) }_{L(\mathbf{q})}\ .
\end{eqnarray}
The second practical consideration concerns the integration intervals. First, we have to compute the proper normalization from the f-sum rule as it has been discussed in detail in Sec.~\ref{sec:moments} above; we will denote the thus normalized convolved XRTS intensity data as $I_A(\mathbf{q},\omega)$.
Second, estimating the area under the ITCF requires us to consider a-symmetric integration boundaries~\cite{schwalbe2025staticlineardensityresponse},
\begin{eqnarray}\label{eq:ITCF_xy}
    F_{ee}^{xy}(\mathbf{q},\tau) = \frac{ \int_{x}^y \textnormal{d}\omega\ I_A(\mathbf{q},\omega)\ e^{-\tau\hbar\omega} }{ \mathcal{L}\left[R(\omega)\right] }\ ;
\end{eqnarray}
the full ITCF is recovered in the limits of $x\to-\infty$ and $y\to\infty$.
Importantly, we have to choose the lower integration boundary $x$ in a way that it captures as much as possible of the up-shifted signal, but, at the same time, it must not allow for the accumulation of noise, which would get enhanced exponentially by the Laplace kernel $e^{-\tau\hbar\omega}$.
The upper integration boundary $y$, on the other hand, can be chosen arbitrarily large.
In practice, Eq.~(\ref{eq:ITCF_xy}) mandates a nested convergence check with respect to both integration boundaries, which is explained in accessible detail in Ref.~\cite{schwalbe2025staticlineardensityresponse}.

\begin{figure}
\centering
\includegraphics[width=0.99\linewidth]{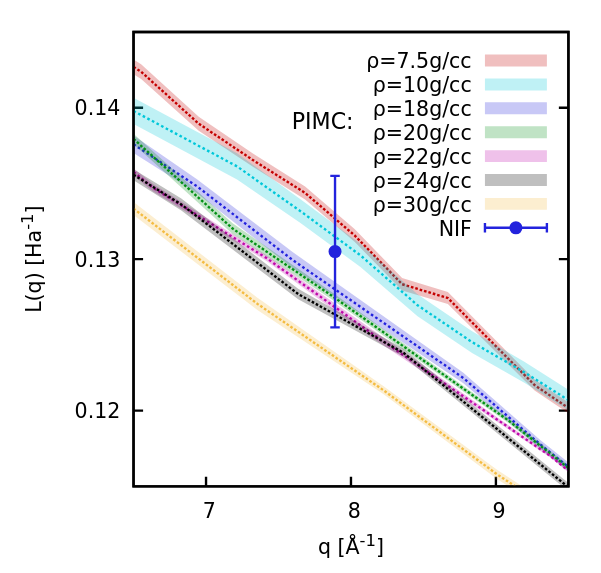}
\caption{\label{fig:full_area} Comparing the area under the full ITCF $L(\mathbf{q})$ [see Eq.~(\ref{eq:static_chi})] extracted from the NIF experiment with warm dense beryllium (blue cross; see also Figs.~\ref{fig:tilo_converges}, \ref{fig:SSF} and \ref{fig:Rayleigh_weight}) to \emph{ab initio} PIMC simulations at the experimental temperature of $T=155.5\,$eV for different values of the mass density $\rho$ (colored curves).
Taken from Ref.~\cite{schwalbe2025staticlineardensityresponse} with the permission of the authors.
}
\end{figure}

\begin{figure}
\centering
\includegraphics[width=0.97\linewidth]{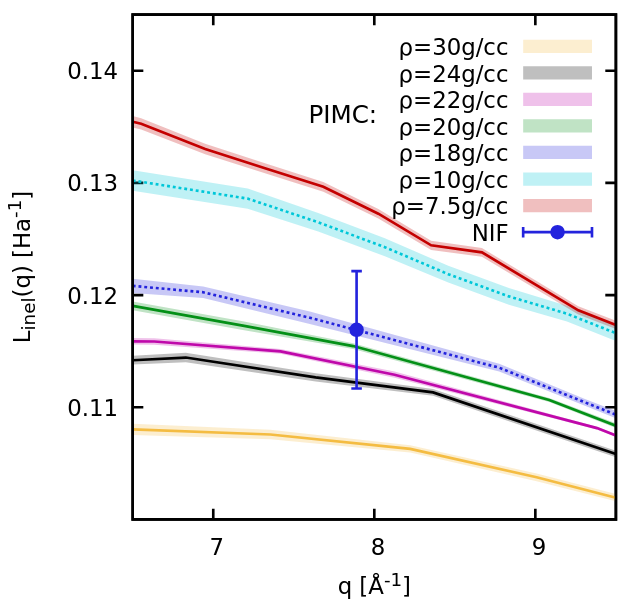}
\caption{\label{fig:inelastic_area} Comparing the area under the inelastic contribution to the ITCF $L_\textnormal{inel}(\mathbf{q})$ [see Eq.~(\ref{eq:ITCF_elastic})] extracted from the NIF experiment with warm dense beryllium (blue cross; see also Figs.~\ref{fig:tilo_converges}, \ref{fig:SSF} and \ref{fig:Rayleigh_weight}) to \emph{ab initio} PIMC simulations at the experimental temperature of $T=155.5\,$eV for different values of the mass density $\rho$ (colored curves).
Taken from Ref.~\cite{schwalbe2025staticlineardensityresponse} with the permission of the authors.
}
\end{figure}

\textbf{Results:} In Fig.~\ref{fig:full_area}, we compare the experimental result for the area under the ITCF $L(\mathbf{q})$ for the same NIF experiment on warm dense beryllium that has already been considered in Figs.~\ref{fig:tilo_converges}, \ref{fig:SSF} and \ref{fig:Rayleigh_weight}, with the blue dot at $q=7.89\,$\AA$^{-1}$ being the experimental data point. The colored curves show \emph{ab initio} PIMC results, and the shaded areas indicate the associated statistical Monte Carlo error bars.
On the on the one hand, Fig.~\ref{fig:full_area} clearly rules out mass densities above $\rho\gtrsim30\,$g/cc, in agreement with the analysis of the Rayleigh weight shown in Sec.~\ref{sec:Rayleigh_weight} above.
One the other hand, the full $L(\mathbf{q})$ is, similar to the full electron--electron static structure factor $S_{ee}(\mathbf{q})$ considered in Sec.~\ref{sec:moments}, rather insensitive to $\rho$, and fails to constrain results towards lower density.

To gain additional insights, Schwalbe \textit{et al.}~\cite{schwalbe2025staticlineardensityresponse} have proposed to consider the decomposed ITCF
\begin{eqnarray}\label{eq:ITCF_elastic}
    F_{ee}(\mathbf{q},\tau) = \underbrace{F_\textnormal{el}(\mathbf{q},\tau)}_{W_\textnormal{R}(\mathbf{q})} + F_\textnormal{inel}(\mathbf{q},\tau)\ ,
\end{eqnarray}
with the elastic component simply given by the Rayleigh weight $W_\textnormal{R}(\mathbf{q})$, see Eq.~(\ref{eq:elastic}) above.
Subtracting $W_\textnormal{R}(\mathbf{q})$ thus also gives us access to the inelastic component of the ITCF, and the corresponding contribution to the area under the ITCF $L_\textnormal{inel}(\mathbf{q})$ is shown in Fig.~\ref{fig:inelastic_area}.
Evidently, the inelastic area / static density response is substantially more sensitive to the mass density, and we find best agreement to the \emph{ab initio} PIMC simulations for $\rho=18\pm6\,$g/cc, consistent with other results based on the ratio of elastic to inelastic scattering~\cite{Dornheim_NatComm_2025} and on the Rayleigh weight~\cite{Dornheim_POP_2025}.

\textbf{Range of application and limitations:} As mentioned above, the estimation of the electronic static linear density response function $\chi_{ee}(\mathbf{q})$ to date constitutes the most challenging aspect of the model-free ITCF paradigm.
In practice, it remains limited to sufficiently hot systems (i.e., with a sufficiently short range of $\tau\in[0,\beta/2]$ that needs to be resolved), and sufficiently light elements (since we need the entire spectral range, including bound-free features from tightly bound K-shell electrons).
Consequently, the only reported practical application of this approach so far is given by its application to the high-quality beryllium data set taken at the NIF by D\"oppner \textit{et al.}~\cite{Tilo_Nature_2023}.

Nevertheless, we highlight the intriguing possibility of high-repetition pump-probe XRTS experiments at modern XFEL facilities, which might allow for sufficient accuracy and photon statistics (and a sufficiently narrow SIF) at sufficiently high temperatures.
Measuring XRTS spectra over a range of separate scattering angles might then allow us to study the electronic density response over a broad range on wavenumbers, and potentially to learn the exact XC-kernel by inverting Eq.~(\ref{eq:define_Kxc}).

\section{Raytracing based analysis of the model-free approach}\label{sec:RT_Examples}

\begin{figure*}
    \centering
    \includegraphics[width=0.32\linewidth,keepaspectratio]{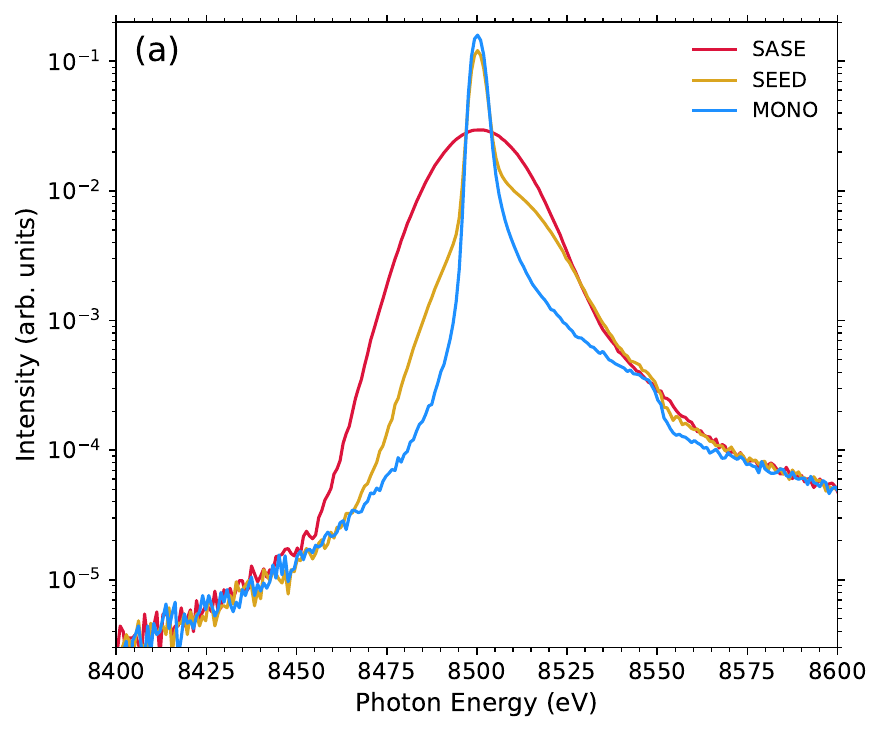}
    ~
    \includegraphics[width=0.32\linewidth,keepaspectratio]{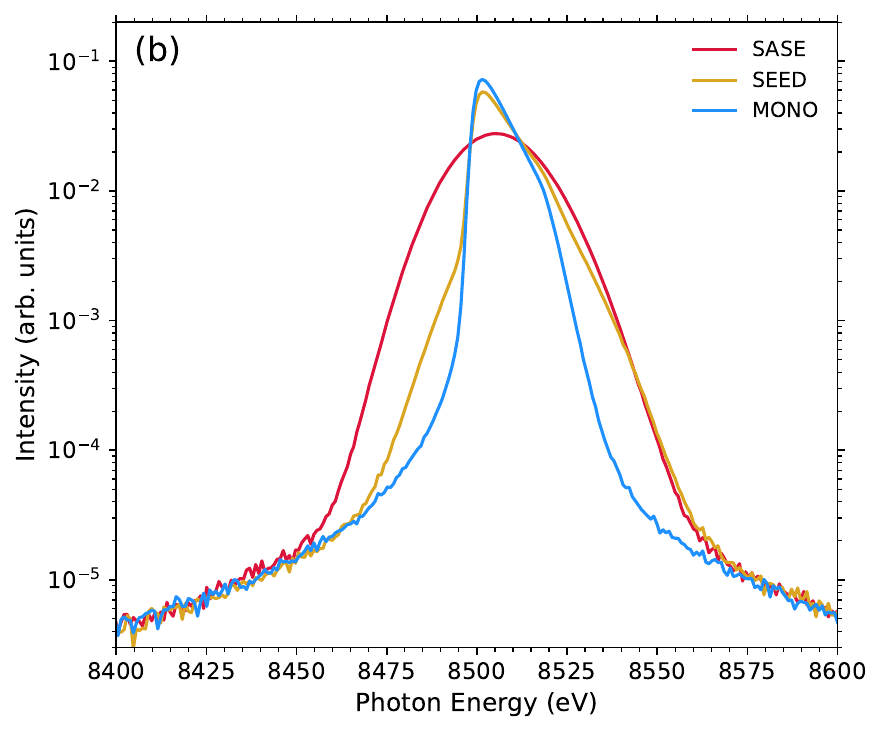}
    ~
    \includegraphics[width=0.32\linewidth,keepaspectratio]{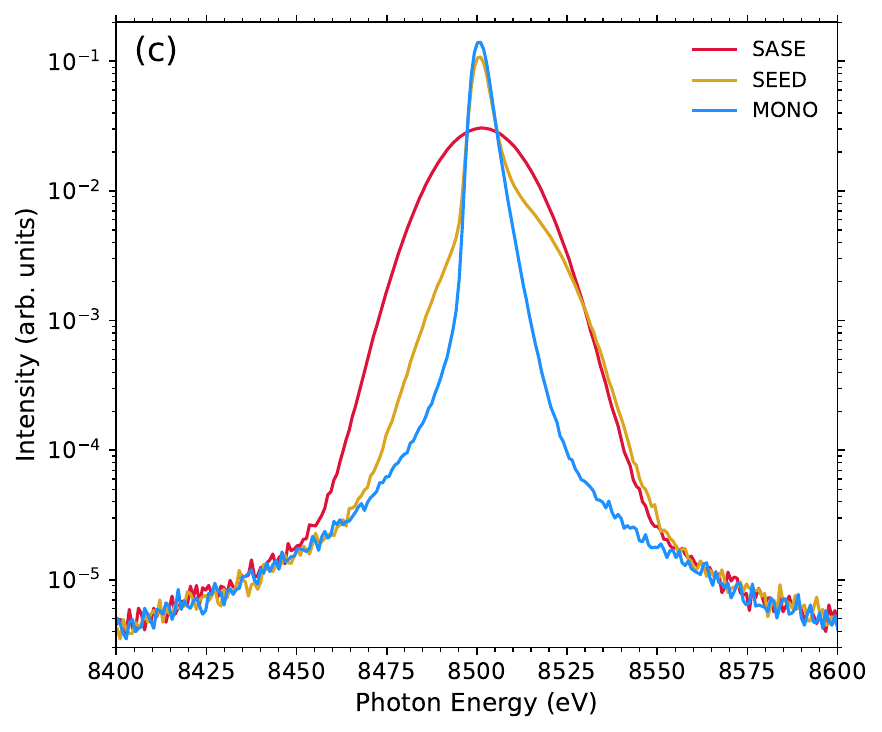}
    
    \caption{
    Ray traced SIFs on (a) 40~$\mu$m HAPG, (b) 200~$\mu$m HOPG, and (c) 40~$\mu$m HOPG for monochromated (blue), seeded (yellow), and SASE (red) XFEL beams.
    }
    \label{fig:RayTraceSIFs}
\end{figure*}

To demonstrate the robustness and applicability of the model-free methodology to realistic experiment setups, here we will apply a number of the methods to analyse ray traced XRTS spectra. As outlined in Section~\ref{sec:RayTracing}, ray tracing simulations can be used to accurately model a spectrometer's instrument function -- including geometric and photon-energy-dependent effects -- which enables us to test whether the assumption of SIF broadening is a good approximation to reality.

The ray tracing simulations were performed using \texttt{HEART}~v2.0.2~\cite{Gawne_CPC_2026}.
We simulate three mosaic crystal von H\'amos spectrometers~\cite{vonHamos_Geometry}, and base them on the spectrometers available at the HED Scientific Instrument at the European XFEL~\cite{Preston_JoI_2020}. Specifically, all three crystals have a radius of curvature of 80~mm, have dimensions of $80 \times 30$~mm$^2$, and are graphite crystals operatred in the (002) reflection.
One of the crystals is 40~$\mu$m thick highly annealed pyrolytic graphite (HAPG), which is the crystal most often employed at the HED Instrument. A previous fit to one of these crystals gave a mosaicity of 0.063$^\circ$~\cite{Gawne_CPC_2026}. The rocking curve modelled as a Voigt profile, with its centre and Gaussian and Lorentzian widths depending on the photon energy as described in Ref.~\cite{Gawne_ITCF_Ratio}.
Another of the crystals is 200~$\mu$m thick highly oriented pyrolytic graphite (HOPG). Such a crystal is available at the HED Instrument but has not been benchmarked, so we estimate the mosaicity as 0.4$^\circ$. The crystallite rocking curve is identical to the HAPG crystal.
Finally, we also consider a 40~$\mu$m thick HOPG crystal which, aside from the thickness, is identical to the 200~$\mu$m thick one. This is done in order to isolate the depth broadening from the mosaic broadening.

For the beam, we consider three different configurations at an XFEL: a SASE beam, a seeded beam, and a monochromated beam. The SASE beam is modelled as a Gaussian centered at 8.5~keV and with a FWHM of 25~eV. The monochromated beam is modelled as a 0.5~eV FWHM Gaussian, also centered on 8.5~keV. Finally, the seeded beam is modelled as the sum of two Gaussians; i.e. a seed atop the SASE pedestal. The seed is a 1.5~eV FWHM Gaussian at 8.5~keV. Since seeding is typically done on one of the sides of the initial SASE beam, the SASE pedestal is a 25~eV FWHM Gaussian centered at 8.507~keV. The peak intensity of the seed is 50$\times$ higher than the pedestal.
The spectrometers are centered on a photon energy of 8.3~keV, which is fairly typical given the XFEL photon energy.

Each of the ray traced spectra were created by randomly sampling 10$^9$ photons from the input scattering spectrum, and sending them towards the crystal. The initial directions of the rays were randomly sampled in polar and azimuthal angles that covered the entire crystal -- this automatically accounts for the solid angle coverage of the crystal, since the regions of crystal further from the source are less likely to be hit by a ray.

Ray traced spectra for just the different beam types on the different crystals are shown in Fig.~\ref{fig:RayTraceSIFs}. These are the SIFs that will be used in the Laplace deconvolution.
The crystal's effect on the instrument function is best observed by the monochromated beam. In all cases, the instrument functions are asymmetric, with the shape strongly dictated by the form of the mosaic distribution function. For example, the Lorentzian-like distribution of HAPG~\cite{Legall_HAPG_2006,Gerlach_JAC_2015,Gawne_CPC_2026} causes a much slower decay in the high energy side of the SIF compared to HOPG, which has a von Mises-Fisher distribution of crystallites~\cite{Beckhoff_SPIE_1996, Freund_SPIE_1996, Gerlach_JAC_2015, Wuttke_Acta_2014, Bornemann_Acta_2020}.
Also visible in the HAPG case is position of the crystal in space -- the drop in intensity near 8.55~keV corresponds to one the physical edges of the crystal, indicating that only part of the crystal is reflecting the 8.5~keV photons to detector positions $>8.55$~keV~\cite{Gawne_JAP_2024}.
The effect of depth broadening is visible between the two HOPG crystals, with the back of the 200~$\mu$m crystal visible as a change in the shape of the instrument function near 8.515~keV.

It is worth noting that while the differences in the shape of the instrument function between the 40~$\mu$m thick HAPG and HOPG crystals are emphasised by the logarithmic scale, the measured FWHM of the monochromated beam are quite similar between the crystals (4.4~eV from HAPG versus 5.2~eV from HOPG). The primary benefit of HAPG over HOPG is the lower available mosaicities which improves the resolution. The ray tracing simulations of the two crystals here (with mosaicities close to the lowest values quoted by Optigraph~GmbH~\cite{optigraph}) show a modest $\sim$15\% improvement in resolution when using HAPG for a photon energy of 8.5~keV.

\begin{figure*}
    \centering
    \includegraphics[width=\linewidth]{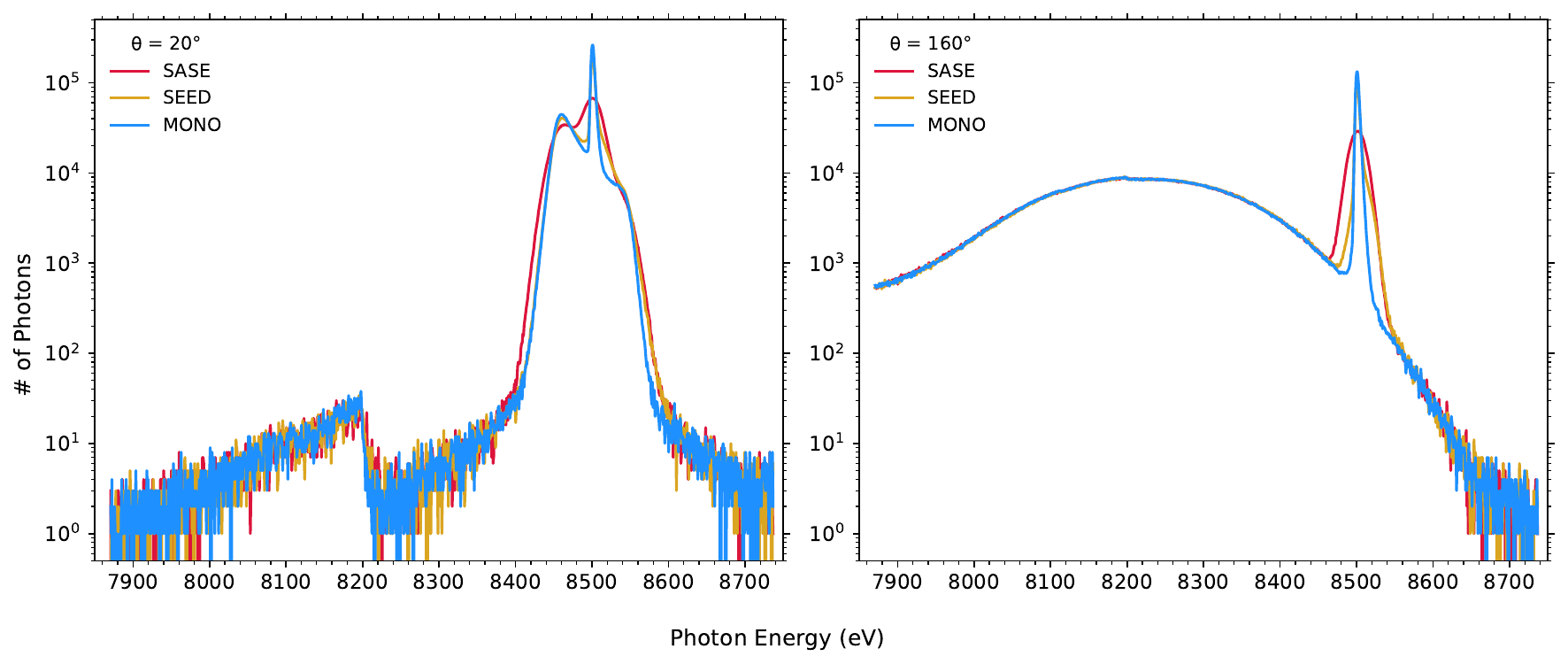}
    \caption{An example of the ray traced XRTS spectra analysed here. Specifically, they are for $T_e = 20$~eV carbon at (a) $\theta = 20^\circ$ and (b) 160$^\circ$, measured on the 40~$\mu$m thick HOPG crystal for SASE (red), self-seeded (yellow) and monochromated (blue) beams.}
    \label{fig:GraphiteSpectra}
\end{figure*}

Finally, the DSFs are modelled using the openly available Chihara code \texttt{xDAVE}~\cite{xdave,Bellenbaum_xdave}. We consider an isochorically heated carbon system, similar to Ref.~\cite{kraus_xrts}, containing 50\% C$^{4+}$ and 50\% C$^{5+}$ ions, a mass density of 2.21~g~cm$^{-3}$, an electron temperature $T_e = 20$~eV, and an ion temperature $T_i = 2$~eV (the Rayleigh weight calculation becomes extremely slow for temperatures any lower). The DSFs are calculated for a beam energy of 8.5~keV and at scattering angles $\theta = 20^\circ$ and 160$^\circ$. 
The free-free feature is calculated using the Lindhard dielectric function using the local field correction interpolation from Ref.~\cite{Gregori_HEDP_2007}. The bound-free and free-bound features are calculated using the impulse approximation~\cite{Schumacher_JPB_1975}. The ionization potential depression of the bound orbitals is calculated using the Crowley model~\cite{Crowley_2014_ipd}. Finally, the ion feature is treated with the Rayleigh weight formalism. The screening cloud within the Rayleigh weight is calculated using a finite-wavelength screening cloud and a Yukawa potential. The ion-ion structure factor is calculated using a hypernetted chain solver~\cite{Wuensch_PRE_2008}, with a Yukawa potential for the ion-ion interaction.

An example of the ray traced XRTS spectra measured using the 40~$\mu$m HOPG crystal are shown in Fig.~\ref{fig:GraphiteSpectra}. The effect of the broadening from the different beams is clearly observed in the plasmon features of the 20$^\circ$ spectrum and the elastic feature at both scattering angles. Conversely, the broadening of the Compton feature in the 160$^\circ$ spectrum and the bound-free features in both spectra are much less apparent.

\subsection{Model-free Thermometry\label{sec:RT_thermometry}}

\begin{figure*}
    \centering
    \includegraphics[width=\linewidth]{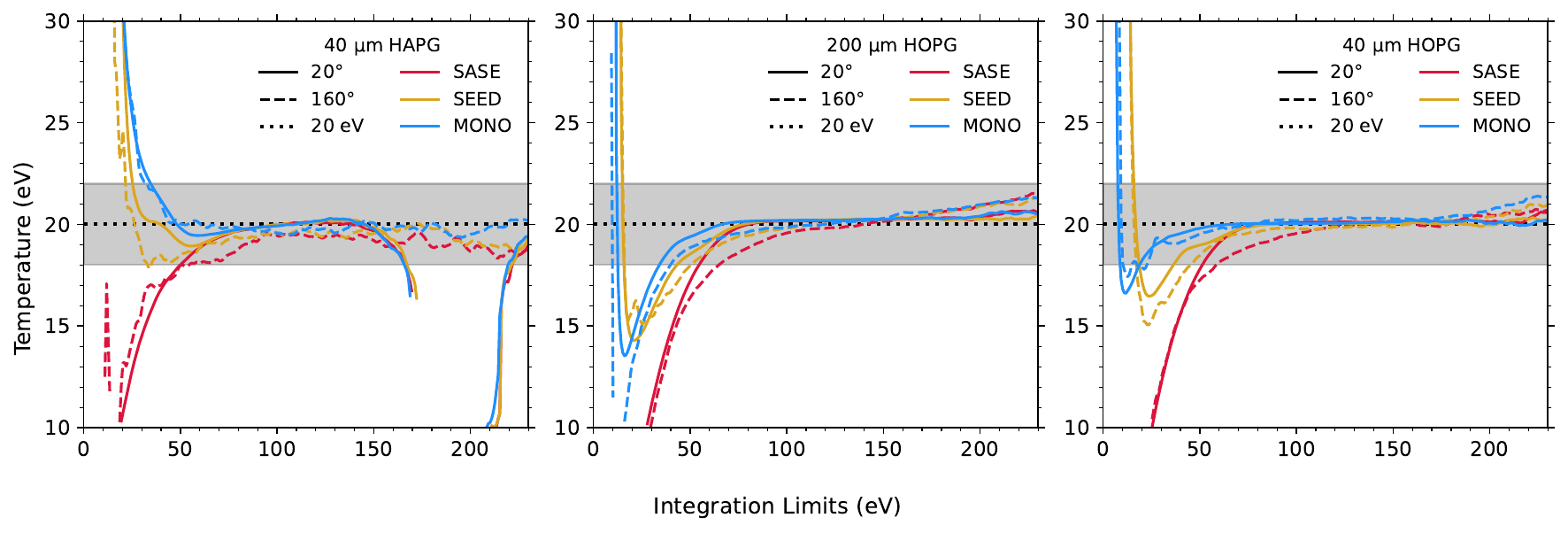}
    \caption{Temperatures inferred from the minimum of the unnormalised ITCF for $T_e = 20$~eV carbon on the different crystals -- 40~$\mu$m HAPG, 200~$\mu$m HOPG, and 40~$\mu$m HOPG -- and the different beam types -- SASE (red), self-seeded (yellow), and monochromated (blue) -- at two scattering angles: 20$^\circ$ (solid) and 160$^\circ$ (dashed). $T_e=20$~eV is indicated by the black dotted line, and the gray region indicates $\pm10\%~T_e$.}
    \label{fig:Graphite20eV}
\end{figure*}

First we apply the model-free thermometry to infer the temperature from the ray traced spectra. Fig.~\ref{fig:Graphite20eV} shows the inferred temperatures versus the symmetric integration limits $x$ for the two scattering angles for each crystal and beam, up to the limits of the spectral range $x \approx 235$~eV.
Note that there are some gaps in the temperature curves, which occur when no minimum in the ITCF could be identified.

Starting with the HAPG crystal, the temperature convergence for all three beams for the 160$^\circ$ DSF settles close to the actual temperature, though the SASE case very slightly underestimates the temperature by $\sim 1$~eV. The 20$^\circ$ cases do not fully converge. Instead, for $x < 150$~eV, the inferred temperature slowly increases, after which the inferred temperature rapidly drops.

Turning to the 200~$\mu$m HOPG crustal, we observe a notably different behaviour to HAPG. Namely, the 20$^\circ$ temperatures converge very close to the correct temperature, and the remain stable over a large integration interval $x \in [70, 190]$~eV. This indicates that the temperature can be derived with high confidence, and that it is both accurate and precise. This improved stability may be partially attributed to the high reflectivity of the HOPG crystal due to its much greater thickness and reflectivity, so more upshifted photons are detected.
Conversely, the 160$^\circ$ spectra do not show such good convergence. Rather, for $x>30$~eV, the temperature slowly increases for all three beam types.

The 40~$\mu$m HOPG crystal shows the best performance of the three. For all three beam types and for both angles, the curves show a region where they are flat and a temperature could be inferred. Moreover, all the temperatures are very close to the exact temperature. 
Compared to HAPG, part of the effect may be due to the higher number of detected photons, though it is obviously fewer than the thicker HOPG crystal.
The main improvement in the stability is attributable to the tails of the instrument function. As observed in Fig.~\ref{fig:RayTraceSIFs} and already commented on, the 40~$\mu$m HOPG crystal has a relatively narrow instrument function, and tails that decay much faster than the HAPG crystal, especially on the upshifted side of the spectrum. Moreover, the photons from a given feature are relatively contained within the width of the instrument function, whereas for HAPG many of the photons will be spread into the slowly-decaying tails. Overall, this means that it is easier to perform the deconvolution for the thin HOPG crystal.

As some general comments, we note that the inferred temperature for the 20$^\circ$ spectrum converges faster than the 160$^\circ$ spectrum, which is unsurprising given the plasmon features are restricted in a narrower spectral range. Similarly, the convergence for the monochromated beam is faster than the SASE or seeded beams, again due to the SIF being narrower.
There is some notable behaviour for the seeded beam, best observed in the HOPG crystals. Namely, for the 20$^\circ$ spectra the convergence follows quite closely to the monochromated beam, but then around $x \approx 50$~eV the convergence begins to follow the SASE beam.
But, for the 160$^\circ$ spectra, the path to convergence follows more closely to the monochromated beam, or between monochromated and SASE. This indicates that for narrow features in the DSF, the relatively weak pedestal still plays a non-negligible role in the temperature inference, and requires the SASE pedestal to be well-resolved and characterised as part of developing the SIF.

The results also highlight the applicability of the ITCF thermometry to real experiments. Even though the crystal instrument functions here depend directly on the energies of the photons hitting the detector, the temperature can nevertheless be inferred via the ITCF method and performing a deconvolution with the SIF measured at the quasi-elastic feature.

\subsection{Extracting the ITCF}\label{sec:RT_ITCF}

\begin{figure*}
    \centering
    \includegraphics[width=\linewidth]{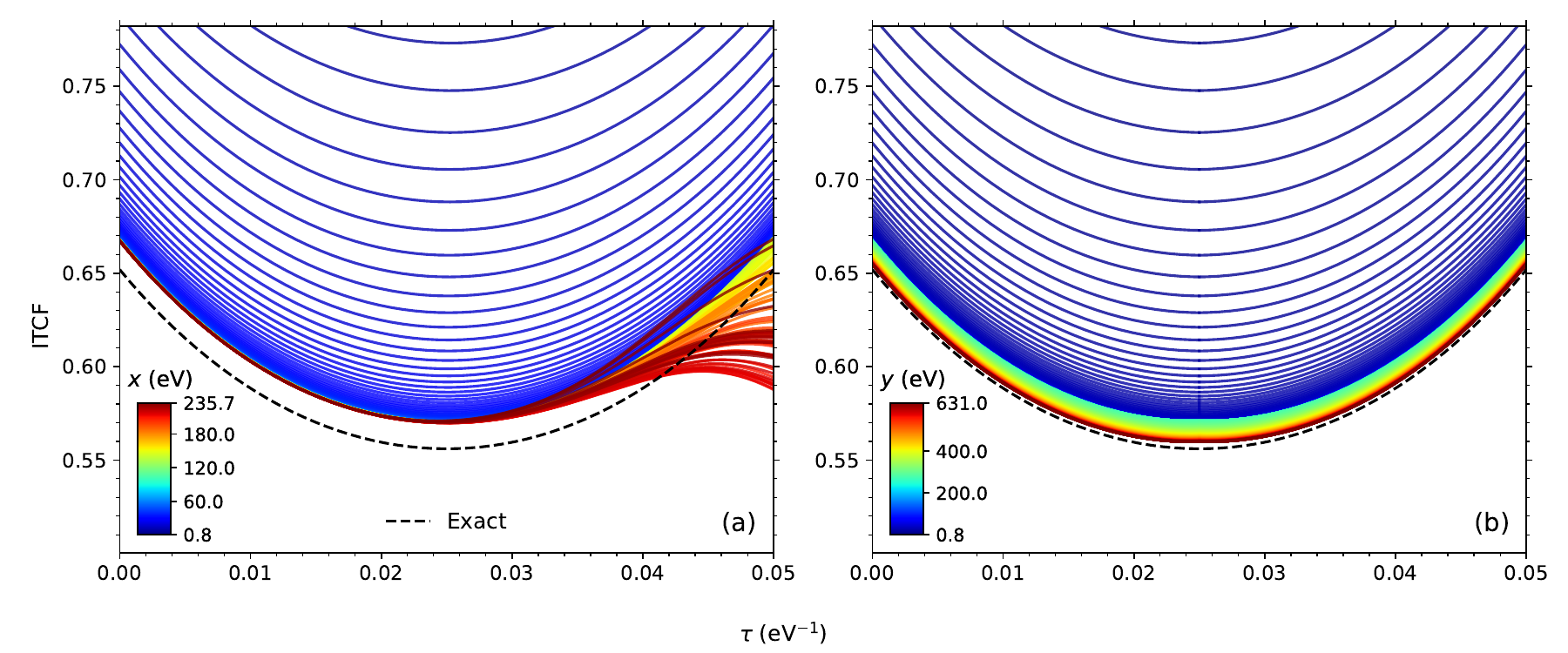}
    \caption{ITCFs extracted from the monochromated $\theta=20^\circ$ spectrum in Fig.~(\ref{fig:GraphiteSpectra}) for different integration limits (coloured solid lines) versus the exact ITCF (black dashed line). (a) The ITCFs are extracted using symmetric integration limits $x$. (b) The ITCF determined using an asymmetric interval: after $y = 235.7$~eV, the downshifted side of the spectrum continues to be integrated over. The ITCF is calculated between $\tau=[0, 0.025]$~eV$^{-1}$, and then mirrored to produce a full ITCF.}
    \label{fig:GraphiteITCF20}
\end{figure*}

\begin{figure*}
    \centering
    \includegraphics[width=\linewidth]{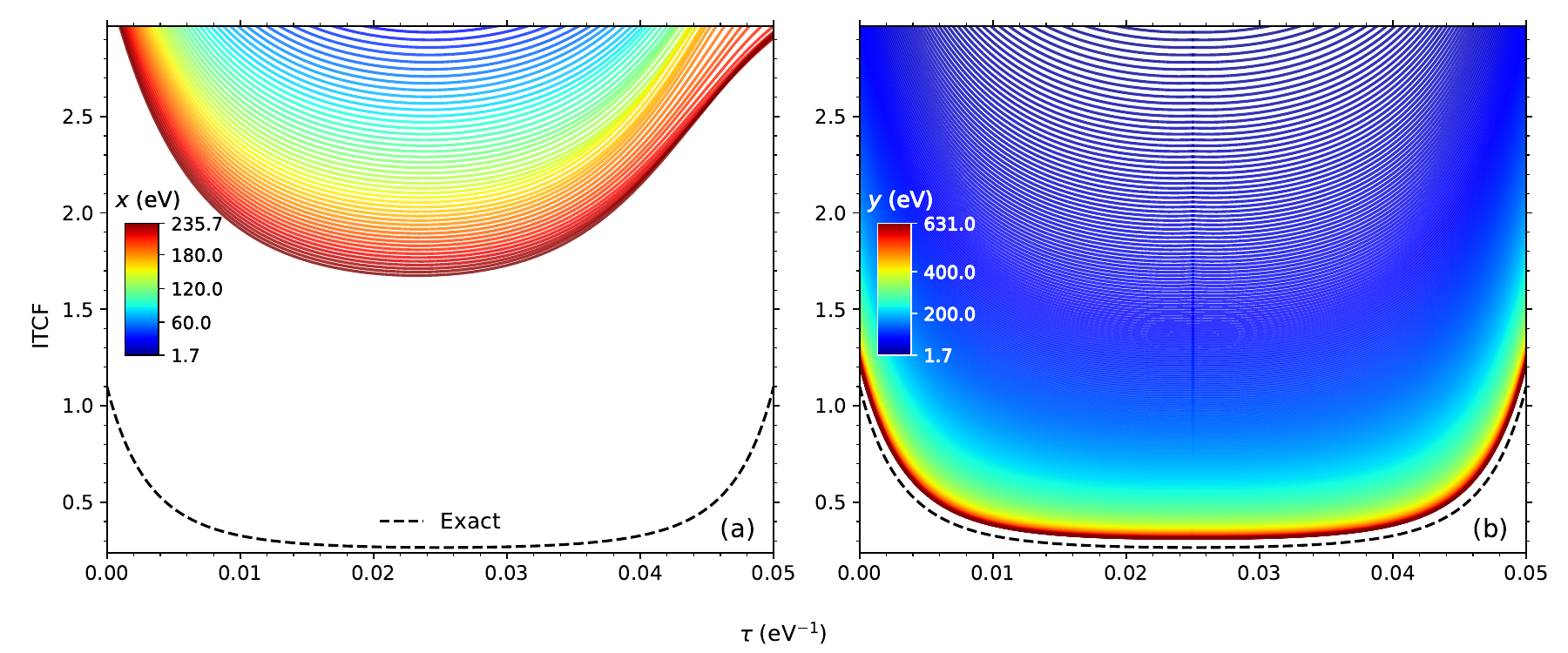}
    \caption{The same as Fig.~\ref{fig:GraphiteITCF20}, but for the $\theta = 160^\circ$ monochromated spectrum in Fig.~\ref{fig:GraphiteSpectra}.}
    \label{fig:GraphiteITCF160}
\end{figure*}

In order to extract the static density response function, the entire ITCF needs to be determined. For simplicity, we only examine calculating the full ITCF for the monochromated spectra measured on the 40~$\mu$m HOPG crystal; see Fig.~\ref{fig:GraphiteSpectra}.

Fig.~\ref{fig:GraphiteITCF20}~(a) shows the convergence of the ITCF for the monochromated versus the symmetric integration limits $x$. The ITCFs are normalised using the f-sum rule at each step integration limit. 
There are two noteworthy issues with the ITCF -- first, on the $\tau>\beta_e/2$ side of the ITCF, the curves lose their symmetric behaviour, which is due to the low number of upshifted photons. This can be averted by increasing the number of detected photons, though this might not be practical due to experimental constraints. Second is that after $x \gtrsim 100$~eV, the convergence of the ITCF slows down substantially. 
Indeed, the spectral range would need to increase by several 100s~eV either side of the elastic feature to directly measure the ITCF directly, which again presents substantial practical issues for the photometrics.
The reason the ITCFs disagree is that the bound-free and free-bound features are yet to be integrated over, so its information is not yet included in the experimental ITCF. Remember that in trying to determine the full ITCF, we are essentially trying to reconstruct the DSF in the Laplace domain. Therefore, missing spectral features in integration will result in discrepancies from the exact form.
However, since the bound-free feature feature is several orders of magnitude weaker than the plasmon feature, the agreement between the ray traced and exact ITCFs is still quite good.

Fig.~\ref{fig:GraphiteITCF20}~(b) shows the ITCF when constructed using asymmetric integration limits. Specifically, we take advantage of the stability of the temperature extracted in Fig.~\ref{fig:Graphite20eV} to restrict the calculation of the ITCF to $\tau = [0, \beta/2] = [0, 0.025]$~eV$^{-1}$. After the symmetric limit $y = 235.7$~eV is reached, the downshifted side of the spectrum continues to be integrated over, providing a larger range up to $y = 631.0$~eV. This spectral range now includes a portion of the bound-free feature..
The ITCF is then mirror to produce a complete ITCF. Compared to Fig.~\ref{fig:GraphiteITCF20}~(a), this new ITCF is in better agreement with the exact ITCF.

A much more dramatic improvement is observed in Fig.~\ref{fig:GraphiteITCF160}, which shows the convergence of the ITCF for the $\theta=160^\circ$ spectrum. Up to the limits of symmetric integration, the ray traced ITCF is substantially differs from the exact ITCF. This is clearly due to a substantial portions of the Compton feature and bound-free/free-bound features being excluded from the ITCF.
Using the asymmetric integration results in a large improvement. These bigger improvements compared to Fig.~\ref{fig:GraphiteITCF20} is due to the larger intensity far from the elastic feature in the 160$^\circ$ spectrum. However, there is still a noticeable disagreement, indicating that a much larger spectral range would be required to fully determine the ITCF.
Alternatively, one could look at the convergence of the area under the ITCF -- or perhaps even the ITCF itself -- with respect to the integration interval in order to determine the converged value for infinite integration limits.

The ITCFs derived using asymmetric integration still show visible disagreement with the exact curve. This is partially because only one side of the spectrum is integrated over; and partially because the integration limits are still too narrow to reproduce the full DSF in the Laplace domain.
For the 20$^\circ$ ITCF, the symmetric integration ITCF overestimates the area under the curve by 3.1\%, while the asymmetric limits reduces this to 0.51\%. Further increasing the spectral range would therefore only confer minor improvements.
For the 160$^\circ$ ITCF, the reduction is from a 434\% overestimation down to 16\%. We can therefore conclude that large scattering vectors require a very large spectral window to fully reconstruct the ITCF. This is not too surprising, given the large spectral intensity of the 160$^\circ$ spectrum seen in Fig.~\ref{fig:GraphiteSpectra}.

Lastly, there are still a couple of points worth noting. First, while the ITCFs in Figs.~\ref{fig:GraphiteITCF20}~(a) and~\ref{fig:GraphiteITCF160}~(a) show large disagreements from the exact case for small integration limits, the minimum of the ITCF is extremely stable. In other words, the detailed balance of the DSF is captured, even though it has not been fully deconvolved.
Second is that we can once again conclude that, despite the crystal response function in the ray tracing being directly dependent on the photon energy, the ITCFs -- particularly at low scattering vectors -- can be accurately reconstructed.

\subsection{Static Structure Factor\label{sec:RT_SSF}}

\begin{figure*}
    \centering
    \includegraphics[width=\linewidth]{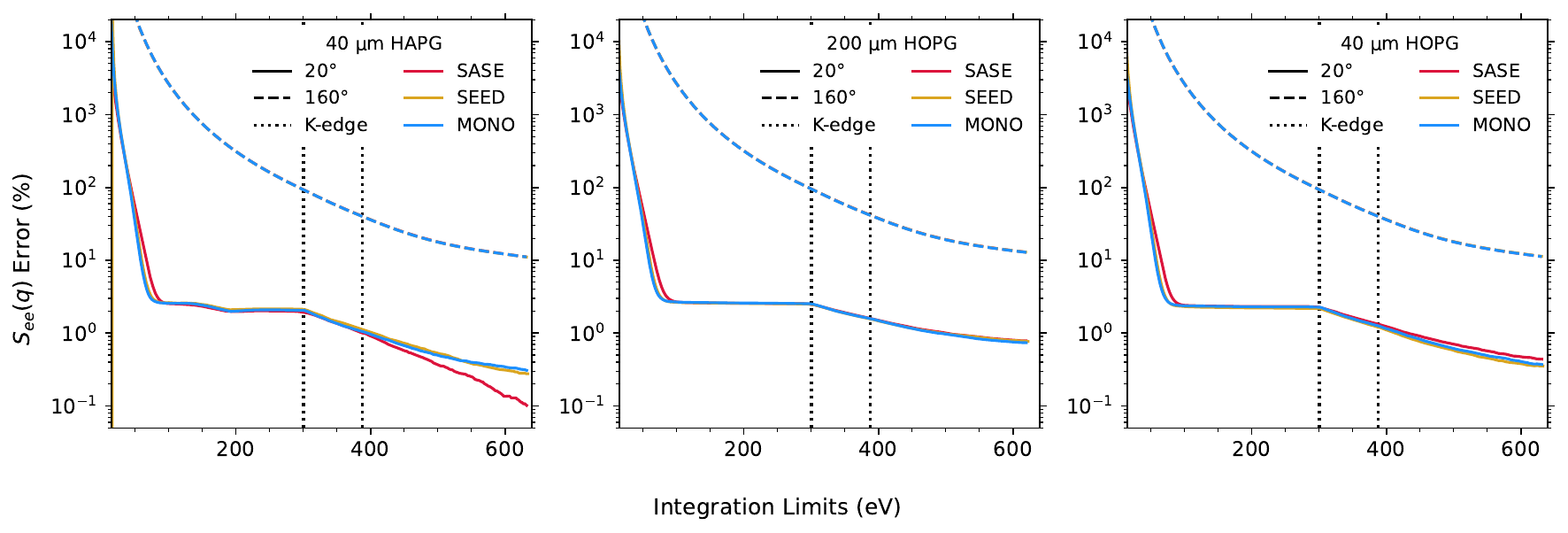}
    \caption{Convergence of the SSF $S_{ee}(q)$ for the ray traced 20$^\circ$ (solid) and 160$^\circ$ (dashed) XRTS spectra, for the SASE (red), seeded (yellow), and monochromated (blue) beams. The K-edges of the carbon ions are indicated with the vertical dashed lines. The y-axis plots the relative error of the inferred SSF in percent. For the 160$^\circ$ degree cases, the SSF convergence appear to overlap on this scale.}
    \label{fig:GraphiteSSF}
\end{figure*}

The SSF $S_{ee}(q)$ can be extracted from the full ITCF by only looking at $F_{ee}(q, \tau=0)$. Fig.~\ref{fig:GraphiteSSF} shows the convergence of the SSF for the different crystals and beams for both the 20$^\circ$ and 160$^\circ$ spectra. The ITCFs here are normalised  to the f-sum rule at each value of $x$.
The error here is calculated here as
\begin{equation}
    {\cal E}(q) = 100\cdot\frac{F^{\rm RT}(q, \tau=0) - S_{ee}^{\rm exact}(q)}{S_{ee}^{\rm exact}(q)} \, ,
\end{equation}
so a systematic overestimation of the SSF is observed in the integration limits shown here.
The integration limits are extended beyond the symmetric limit $x_{\rm max} \approx 235$~eV because, as observed in the previous subsection, this can be used to extract more accurate values from the ITCF curve at a given $\tau$.

The SSF extraction for 20$^\circ$ offers very useful insight into the way information from the DSF is encoded in the ITCF. For all crystals and beams, the SSF shows fairly rapid convergence to an error around 2--3\%. However, once the K-edge of the C$^{4+}$ ion is reached, the SSF begins to reduce again, which in turn reduces the error. Passing the K-edge of the C$^{5+}$ ion has a similar, albeit substantially smaller, effect on the convergence.

Given the definition of the ITCF from the DSF, it is not surprising that the spectral information of a given spectral feature is encoded in specific value of the ITCF at a given $\tau$. Nevertheless, this provides a visual demonstration of how these features affect the values of the ITCF at a given $\tau$.

By comparison to the 160$^\circ$ case, we gain insight into the relative importance of a given feature to the ITCF. For $F_{ee}(q, \tau=0)$, the importance of the K-edge feature is much reduced at this scattering vector, with the SSF appearing to smoothly reduce across the spectral range, without obvious signs of the K-edges. This may be explained by the much higher signal from the Compton feature compared to the bound-free feature at any given photon energy.  
The decay is very slow across the spectral range, and shows a substantial error of 10\% at the end of the spectral range, without any convergence being reached. This necessitates extrapolating the SSF in order to obtain a value. Using Eq.~(\ref{eq:extrapolate_SSF}), the error in the SSF can be reduced from $\sim$11--13\% down to $\sim$7--11\%, depending on the spectrum and the lower limit of the fitting region. These improvements are rather modest, and highlight the requirements of large spectral ranges when attempting to extract the SSF for DSFs with substantial contributions at large energy shifts.

\subsection{Detecting Non-Equilibrium\label{sec:RT_neq}}

\begin{figure*}
    \centering
    \includegraphics[width=0.30\linewidth,keepaspectratio]{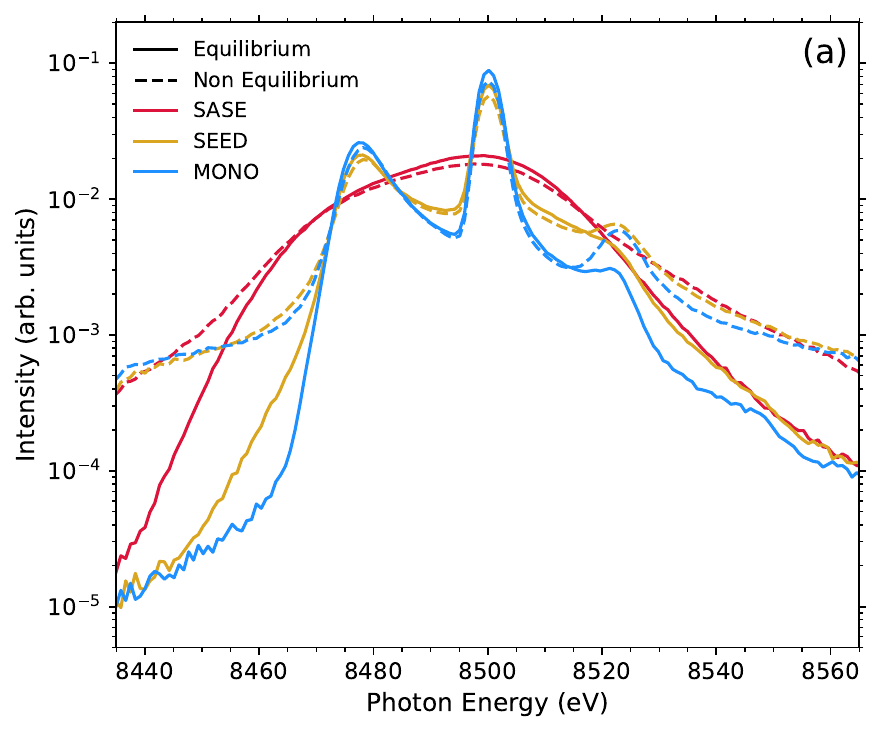}
    ~
    \includegraphics[width=0.30\linewidth,keepaspectratio]{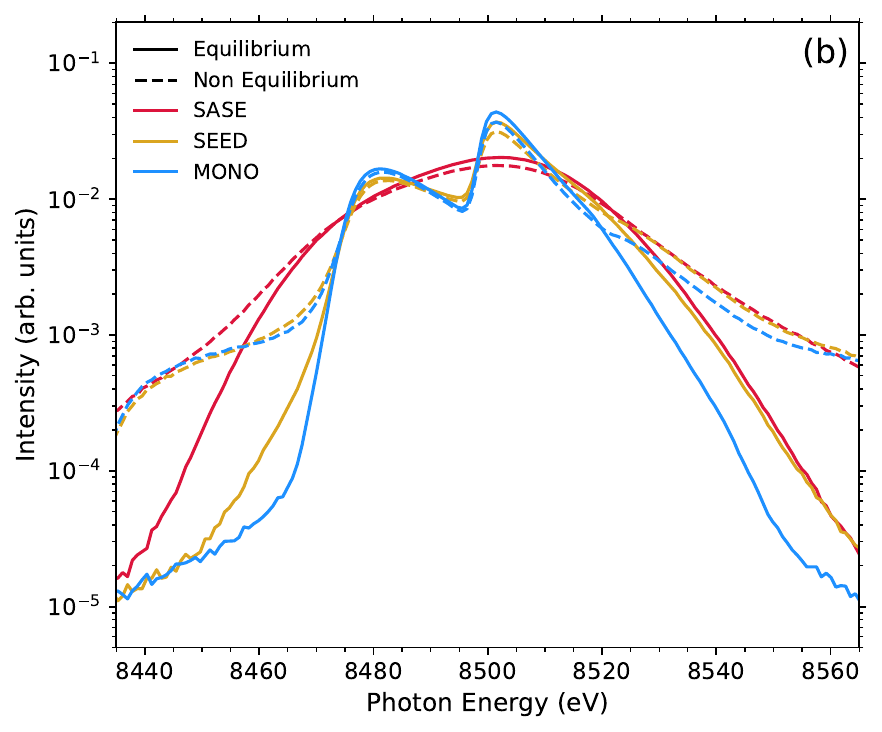}
    ~
    \includegraphics[width=0.30\linewidth,keepaspectratio]{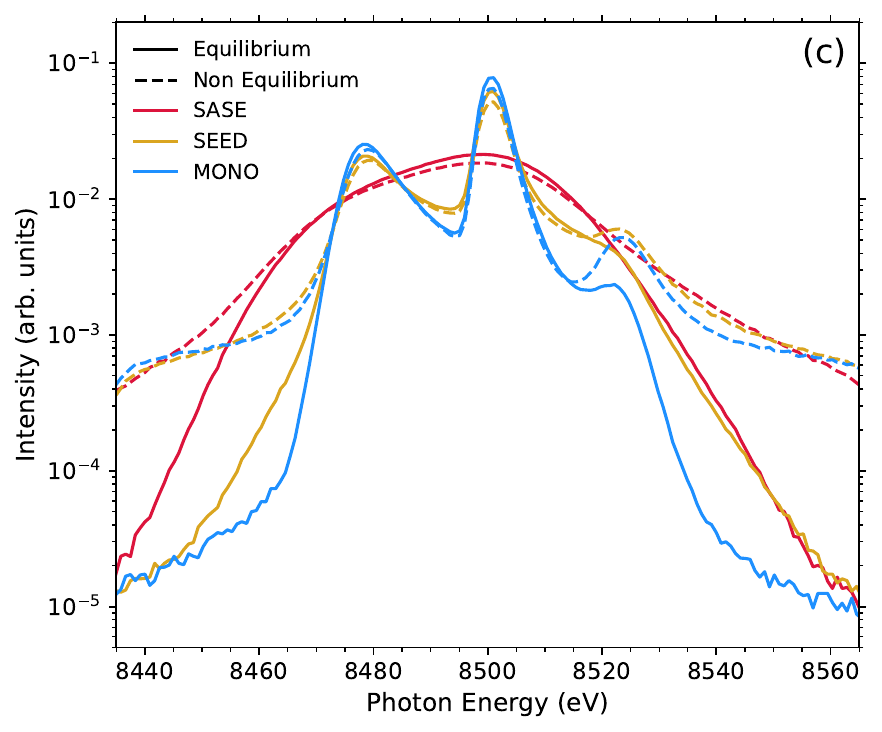}
    \vspace{0.1mm}
    \includegraphics[width=0.30\linewidth,keepaspectratio]{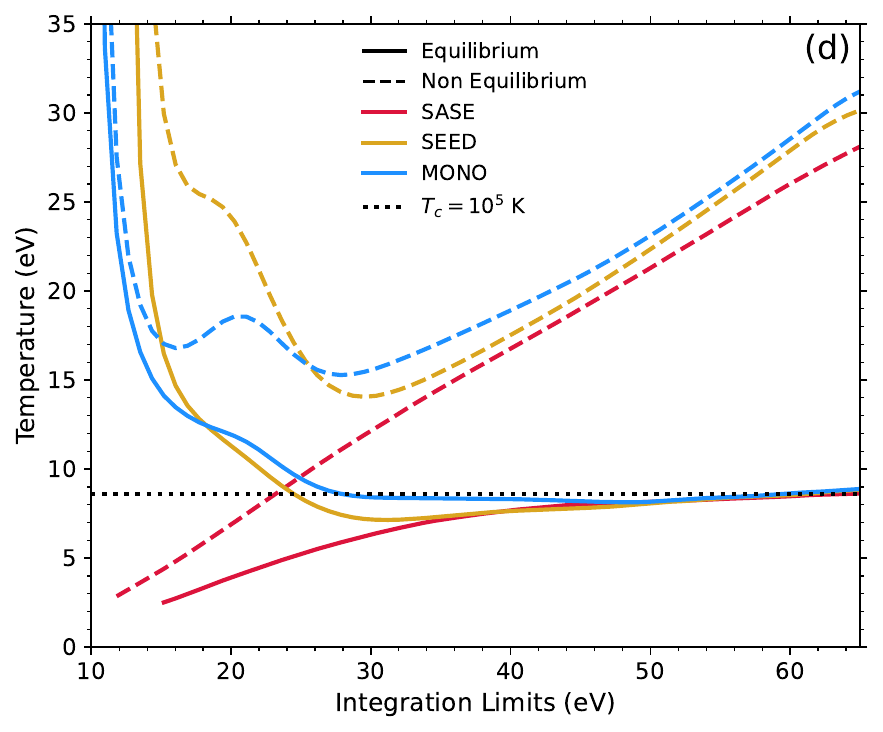}
    ~
    \includegraphics[width=0.30\linewidth,keepaspectratio]{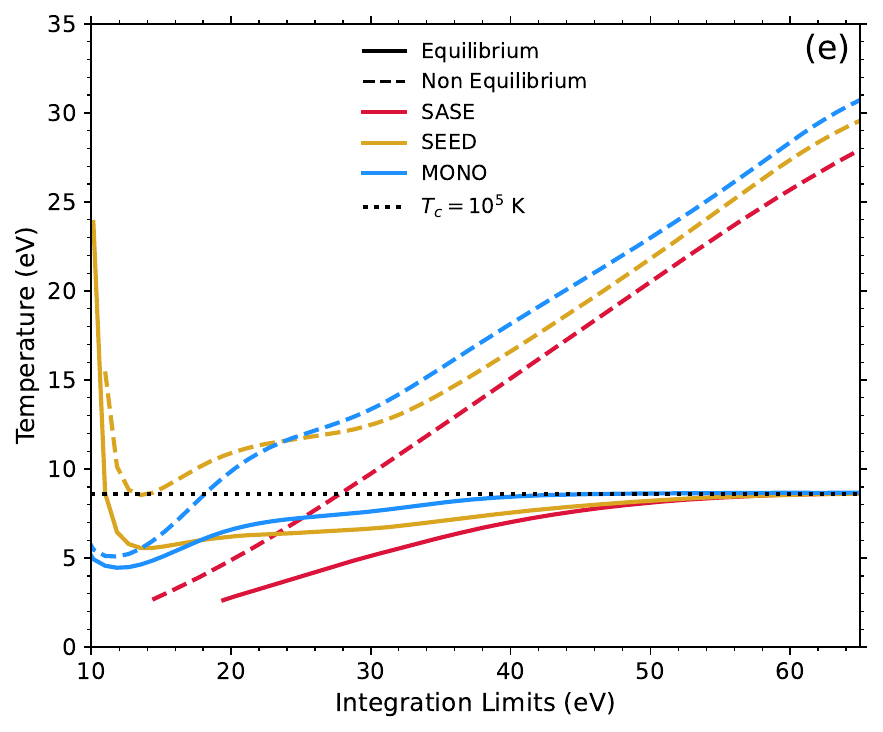}
    ~
    \includegraphics[width=0.30\linewidth,keepaspectratio]{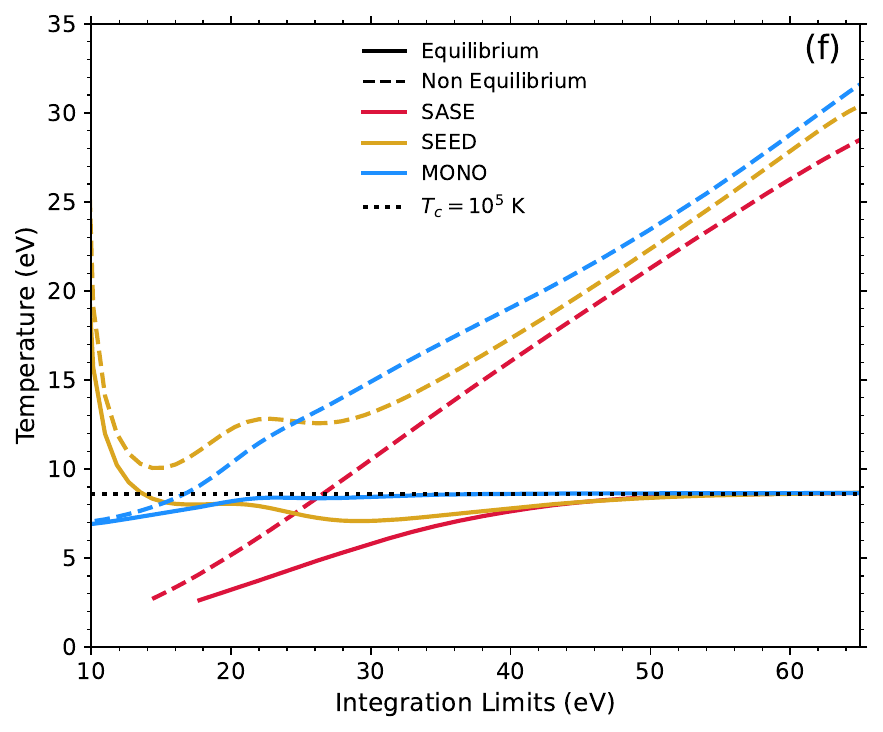}
    
    \caption{
    Ray traced spectra of equilibrium (solid lines) and non-equilibrium (dashed lines) DSFs from Ref.~\cite{Vorberger_PLA_2024} for different crystals -- (a) 40~$\mu$m thick HAPG, (b) 200~$\mu$m HOPG, and (c) 40~$\mu$m HOPG -- when probed with a monochromated (blue), seeded (yellow) and SASE (red) XFEL beam. (d)--(f) show the corresponding temperatures versus integration limits using the model-free ITCF method.
    }
    \label{fig:ITCFsNE}
\end{figure*}

Next we consider how non-equilibrium effects may be inferred from experimental XRTS measurements. Plotted in Fig.~\ref{fig:ITCFsNE}~(a)--(c) are ray traced XRTS spectra for two Al systems from Ref.~\cite{Vorberger_PLA_2024} at $q = 0.5~a_0^{-1}$: one is an equilibrium system with a temperature $T_c = 10^5~{\rm K} \approx 8.6~{\rm eV}$; the second is a system where 10\% of the electrons (`hot' electrons) have been driven out of equilibrium with the `cold' electrons.

Another difference from the equilibrium Chihara carbon spectra examined so far is that the ion feature for these Al simulations is frequency-resolved rather than treated in the Rayleigh weight formalism. Clearly none of the detail of the ion feature is resolved in the spectra, though the quasi-elastic feature is slightly, but measurably, broader than the SIF. 

The contribution of the non-equilibrium electrons to the spectra is a flat plateau that decays very slowly. This is most noticeable for the monochromated beam spectra on the 40~$\mu$m HAPG and HOPG crystals: the equilibrium XRTS spectra decay quickly beyond the plasmons at $\omega = \pm 22$~eV, while the non-equilibrium spectra show substantial scattering. 

Ref.~\cite{Vorberger_PLA_2024} suggested a couple of methods for determining whether a system is in equilibrium or not. First, is to check whether the ITCF is symmetric; though Figs.~\ref{fig:GraphiteITCF20}~(a) and Figs.~\ref{fig:GraphiteITCF160}~(a) indicate this may be difficult in practice.
The second method is to measure the temperature at at least two scattering angles measured simultaneously, with equilibrium systems giving a consistent temperature, while non-equilibrium systems showing very different temperatures. This occurs since a non-equilibrium system does not have a well-defined concept of temperature.

Here we find a very clear signature of non-equilibrium measured at a single scattering angle. Fig.~\ref{fig:ITCFsNE}~(d)--(f) shows the temperatures extracted for the different crystals and beam types for both the equilibrium and non-equilibrium DSFs. Since the DSF is only calculated to $\lvert \omega \rvert \le 68$~eV, we only consider the integration limits $x \le 65$~eV.
Nevertheless, for the equilibrium system, this is clearly a sufficient range to accurately infer the actual temperature.
The non-equilibrium system shows a starkly different behaviour: the inferred temperature simply does not converge as the flat plateau drives the minimum of the ITCF towards smaller $\tau$ (i.e. higher temperatures).

In the spectral range considered here, the form of the non-equilibrium DSF prevents a temperature being inferred at all. While there may be other causes of this behaviour (e.g. the SIF being too inaccurate), if they can be ruled out, the lack of convergence in the temperature may in and of itself may be another indicator of non-equilibrium.

\subsection{Thermometry without the SIF\label{sec:noSIF}}

\begin{figure*}
    \centering
    \includegraphics[width=\linewidth,keepaspectratio]{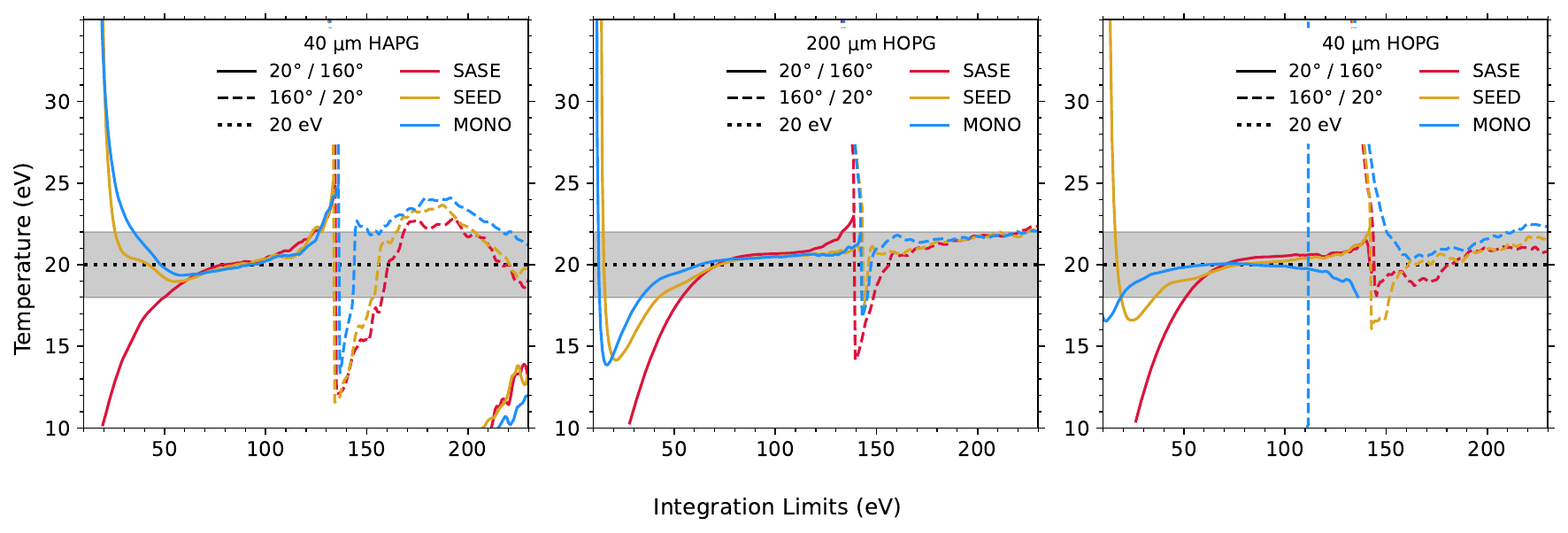}
    \vspace{0.1mm}
    \includegraphics[width=\linewidth,keepaspectratio]{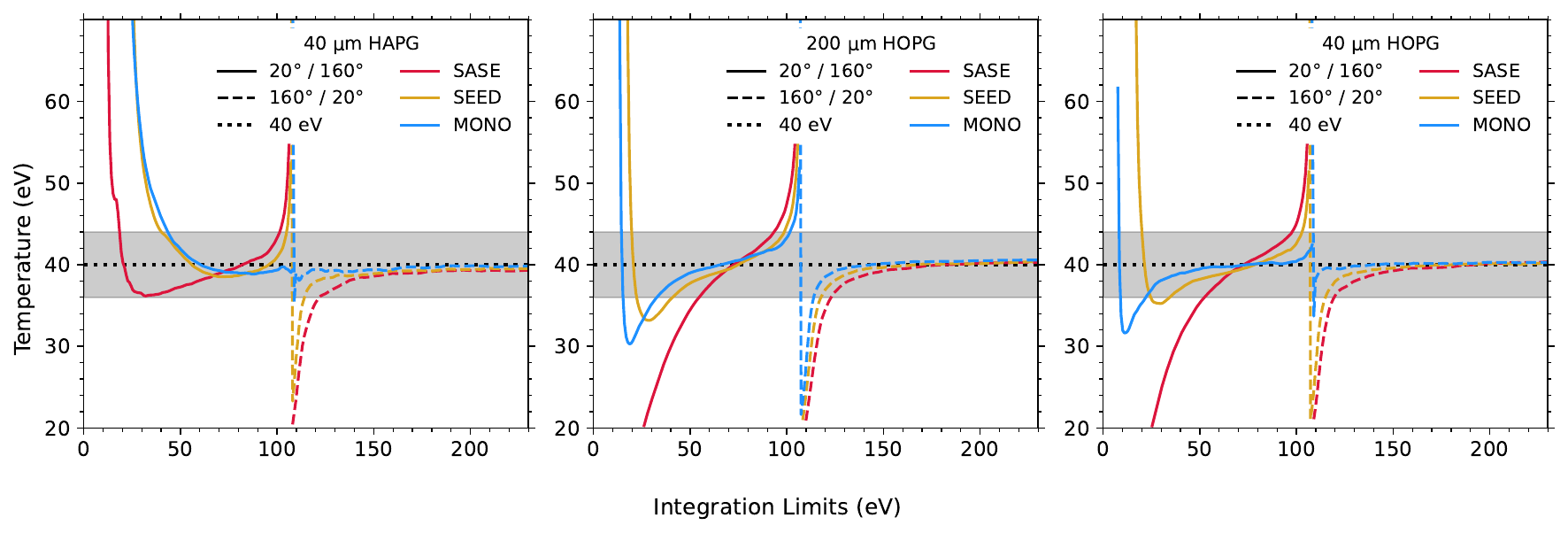}    
    \caption{
    Inferred temperatures versus integration limits using the ITCF ratios method for the SASE (red), seeded (yellow), and monochromated (blue) beams. The temperature is inferred from the minimum of the ratio. The solid lines indicate the minimum for the $20^\circ / 160^\circ$ ratio, and the dashed line for the $160^\circ / 20^\circ$ ratio. The top three plots are for $T_e=20$~eV, while the bottom three are for $T_e=40$~eV. The black dotted line indicates the exact temperature, and the gray region indicates $\pm10\%~T_e$.
    }
    \label{fig:ITCF_Ratio}
\end{figure*}

The ITCF methodology described so far is an in-principle model-free approach to analysing XRTS spectra. However, performing a fully model-free deconvolution assumes the SIF of the experiment is known exactly.
In practice, this is challenging to accomplish. First, as discussed already, the SIF is a simplification to the actual source and instrument function broadening. 
Second, is that measuring a SIF is difficult. Due to its geometry-dependence, it needs to be measured in every experiment, but there are only a few methods to do so.

At XFEL facilities, the XFEL beam spectrum can be measured with high resolution up- and downstream of the target, so the source broadening is well-characterised.
The spectrometer instrument function needs to be inferred from features with known lineshapes; e.g. emission lines or the quasi-elastic feature. However, the fact that the instrument function can have features covering several orders of magnitude (see Fig.~\ref{fig:RayTraceSIFs}), and the presence of other features (e.g. satellite emission, or inelastic scattering features) make it difficult to fully and accurately characterise.

For laser facilities, generally neither the source spectrum nor the instrument function are measured on a shot-to-shot basis. Indeed, the analysis of XRTS spectra from laser facilities can involve fitting the SIF (e.g. which emission lines are present, and their widths) to the quasi-elastic feature of the XRTS spectrum being analysed.

Generally, the SIF is estimated using simple functions such as Voigt profiles~\cite{Tilo_Nature_2023} or Gaussians~\cite{Martin_POP_2025}. While ray tracing simulations can provide more physical SIFs, these are nevertheless still models with built-in assumptions on how x-rays interact with the spectrometer.
All this invariably means that modelling can enter the model-free ITCF method via the SIF.

Fortunately, there is a way to bypass the SIF in the ITCF analysis. From Eq.~(\ref{eq:deconvolution}), it is clear that if two XRTS spectra are recorded simultaneously at two different scattering vectors, and they share the same SIF $R(E)$, then the ratio of the Laplace transformed spectra is equivalent to the ratio of their ITCFs~\cite{Gawne_ITCF_Ratio}:
\begin{equation}\label{eq:ratio}
    \frac{\mathcal{L}\left[S_{ee}(\mathbf{q}_1,E)\circledast R(E)\right]}{\mathcal{L}\left[S_{ee}(\mathbf{q}_2,E)\circledast R(E)\right]} = \frac{F_{ee}(\mathbf{q}_1,\tau)}{F_{ee}(\mathbf{q}_2,\tau)}\ .
\end{equation}
In effect, a deconvolution is performed here without needing explicit knowledge of the SIF.
Moreover, it is obvious from Eq.~(\ref{eq:ITCF_symmetry}) that the ratio~(\ref{eq:ratio}) fulfills the same symmetry condition, attaining either a maximum or a minimum at $\tau=\beta/2$ in thermal equilibrium.
Whether it is a maximum or minimum depends on the relative curvature of the numerator and denominator.
Therefore, the temperature can be extracted from the experiment, without needing to include the SIF.

For the demonstration here, we extract the temperature using the minimum of $20^\circ / 160^\circ$ ratio and the $160^\circ / 20^\circ$ ratio (equivalent to the maximum of the $20^\circ / 160^\circ$ ratio). These are shown in the top plots of Fig.~\ref{fig:ITCF_Ratio} for the spectra in Fig.~\ref{fig:GraphiteSpectra}.
For the HAPG crystal, it is still difficult to identify a clear region of convergence. The 40~$\mu$m HOPG crystal looks more stable than the HAPG crystal, though convergence is not fully established.
The 200~$\mu$m HOPG crystal shows some short regions of convergence for the three beams for the $20^\circ / 160^\circ$, thought the temperature is slightly overestimated by $\sim1$~eV.

This performance appears to be worse than the standard ITCF approach in Fig.~\ref{fig:Graphite20eV}. The main challenge is that there needs to be a sufficient number of detected upshifted photons in both the spectra in order to detect detailed balance. Compared to having a known SIF and a single spectrum, this means the ratios method has more stringent requirements on the photometrics.

To demonstrate the effectiveness of the method when more upshifted photons are detected, we performed identical simulations with the same parameters, except that $T_e = 40$~eV. The higher temperature means a larger portion of the photons will be upshifted. The inferred temperatures are shown in the bottom three plots of Fig.~\ref{fig:ITCF_Ratio}. All three crystals now show good convergence to the correct temperature.

Recent work~\cite{Gawne_ITCF_Ratio} has examined the robustness of the ITCF ratio thermometry to a variety of experimental considerations. The main conclusion is that provided sufficient upshifted photons are detected, the method is quite robust to differences in the spectrometer geometries and the crystal properties. The ratios method therefore a provides a truly model-free approach to analysing XRTS spectra, and is realisable in experiments.

Furthermore, the ratio of two ITCFs contains more information than just the temperature. For example, the proper normalisation of the ratio, the ratio of the SSFs, and the Rayleigh weights would also be contained. Future works should therefore explore how these properties may be extracted.
Moreover, disagreements in temperatures between the ratios of three or more ITCFs would also be a strong indicator of non-equilibrium, if the ratios even converge on a temperature. Conversely, using at least three scattering angles provides a consistency check on the temperature even in equilibrium, and can be used to reduce uncertainties in the inferred temperature~\cite{Gawne_ITCF_Ratio}.

\section{Summary and Outlook\label{sec:summary}}

In this work, we have presented a comprehensive overview of the model-free analysis of XRTS experiments with extreme states of matter in terms of the recently proposed ITCF framework.
The key advantage of the ITCF method is the combination of the remarkable stability of the two-sided Laplace transform $\mathcal{L}\left[\dots\right]$ [cf.~Eq.~(\ref{eq:ITCF})] with the deconvolution theorem (\ref{eq:deconvolution}), which allows one direct access to the physics information of interest in the form of $F_{ee}(\mathbf{q},\tau)$.
Indeed, the latter contains, by definition, exactly the same information as the more familiar dynamic electron structure factor $S_{ee}(\mathbf{q},\omega)$, only in a different representation.
Dornheim \textit{et al.}~\cite{Dornheim_MRE_2023} have argued that $F_{ee}(\mathbf{q},\tau)$ might even be the preferable observable to extract temperature, static density response and the degree of non-equilibrium, whereas information about the width of a particular spectral feature---e.g., the plasmon damping---hardly affect the ITCF but are obvious in $S_{ee}(\mathbf{q},\omega)$.

Since its original presentation as a model-free thermometry technique in 2022~\cite{Dornheim_T_2022,Dornheim_T2_2022}, the ITCF method has been further developed to also yield the normalization $S_{ee}(\mathbf{q})=F_{ee}(\mathbf{q},0)$ and the Rayleigh weight $W_R(\mathbf{q})$ via the f-sum rule~\cite{Dornheim_SciRep_2024,Dornheim_POP_2025}, the static linear density response function $\chi_{ee}(\mathbf{q},0)$ via the imaginary-time version of the fluctuation--dissipation theorem~\cite{schwalbe2025staticlineardensityresponse}, and the degree of non-equilibrium by quantification of the violation of the symmetry relation (\ref{eq:ITCF_symmetry})~\cite{Vorberger_PLA_2024,Bellenbaum_APL_2025}.
Subsequently, Gawne \textit{et al.}~\cite{Gawne_ITCF_Ratio} have even eliminated the need for any knowledge about the SIF if suitable XRTS data are obtained for multiple scattering angles.
Despite being a relative newcomer to the field, the ITCF method has already been used for the interpretation of a diverse set of measurements, both at state-of-the-art XFEL facilities~\cite{Smid_SciRep_2026,Dornheim_SciRep_2024,bohme2026correlationfunctionmetrologywarm,Dornheim_T2_2022,Dornheim_T_2022,Bellenbaum_APL_2025} and more traditional backlighter facilities~\cite{Dornheim_NatComm_2025,schwalbe2025staticlineardensityresponse,Schoerner_PRE_2023,Dornheim_T2_2022,Dornheim_T_2022,shi2025firstprinciplesanalysiswarmdense,Dornheim_POP_2025}.

\begin{figure}
    \centering
    \includegraphics[width=0.99\linewidth]{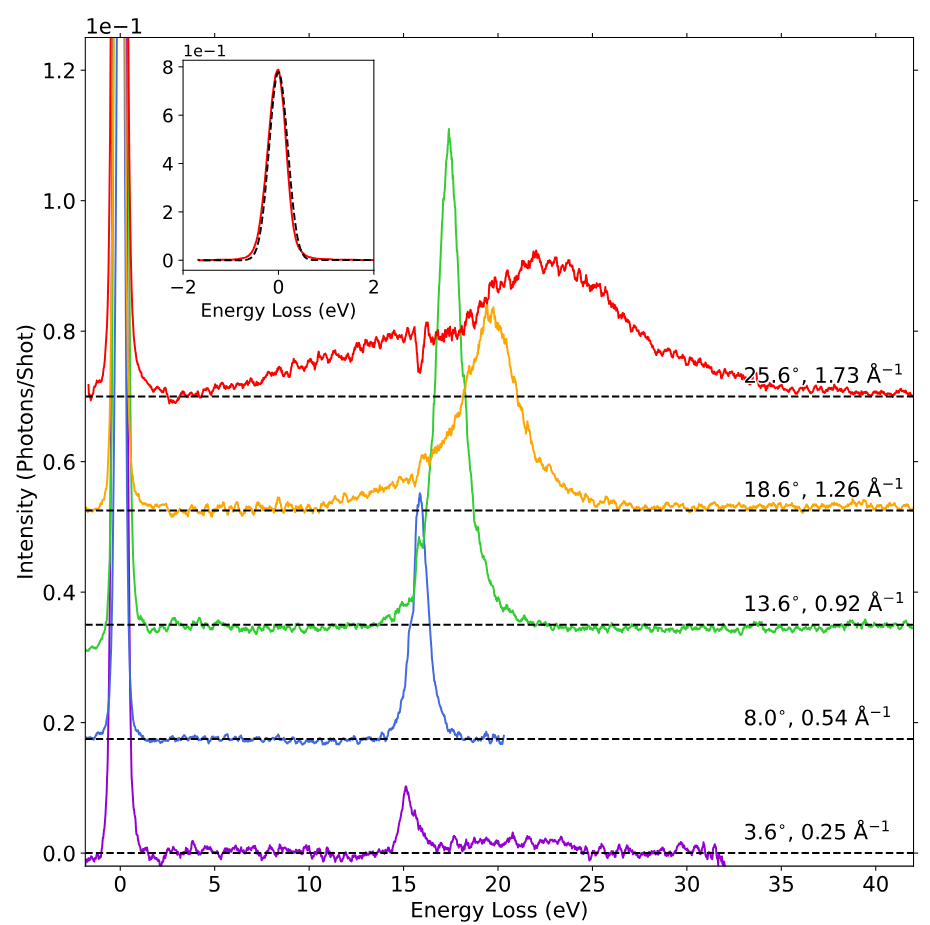}
    \caption{XRTS plasmon spectra of aluminum at ambient conditions measured at the European XFEL for a range of scattering angles $\theta=3.6^\circ-25.6^\circ$. Using a diced crystal analyzer set-up, it is feasible to achieve a spectral resolution of $\sim 10\,$meV over electronic energy scales of several tens of eV.
    Taken from Ref.~\cite{Gawne_PRB_2024} with the permission of the authors.
    \label{fig:DCA}}
\end{figure}

In the following, we will conclude this review with a brief overview of the relevant limitations and of potentially rewarding avenues for impactful future work.

The most stifling limitation of the original thermometry approach is the broadening due to the SIF $R(\omega)$, which limits its applicability to relatively high temperatures, $T\gtrsim 5\,$eV.
In this regard, the advent of XRTS set-ups with diced crystal analyzers (DCAs)~\cite{McBride_RSI_2018,Wollenweber_RSI_2021,Gawne_PRB_2024,Gawne_ElectronicStructure_2025,gawne2025orientationaleffectslowpair} have the potential to be a true game changer.
Indeed, White \textit{et al.}~\cite{White_Nature_2025} have used a DCA set-up to carry out XRTS measurements with a $\sim\,$meV resolution in backscattering geometry to infer the ion temperature from the Doppler broadening of the ion feature.
In addition, Gawne \textit{et al.}~\cite{Gawne_PRB_2024} have demonstrated that it is possible to resolve the plasmon in ambient aluminum (as well as silicon~\cite{Gawne_ElectronicStructure_2025}) with a resolution of $\sim10\,$meV over a spectral window of tens of eV; see Fig.~\ref{fig:DCA} for corresponding result for a range of relevant scattering angles $\theta$.
The combination of this set-up with the unprecedented capabilities for high-repetition pump-probe experiments, e.g., at the high energy density instrument at the European XFEL~\cite{Zastrau_JSI_2021} has the potential to extend the model-free ITCF thermometry approach to electronic temperatures on the order of $T\sim1\,$eV, making it substantially more relevant for laboratory astrophysics, material science, and for the study of ICF ablator materials during the initial segment of the compression path~\cite{vorberger2025roadmapwarmdensematter}.

\begin{figure}
    \centering
    \includegraphics[width=0.99\linewidth]{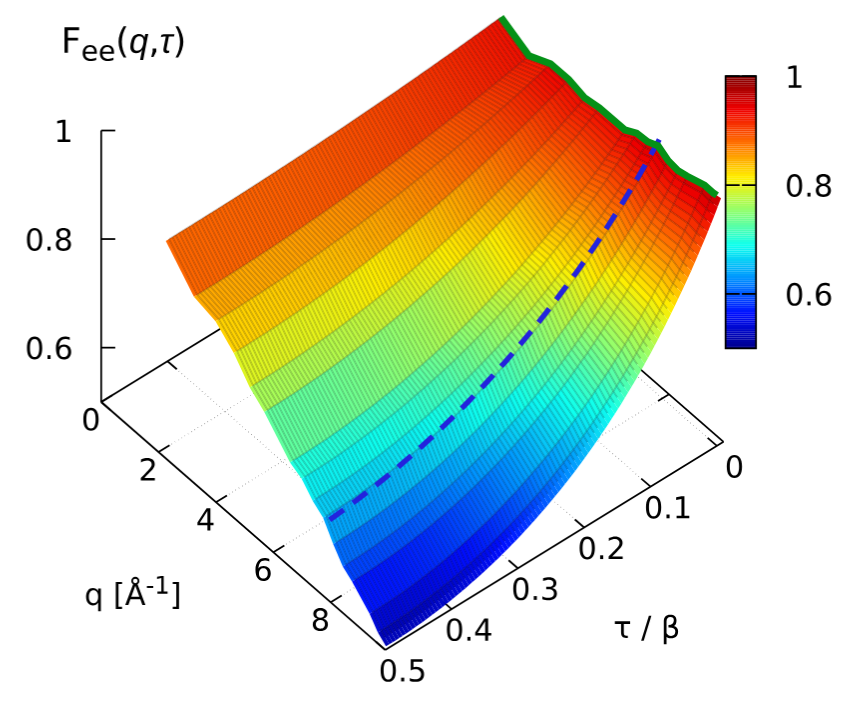}
    \caption{Full wavenumber- and frequency-dependence of the ITCF $F_{ee}(\mathbf{q},\tau)$ of warm dense beryllium. Colored surface: PIMC; dashed blue: NIF experiment~\cite{Tilo_Nature_2023}.
    Taken from Ref.~\cite{Dornheim_NatComm_2025} with the permission of the authors.
    \label{fig:3DITCF}}
\end{figure}

Even though the ITCF method is model-free per se, it still benefits from emerging improved modelling capabilities in a variety of ways.
For example, improved models for $S_{ee}(\mathbf{q},\omega)$~\cite{bohme2023evidencefreeboundtransitionswarm,Baczewski_PRL_2016,Moldabekov_MRE_2025,Moldabekov_filter,Bellenbaum_xdave} are important input for raytracing (and in the future potentially also even-generation~\cite{Uwe_Events_2026}) simulations that provide realistic synthetic data sets.
These are indispensable to further explore the limitations and capabilities of existing approaches and set-ups, and, in particular, to suggest new set-ups that are explicitly designed with the ITCF method in mind; the synthetic analysis of a hypothetical $40\,$$\mu$m HOPG crystal set-up presented in Sec.~\ref{sec:RT_Examples}) constitutes a case in point.
In addition, improved models for $S_{ee}(\mathbf{q},\omega)$ will also be important to explore hitherto unaddressed challenges, such as inhomogeneity effects in extended targets~\cite{Thiele_PRE_2010,Golovkin_HEDP_2013,Chapman_POP_2014,Poole_PPCF_2025}.
On the other hand, we note that the ITCF method has also inspired conceptual improvements in the other direction, such as the recent enhancement of linear-response TDDFT simulations that utilizes $\mathcal{L}\left[\dots\right]$ as a robust and physically motivated convergence criterion~\cite{Moldabekov_filter}.

Finally, we mention the important ongoing interplay between the ITCF method and emerging \emph{ab initio} PIMC simulation capabilities~\cite{Dornheim_MRE_2024,Dornheim_NatComm_2025,schwalbe2025staticlineardensityresponse}. 
As mentioned above, PIMC simulations give one access to the full ITCF, see Fig.~\ref{fig:3DITCF} for corresponding results for strongly compressed beryllium measured at the NIF~\cite{Dornheim_NatComm_2025,Tilo_Nature_2023}.
First, the availability of highly accurate PIMC results has been pivotal for the development of many aspects of the ITCF method, such as the idea of extracting the frequency moments of $S_{ee}(\mathbf{q},\omega)$ directly from the TICF~\cite{Dornheim_moments_2023} or the idea to use Eq.~(\ref{eq:static_chi}) to obtain $\chi_{ee}(\mathbf{q},0)$~\cite{Dornheim_MRE_2023}. 
Extending these ideas, e.g., to higher-order moments of the dynamic structure factor is an important topic for future work even though achieving convergence might be challenging in practice.
Second, PIMC results for $F_{ee}(\mathbf{q},\tau)$ have already given us important insights into the manifestation of physical effects onto the ITCF, see, e.g., the recent works onto the roton-type feature of the strongly coupled uniform electron gas~\cite{Dornheim_MRE_2023,Chuna_JCP_2025,Chuna_PRB_2025} for a particularly relevant example.
Further important insights come from the burgeoning field of analytic continuation, i.e., the numerical inversion of the two-sided Laplace transform to reconstruct $S_{ee}(\mathbf{q},\omega)$ from $F_{ee}(\mathbf{q},\tau)$, where recent developments~\cite{dornheim_dynamic,Chuna_PRB_2025,BENEDIXROBLES2026109904,chuna2025noiselesslimitimprovedpriorlimit} hint at the enticing possibility to obtain approximation-free results for the dynamic structure factor of real materials.




\section*{Data Availability}

All new data will be made freely available upon publication.





\bibliography{sn-bibliography}

\section*{Acknowledgments}
This work has received funding from the European Union's Just Transition Fund (JTF) within the project \emph{R\"ontgenlaser-Optimierung der Laserfusion} (ROLF), contract number 5086999001, co-financed by the Saxon state government out of the State budget approved by the Saxon State Parliament. This work has received funding from the European Research Council (ERC) under the European Union’s Horizon 2022 research and innovation programme (Grant agreement No. 101076233, "PREXTREME"). 
Views and opinions expressed are however those of the authors only and do not necessarily reflect those of the European Union or the European Research Council Executive Agency. Neither the European Union nor the granting authority can be held responsible for them.
Tobias Dornheim gratefully acknowledges funding from the Deutsche Forschungsgemeinschaft (DFG) via project DO 2670/1-1.

\section*{Author Contributions Statement}

TG and TD have contributed equally to this work.

TG carried out all raytracing simulations and wrote substantial parts of the ms.
TD has led the conceptualization of this work and wrote substantial parts of the ms.
JV has computed non-equilibrium dynamic structure factors and wrote substantial parts of the ms.
ZM and HMB have written selected sections of the ms.
All authors contributed to the review and finalization process.

\section*{Competing Interests Statement}

The authors declare no competing interests.

\end{document}